\documentclass[
    aps,
    prl,
    twocolumn,
    10pt,
    a4paper,
    superscriptaddress,
    notitlepage,
    longbibliography,
    floatfix
]{revtex4-2}

\usepackage{amsmath}
\usepackage{amssymb}
\usepackage{textcomp}
\usepackage{nicefrac}
\usepackage{graphicx}
\usepackage{microtype}
\usepackage{enumitem}
\usepackage{xcolor}
\usepackage{multirow}   
\usepackage{physics}
\usepackage{siunitx}
\usepackage{booktabs}
\usepackage{csquotes}

\usepackage{hyperref}
\hypersetup{colorlinks=true, citecolor=blue, urlcolor=blue, linkcolor=blue}

\usepackage{txfonts}

\newcommand{\vect}[1]{\ensuremath{\boldsymbol{#1}}}

\newcommand{\mub}{\mu_\mathrm{B}}
\newcommand{\kb}{k_\text{B}}

\newcommand{\mathpi}{\piup}

\newcommand{\hamil}{\mathcal{H}}

\newcommand{\unitvect}[1]{\hat{\vect{#1}}}

\usepackage{pifont}

\newcommand{\mode}[1]{\textbf{(#1)}}

\begin{document}
\title{Electrical Activity of Topological Chiral Edge Magnons}

\author{Robin R.~Neumann}
\affiliation{Institut f\"ur Physik, Martin-Luther-Universit\"at Halle-Wittenberg, D-06099 Halle (Saale), Germany}
\author{J\"urgen Henk}
\affiliation{Institut f\"ur Physik, Martin-Luther-Universit\"at Halle-Wittenberg, D-06099 Halle (Saale), Germany}
\author{Ingrid Mertig}
\affiliation{Institut f\"ur Physik, Martin-Luther-Universit\"at Halle-Wittenberg, D-06099 Halle (Saale), Germany}
\author{Alexander Mook}
\affiliation{Institut f\"ur Physik, Johannes Gutenberg-Universit\"at Mainz, D-55128 Mainz, Germany}

\begin{abstract}
Topological magnon insulators support chiral edge excitations, whose lack of electric charge makes them notoriously difficult to detect experimentally.
We show that relativistic magnetoelectric coupling universally renders chiral edge magnons electrically active, thereby facilitating \emph{electrical} probes of magnon topology.
Considering a two-dimensional out-of-plane magnetized topological magnon insulator, we predict a fluctuation-activated electric polarization perpendicular to the sample edges.
Furthermore, the chiral topological electromagnons give rise to a unique in-gap signal in electrical absorption experiments.
These results suggest THz spectroscopy as a promising probe for topological magnons.
\end{abstract}

\date{\today}

\maketitle

\paragraph{Introduction.}
Topology has become a key concept in condensed matter physics, with the quantized Hall conductance being a prominent example
\cite{
    klitzing_new_1980,
    haldane_model_1988%
}.
Although topological band structure theory can be carried over to magnons~%
\cite{
    matsumoto_theoretical_2011,
    zhang_topological_2013,
    shindou_topological_2013,
    mook_edge_2014,
    owerre_first_2016,
    kim_realization_2016%
}, i.e.\ the elementary excitations of magnetic order~\cite{bloch_zur_1930}, their bosonic statistics does not give rise to quantized transport~%
\cite{
    tao_quantum_1986,
    katsura_theory_2010,
    onose_observation_2010,
    matsumoto_thermal_2014,
    hirschberger_thermal_2015,
    nakata_magnonic_2017,
    mook_taking_2018,
    neumann_thermal_2022%
}.
Furthermore, magnons lack electric charge, which, although being an attractive trait for technologies free of Joule heating \cite{Chumak2015}, renders them ``dark'' in charge-probing spectroscopies.
In addition, inelastic neutron scattering, which is the conventional probe of magnons, reveals their bulk band gaps, but fails to detect edge states~%
\cite{
    chisnell_topological_2015,
    zhu_topological_2021%
}.
In short, the state of the art does not offer an appropriate tool for the detection of topological magnons and new ideas are needed
\cite{Perreault2016,malz_topological_2019,Feldmeier2020,Rustagi2020,Guemard2022, Hetenyi2022, Mitra2023, Bostrom2023}.

Despite their charge neutrality, magnons can respond to external electric fields. Those electrically active magnons, so-called \textit{electromagnons}, have been studied by THz spectroscopy~%
\cite{
    moriya_far_1966,
    moriya_theory_1968,
    moriya_light_1970,
    pimenov_possible_2006,
    pimenov_terahertz_2008,
    kida_electrically_2008%
},
by magnon-photon coupling in cavities~%
\cite{
    hirosawa_magnetoelectric_2022,
    Curtis2022,
    toklikishvili_electrically_2023%
},
and by parametric amplification of topological magnons~\cite{malz_topological_2019}.
The experimental proof of principle for driving magnons electrically has already been provided~\cite{kubacka_large-amplitude_2014}.

Herein, we investigate the electrical activity of topological chiral edge magnons in ferromagnets in order to explore their spectroscopic signatures.
Knowing that a flow of magnons induces electric fields \cite{meier_magnetization_2003, schutz_persistent_2003} by virtue of the vacuum magnetoelectric (VME) effect~%
\cite{
    hirsch_overlooked_1999,
    hnizdo_magnetic_2012,
    griffiths_mansuripurs_2013%
},
we first show that a flow of chiral edge magnons universally causes an electric edge polarization. 
Motivated by this result, we consider the Katsura-Nagaosa-Balatsky (KNB) mechanism~\cite{katsura_spin_2005} to study the electric polarization of a two-dimensional topological magnon insulator (TMI) and to disentangle contributions from topologically trivial and nontrivial magnons.
Second, we investigate the response of topological magnons to external alternating electric fields in TMI nanoribbons and flakes, in  which edge magnons cause  electric absorption peaks within the magnon bulk band gap.
Our results suggest that chiral edge magnons are electrically active and that terahertz spectroscopy could evidence their existence experimentally. 

\paragraph{Intuitive expectations.}
\begin{figure}
    \centering
    \includegraphics[width=\linewidth]{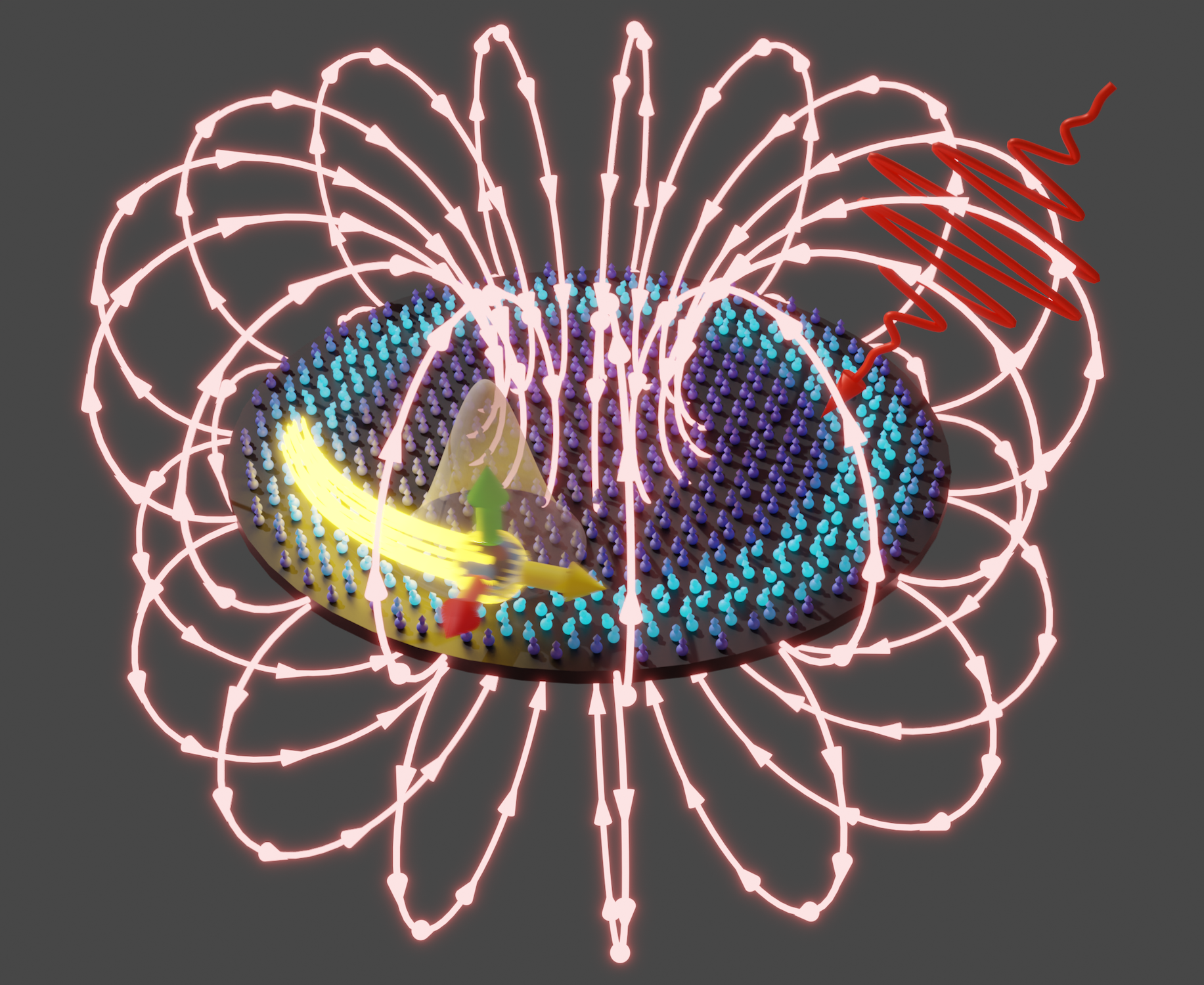}
    \caption{
        Propagating chiral edge magnon in a disk-shaped topological magnon insulator.
        Arrows indicate velocity (yellow), magnetic dipole moment (green), and electric dipole moment (red, due to the vacuum magnetoelectric effect) of the wave packet (transparent sphere).
        The electric field (light red lines) follows from the generalized Biot-Savart law.
        The dark (light) blue arrows represent localized spins in their ground state (excited) state.
        The red oscillating curve illustrates an external alternating electric field, which excites and probes chiral edge magnons.
    }
    \label{fig:abstract}
\end{figure}

\begin{figure}
    \centering
    \includegraphics[width=\linewidth]{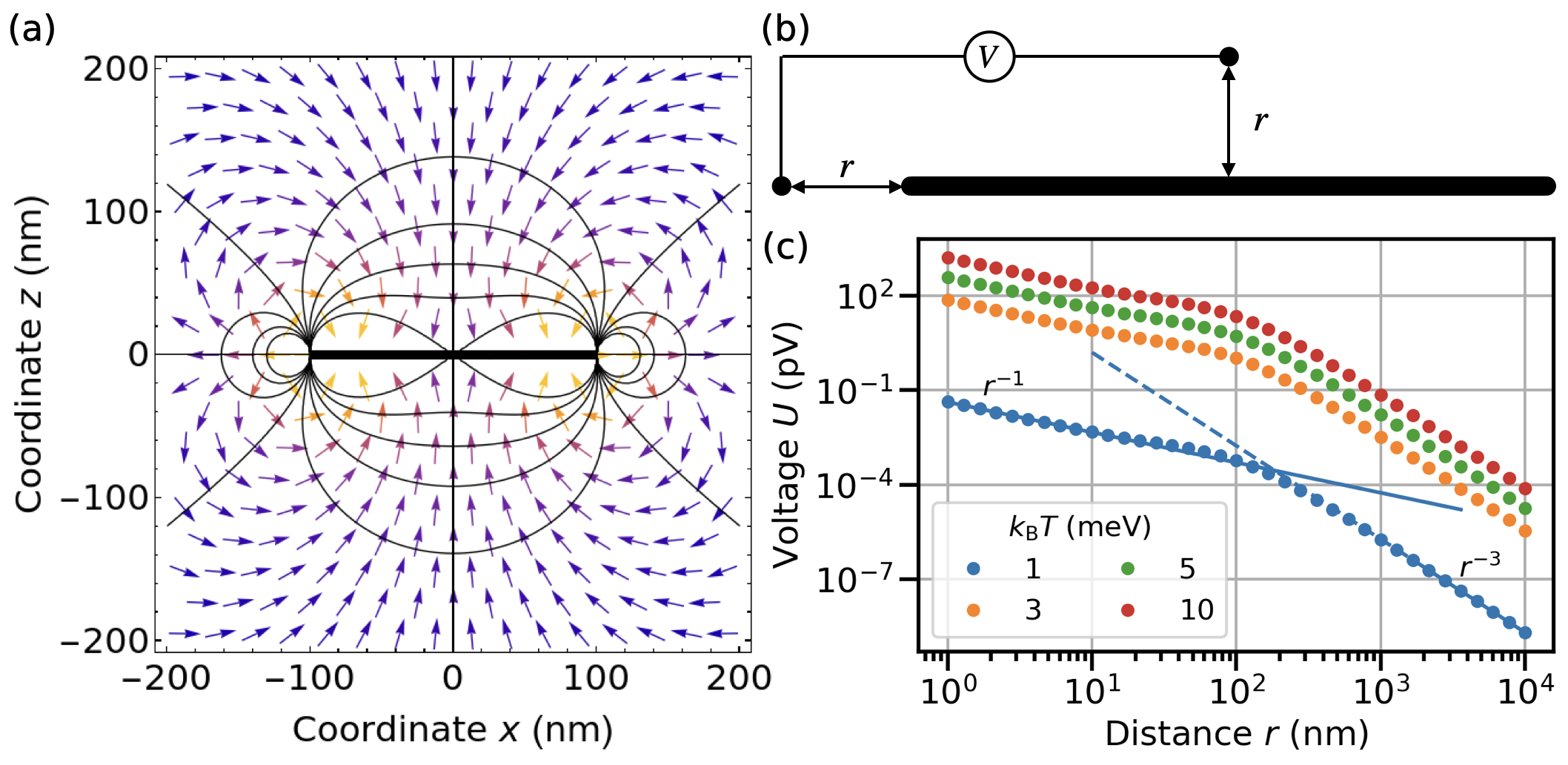}
    \caption{Electric field of a magnetic disk in the $xy$ plane and with a radius of $\SI{100}{\nano\meter}$.
        (a)~Field distribution in the $xz$-plane ($y = 0$).
        Arrows and arrow colors indicate directions and magnitude of the electric field, respectively, while black lines indicate equipotential lines.
        The disk is indicated by the black rectangle at $z = 0$.
        (b)~Schematic of a voltage measurement setup.
        (c)~Estimated distance dependence of the voltage due to the vacuum magnetoelectric effect of topological magnons for the setup of panel~(b).
        Dots represent numerical data, while straight and dashed lines are linear fits to $U(r)$ for $\kb T = \SI{1}{\milli\electronvolt}$ ($\SI{11.6}{\kelvin}$).
        For parameters, see text.
    }
    \label{fig:acfield}
\end{figure}
Magnons carry a magnetic moment~$\vect{m}$, giving rise to a relativistic electric dipole $\vect{p} = \vect{v} \times \vect{m} / c^2$ ($\vect{v}$ magnon velocity, $c$ speed of light), which is the VME effect resulting from Lorentz transformation from the magnon's rest frame to the lab frame~%
\cite{
    hirsch_overlooked_1999,
    meier_magnetization_2003,
    hnizdo_magnetic_2012,
    griffiths_mansuripurs_2013%
}.
For chiral edge magnons in two-dimensional and out-of-plane magnetized TMI, $\vect{m} \parallel \hat{\vect{z}}$ and $\vect{p} \parallel \vect{v} \times \hat{\vect{z}}$ points along the local edge normal, as indicated in Fig.~\ref{fig:abstract}.
The sign of $\vect{p}$ depends on the magnetization direction and on the velocity (hence, on the chirality) of the edge magnons.
The chiral magnon edge current causes an electric field $\vect{E} = -\grad \phi$, whose scalar potential 
\begin{align}
    \phi(\vect{r})
    =
    \frac{\mu_0 I_\text{m}}{4 \mathpi}
    \oint 
    \qty[
        \dd{\vect{r}'}
        \times
        \unitvect{m}(\vect{r}')
    ]
    \cdot
    \frac{\vect{r} - \vect{r}'}{\abs{\vect{r} - \vect{r}'}^3},
\end{align}
is obtained from a generalized Biot-Savart law \cite{meier_magnetization_2003, schutz_persistent_2003} ($\mu_0$ vacuum permeability, $I_{\text{m}}$ magnetization current, $\unitvect{m}$ direction of the magnetic dipole).

For a magnetic current carried by chiral magnons on a circular trajectory of radius $R$, we write $\vect{r} = \rho \unitvect{e}_\rho(\varphi) + z \unitvect{e}_z$ in cylindrical coordinates; $\unitvect{e}_z$ ($\unitvect{e}_\rho$) is out-of-plane (radial) to the magnetic current loop.
If $R \ll \rho$, we approximate
\begin{align}
    \phi(\rho, \varphi, z)
    \approx
    \frac{\mu_0 I_{\text{m}}}{4}
    \frac{
        R^2 (\rho^2 - 2 z^2)
    }{
        (\rho^2 + z^2)^{\nicefrac{5}{2}}
    }
    ,
\end{align}
while for $\rho = 0$ the exact potential reads
\begin{align}
    \phi(0, \varphi, z)
    =
    -\frac{\mu_0 I_{\text{m}}}{2}
    \frac{R^2}{(R^2 + z^2)^{\nicefrac{3}{2}}}
\end{align}
[see Supplemental Material (SM)~\cite{supplement}].
Thus, the potential drops with distance as $z^{-3}$ in the far field limit.
The sign of $I_{\text{m}}$, i.e., the chirality of the edge magnons, determines the direction of the electric field $\vect{E}$. As expected, the $\vect{E}$ field lines point radially outward from the edge of the disk and resemble a dipole field close to the edge [see Fig.~\ref{fig:acfield}(a)]. The largest electric field is found in the vicinity of the edges hosting the chiral edge magnons.

In order to estimate $\vect{E}$, we compute
$
    I_{\text{m}}
    =
    \frac{g \mub v}{2 \mathpi}
    \int_{-\mathpi/a}^{\mathpi/a} \dd{k} \rho(\varepsilon_{k})
$
($g$ Land\'{e} $g$-factor, $\mub$ Bohr magneton, $a$ lattice constant, $v$ edge magnon's group velocity). The occupation is given by 
$
    \rho(\varepsilon)
    =
    [\exp(\varepsilon/(k_\text{B} T))-1]^{-1}
$
($k_\text{B}$ Boltzmann constant, $T$ temperature).
We assume $a = \SI{1}{\nano\meter}$ and, in order to describe van-der-Waals magnets \cite{zhu_topological_2021, chen_magnetic_2021}, $v = \SI{1000}{\meter\per\second}$, $\varepsilon_k = \hbar v k + \SI{12}{\meV}$ (i.e., the topological band gap is \SI{4}{\meV}).
The voltage $U(r) = \phi(r, 0, 0) - \phi(0, 0, r)$ between two leads at a distance $r$ from the edge and the center of the disk is evaluated numerically [Fig.~\ref{fig:acfield}(b)].  Its $r$-dependence, shown in Fig.~\ref{fig:acfield}(c) for selected temperatures,
exhibits two regimes: $r \ll R$ with a $r^{-1}$-dependence and $r \gg R$ with a $r^{-3}$-dependence. The crossover is around $R =\SI{100}{\nano\meter}$.
These results suggest that a nanovolt-sensitive measurement could prove the existence of chiral edge magnons.
We now contrast this intuitive picture with a microscopic theory.

\paragraph{Microscopic theory.}

\begin{figure*}
    \centering
    \includegraphics[width=\linewidth]{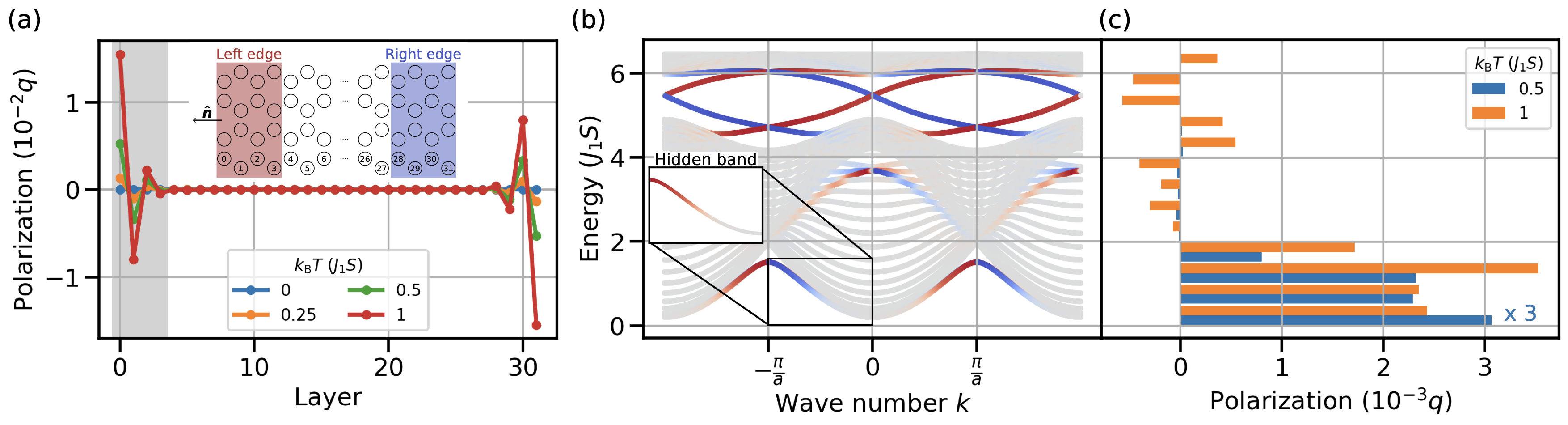}
    \caption{Electric polarization and magnons of a nanoribbon in armchair nanoribbon geometry with 32 layers.
        (a)~Layer-resolved electric polarization due to the Katsura-Nagaosa-Balatsky effect for various temperatures (as indicated) projected onto the outward-facing in-plane normal vector $\unitvect{n}$ of the left edge.
        Inset: section of the nanoribbon including layer labels and the normal vector $\unitvect{n}$.
        The system is finite (infinite) along the horizontal (vertical) direction.
        (b)~Magnon spectrum. Probability amplitudes of the left (red) and right (blue) edge [defined as the 4 outermost layers at each edge side; cf.~the inset in panel~(a)] are encoded by color.
        Inset: hidden band covered by the lowest band.
        (c)~Energy-resolved contributions of the magnons to the electric polarization at the left edge [highlighted in panel~(a) by a gray background] for two selected temperatures (as indicated).
        Each bar comprises contributions accumulated in an energy interval of $J_1 S / 2$.
        The blue bars are multiplied by 3 for better visibility.
        Parameters read $J_1 = 1, J_2 = 0.25, J_3 = 0, D_z = -0.1, A = 0.1, \text{and } S = 1$.
        Results for other terminations can be found in the SM~\cite{supplement}.
    }
    \label{fig:slab_pol}
\end{figure*}
We consider a two-dimensional TMI on a honeycomb lattice, which is an effective model for van-der-Waals magnets. The Hamiltonian
\begin{align}
    \begin{split}
    \hamil
    &=
    -\sum_{r=1}^3 \frac{J_r}{2 \hbar^2}
    \sum_{{\langle ij \rangle}_r} \vect{S}_i \cdot \vect{S}_j
    +
    \frac{1}{2 \hbar^2}
    \sum_{{\langle ij \rangle}_2}
    \vect{D}_{ij}
    \cdot
    \qty(\vect{S}_i \times \vect{S}_j)
    \\
    & \quad
    -
    \frac{A}{\hbar^2}
    \sum_i
    \qty(S_i^z)^2
    \label{eq:hamil}
    \end{split}
\end{align}
includes Heisenberg exchange interactions $J_r$ up to 3$^\text{rd}$ nearest neighbors, out-of-plane Dzyaloshinskii-Moriya interaction (DMI) $\vect{D}_{ij} = \pm D_z \unitvect{z}$ between 2$^\text{nd}$ nearest neighbors, and out-of-plane anisotropy $A$ ($\hbar$ reduced Planck constant).
In the following we choose relative parameters: $J_1 = 1, J_2 = 0.25, J_3 = 0, D_z = -0.1, A = 0.1,$ and $S = 1$.
The ground state is an out-of-plane collinear ferromagnet.

The model~\eqref{eq:hamil} is known to yield topological magnons in linear spin-wave theory~%
\cite{
    owerre_first_2016,
    kim_realization_2016%
},
which are shown for the armchair nanoribbon geometry in Fig.~\ref{fig:slab_pol}(b).
The in-gap states have positive (negative) group velocity and are localized on the left (right) edge.
The relation between velocity and localization depends on the Chern number, which can be reversed with the magnetization or the sign of $D_z$.

The relativistic electric dipole between two spins at sites $i$ and $j$ reads 
$
    \vect{p}_{ij} = q_{ij} \, \vect{e}_{ij} \times \qty(\vect{S}_i \times \vect{S}_j) / \hbar^2
$
according to the spin-current [or Katsura-Nagaosa-Balatsky (KNB)] mechanism~%
\cite{
    katsura_spin_2005,
    tokura_multiferroics_2014%
}
($q_{ij}$ effective charge, $\vect{e}_{ij}$ bond vector between site $i$ and $j$,  $\vect{S}_{i}$ and  $\vect{S}_{j}$ spin operators).
As shown in the SM \cite{supplement}, the VME and the KNB effects are equivalent for magnons in Heisenberg ferromagnets, but the KNB effect can be around 5 to 6 orders of magnitude larger.

We expand
$
    \vect{p}_{ij}
    =
    \vect{p}_{ij}^{(0)}
    +
    \vect{p}_{ij}^{(1)}
    +
    \vect{p}_{ij}^{(2)}
    +
    \cdots
$
by means of the Holstein-Primakoff transformation~\cite{holstein_field_1940}, where the superscript denotes the number of bosons (explicit expressions for the operators are provided in the SM~\cite{supplement}).
$
    \vect{p}_{ij}^{(0)}
$
is the classical ground state polarization, which is zero in our case.
$
    \vect{p}_{ij}^{(1)}
$
vanishes in equilibrium, but encodes the dynamic electric dipole moment associated with spin dynamics, and the bilinear
$
    \vect{p}_{ij}^{(2)}
$
tells about the expectation value per magnon. 
In equilibrium, the fluctuation-induced
$
    \vect{p}_{ij}^{(2)}
$
is the dominant contribution to the macroscopic polarization. 

Below, we consider the layer-resolved polarization
\begin{align}
    \vect{P}_n
    =
    \frac{q}{\hbar^2 L}
    \sum_{\underset{i \in L_n}{\langle ij \rangle}}
    \vect{e}_{ij}
    \times
    \qty(\vect{S}_i \times \vect{S}_j)
    ,
\end{align}
of layer $n$ in the nanoribbon, which is a sum over all inter-site electric dipole moments $\vect{p}_{ij}$ in that layer ($L_n$ set of all sites in layer $n$, $L$ circumference of the nanoribbon).
The thermal equilibrium expectation value of $\vect{P}_n$ (originating from $\vect{p}_{ij}^{(2)}$) projected onto the in-plane normal vector $\unitvect{n}$ of the left edge, shown in Fig.~\ref{fig:slab_pol}(a), features nonzero values at the edges of the nanoribbon, while it vanishes in the bulk.
Inversion symmetry dictates that the polarizations at opposite edges are antiparallelly oriented.
However, based on the intuitive expectations following from the VME effect and the chirality of the edge magnons [cf.~Fig.~\ref{fig:slab_pol}(b)], one would expect a \emph{negative} projected polarization at the left edge, which is opposite to the numerical results.

This discrepancy is understood by analyzing the energy-resolved contributions to the left-edge polarization
$
    \sum_{n = 0}^3 \vect{P}_{n}
$
[see Fig.~\ref{fig:slab_pol}(c); layers~0--3 are highlighted in Fig.~\ref{fig:slab_pol}(a)].
There exist not only contributions from within the topological band gap, but also much larger ones from energies below the gap.
The latter arise from trivial sub-gap edge modes [see Fig.~\ref{fig:slab_pol}(b)], whose thermal occupation is larger than that of the nontrivial in-gap states.
The existence of trivial sub-gap edge modes is unavoidable: these arise from the weaker effective internal magnetic field for spins at the edges (missing neighbor sites).
The sub-gap states highlighted in the inset of Fig.~\ref{fig:slab_pol}(b) (these are hidden below the lowest band) have a velocity opposite to that of the nontrivial mode localized at the same edge, and, therefore, an opposite electric dipole moment. The contributions of trivial sub-gap modes to
$
    \sum_{n = 0}^3 \vect{P}_{n}
$
dominate over those of the nontrivial in-gap modes at all temperatures. Furthermore, the trivial contributions proved robust against disorder and manipulations of the edges (see SM~\cite{supplement}).

In short, the equilibrium properties of the topological magnons are overshadowed by those of trivial magnons, and the intuitive picture is incomplete.

\paragraph{Absorption of alternating electric fields.}
The above discussion demonstrates the need to go beyond thermal equilibrium, in which sub-gap states are favored over in-gap states.
As we show in the SM~\cite{supplement}, the in-gap states do not respond to alternating magnetic fields.
Therefore, we study the possibility to address resonantly the magnons with alternating \emph{electric} fields $\vect{E}(t)$ by including a perturbation
\begin{align}
    \hamil'
    =
    -V \vect{P} \cdot \vect{E}(t)
    =
    \frac{q}{\hbar^2}
    \sum_{\langle ij \rangle}
    \qty(\vect{e}_{ij} \times \vect{E}(t))
    \cdot
    \qty(\vect{S}_i \times \vect{S}_j)
    \label{eq:pert}
\end{align}
to the Hamiltonian~\eqref{eq:hamil} ($\vect{P}$ total electric polarization). This form suggests that the external electric field induces a time-varying DMI, the corresponding DMI vector of which is out-of-plane (in-plane) for in-plane (out-of-plane) fields.

The linear response of $\vect{P}$ to the perturbation, 
$
    \varDelta \expval{P_\mu(\omega)}
    =
    \chi_{\mu \nu}(\omega) \, E_\nu(\omega)
$,
is quantified by the electric susceptibility $\chi_{\mu \nu}(\omega)$
($\mu, \nu = x, y, z$; implicit summation over $\nu$). 
In Kubo's formalism \cite{kubo_statistical-mechanical_1957, bruus_many-body_2004},
$
    \chi_{\mu\nu}(\omega)
    =
    -V
    C_{P_\mu P_\nu}^{\mathrm{R}}(\omega)
$
is obtained from the retarded polarization autocorrelation function
$
    C_{P_\mu P_\nu}^{\mathrm{R}}(\omega)
$,
which is evaluated in the SM~\cite{supplement}.
There are various types of contributions $\chi^{(i)}_{\mu\nu}$ to $\chi_{\mu\nu}$, among them one- ($i = 1$; leading order derived from $\vect{p}_{ij}^{(1)}$) and two-magnon processes ($i = 2$; leading order derived from $\vect{p}_{ij}^{(2)}$).
While the one-magnon processes are governed by the (out-of-plane or transversal) fluctuations of the magnons' electric dipole moments about their mean value, the two-magnon processes are governed by the bilinear part of the total dipole moment that is also responsible for the finite mean value in equilibrium (longitudinal fluctuations).
Therefore, one-magnon processes appear for out-of-plane, while two-magnon processes appear for in-plane electric fields.
Since according to Eq.~\eqref{eq:pert} only the former induces in-plane DMI, which breaks magnon-number conservation, only one-magnon processes may change the magnon number, while two-magnon processes can only cause interband transitions of thermally excited magnons.
Thus, contrary to $\chi_{\mu\nu}^{(1)}$, $\chi_{\mu\nu}^{(2)}$ can be frozen out.
We therefore focus on one-magnon processes in the rest of the paper, while delegating further details, mathematical expressions, derivations, and results for two-magnon processes to the SM~\cite{supplement}.

The imaginary part 
\begin{align}
    \Im \chi^{(1)}_{\mu\mu}(\omega)
    =
    \mathpi V
    \sum_{n = 1}^{N}
    \abs{\qty(\mathcal{P}^{(1)}_{\mu})_n}^2
    \delta(\hbar \omega - \varepsilon_{n,\vect{k}=\vect{0}})
\end{align}
of the diagonal electric one-magnon susceptibility ($\omega > 0$, $N$ number of bands)
contains the linear electric dipole element
$
    \qty(\mathcal{P}^{(1)}_\mu)_n
$
for component $\mu$ and band $n$, whose general expression is derived in the SM~\cite{supplement}. 
The $\delta$-distribution, which we replace by a Lorentzian of width $\eta$ for numerical calculations, ensures energy conservation, such that resonance frequencies appear at the eigenfrequencies of the system, while only magnons at $\vect{k} = \vect{0}$ can be probed due to momentum conservation.

\begin{figure}
    \centering
    \includegraphics[width=\linewidth]{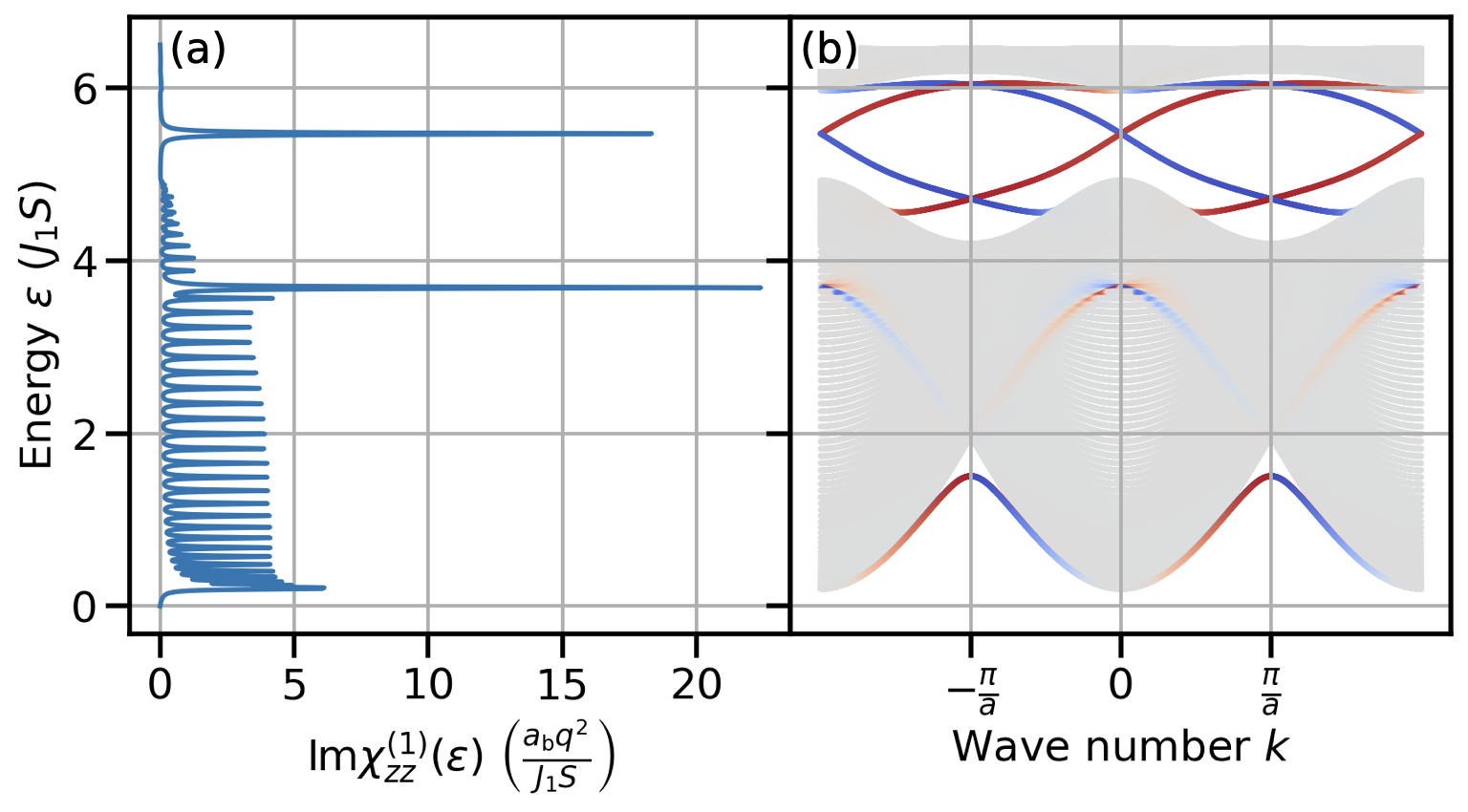}
    \caption{Electric susceptibility of a 120 layers wide nanoribbon with armchair terminations.
        (a)~Imaginary part of the one-magnon susceptibility vs. energy.
        (b)~Magnon spectrum with localization at the left (red) and right (blue) edge encoded by color.
        $a_{\text{b}}$ is the bulk lattice constant of the underlying honeycomb lattice, while $a = \sqrt{3} a_{\text{b}}$ is the lattice constant of the nanoribbon.
        The width of the Lorentzians is $\eta = 0.01 J_1 S$.
        Parameters as for Fig.~\ref{fig:slab_pol}.
        Results for other terminations can be found in the SM~\cite{supplement}.
    }
    \label{fig:slab_plb}
\end{figure}

Returning to the honeycomb model, absorption appears only for $\mu = z$: it shows a pronounced in-gap peak just below $\varepsilon/(J_1 S) = 6$, which is attributed to topological magnons (see Fig.~\ref{fig:slab_plb}).
In the SM~\cite{supplement}, we show that \emph{the absorption only takes place at the edges}, therefore, only modes with nonzero probability amplitude at the edges may contribute.
Bulk modes have nonzero probability amplitudes at the edges as well, but cause peaks above and below the gap.

Together with inelastic neutron scattering, which can locate gaps in the bulk magnon spectrum, in-gap peaks of $\Im \chi^{(1)}_{zz}(\omega)$ could allow to detect topological magnons in principle.
Momentum conservation tells that only topological magnons with $\vect{k} = \vect{0}$ contribute to the signal; however, it is not guaranteed that topological magnons exist at this particular $\vect{k}$ (see SM for a counterexample~\cite{supplement}).
Thus, in-gap peaks are no necessary consequence of topological edge modes.
Furthermore, absorption at a sample's edge could be overshadowed by other sources and hence might be hard to resolve in experiment.

\begin{figure}
    \centering
    \includegraphics[width=\linewidth]{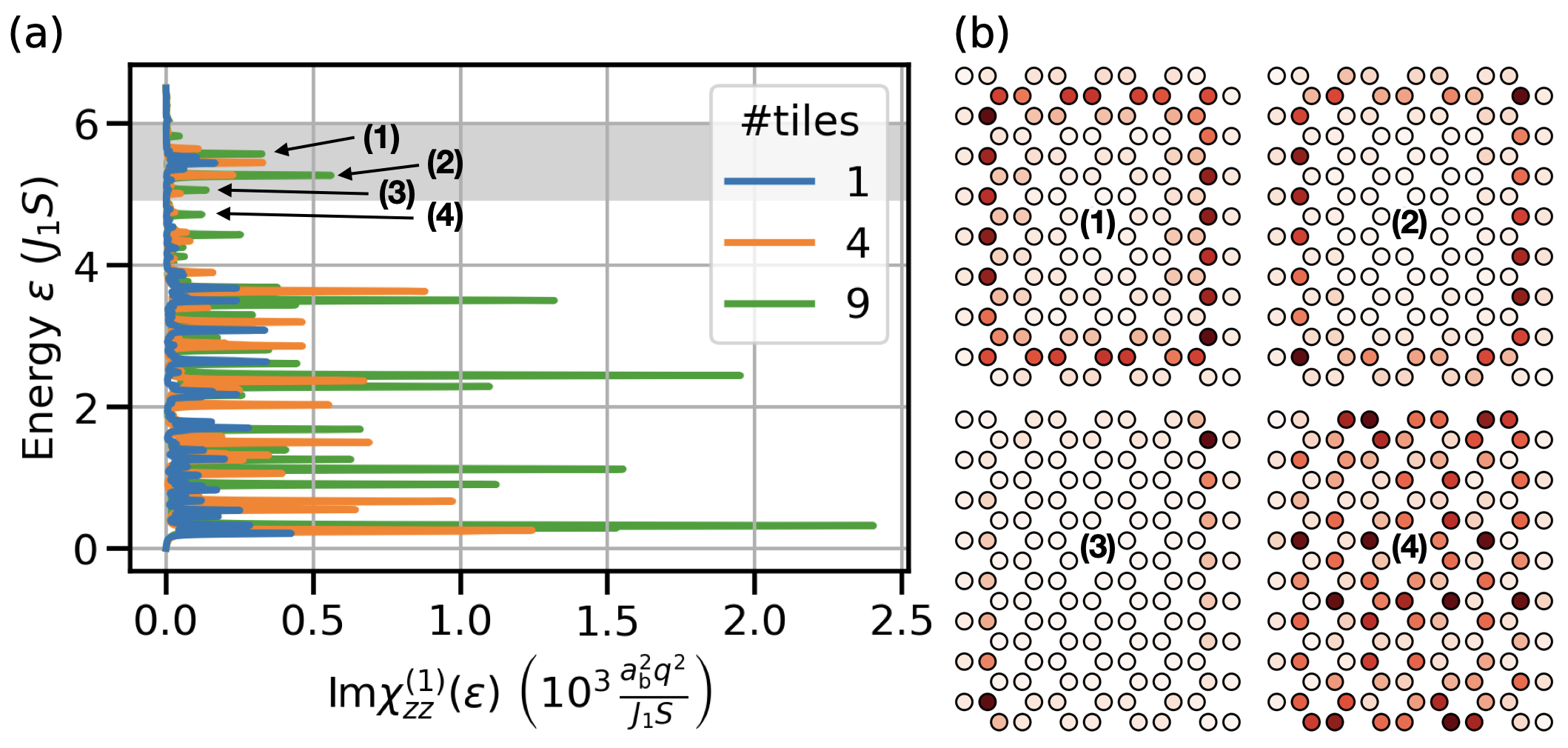}
    \caption{
        Electric susceptibility of a flake of 1152 spins cut into four (nine) smaller equally sized tiles [see legend in panel~(a)].
        (a) Energy-resolved imaginary part of the electric one-magnon susceptibility. 
        For the 9-tile spectrum (green line), selected in-gap resonances \mode{1}--\mode{4} are marked with arrows.
        The gray background indicates the topological bulk band gap.
        (b)~Real-space probability distributions of the four electrically active magnon modes \mode{1}--\mode{4} marked in panel~(a), one tile for each mode (as indicated).
        Nine of these tiles make up the green absorption spectrum in panel~(a).
        The width of the Lorentzians is $\eta = 0.01 J_1 S$.
        Parameters as for Fig.~\ref{fig:slab_pol}.
    }
    \label{fig:flake_plb}
\end{figure}

The above suggests that momentum conservation has to be lifted.
We therefore consider the electric absorption of flakes instead of nanoribbons.
While a large flake is roughly similar to a nanoribbon, increasing deviations are expected upon shrinking the flake.
We have computed the electric one-magnon absorption of a flake encompassing 1152 spins [blue line in Fig.~\ref{fig:flake_plb}(a)] that is \enquote{cut} into 4 (orange line) and 9 equally sized smaller tiles (green line).
This cutting increases the signal magnitude of infinite-wavelength peaks due to the introduction of internal edges and leads to additional peaks from in-gap states with smaller wavelengths.

To prove the topological origin of the in-gap peaks for the green line, we have selected four peaks [labeled \mode{1}--\mode{4} in Fig.~\ref{fig:flake_plb}(a)], for which the real-space probability distribution of the corresponding magnon states is shown in Fig.~\ref{fig:flake_plb}(b).
Each flake depicts one of the 9 tiles responsible for the green absorption spectrum in Fig.~\ref{fig:flake_plb}(a).
The darker the color, the stronger the localization of the corresponding magnon mode at that site.
In the cases \mode{1}, \mode{2}, and \mode{3}, the electrically active modes have vanishing weights in the bulk.
In contrast, mode \mode{4} is delocalized throughout the bulk.
The topological bulk gap, indicated by a horizontal gray stripe in the background of Fig.~\ref{fig:flake_plb}(a), includes modes \mode{1}--\mode{3}, while \mode{4} falls outside this energy window, demonstrating the topological origin of the \emph{in-gap} absorption peaks.

\paragraph{Quantitative estimate.}
The effective charge is estimated as $q \approx \num{e-4} |e|$ to $\num{e-3} |e|$ for GaV$_4$S$_8$~\cite{nikolaev_microscopic_2019}, CrBr$_3$~\cite{fumega_moire-driven_2023}, and YIG~\cite{liu_electric_2011} (cf.~SM~\cite{supplement}; $|e|$ elementary charge).
Here, we present calculations based on Heisenberg-DMI and Heisenberg-Kitaev models for the experimental parameters of CrI$_3$~\cite{chen_topological_2018} (cf.~SM~\cite{supplement}), a putative TMI\@.
The three-dimensional electric edge polarization, which depends on the weight of the edges in the probed volume, is estimated to about \SI{10}{\micro\coulomb\per\meter\squared} within the first four layers.
For the imaginary part of the three-dimensional electric susceptibility, which is inversely proportional to the linewidth $\eta$, we obtain $\num{6e-3} \varepsilon_0$ for $\eta = \SI{0.1}{\milli\electronvolt}$  ($\varepsilon_0$ vacuum permittivity).
This value decreases with increasing size of the nanoribbon, as is expected for edge effects.
We expect that our prediction qualitatively applies as well to other ferromagnetic TMI, such as Lu$_2$V$_2$O$_7$~%
\cite{
    onose_observation_2010%
},
Cu(1,3-benzenedicarboxylate)~%
\cite{
    chisnell_topological_2015,
    hirschberger_thermal_2015%
},
CrSiTe$_3$ and CrGeTe$_3$~%
\cite{
    zhu_topological_2021%
},
and VI$_3$~%
\cite{
    zhang_anomalous_2021%
}.


\paragraph{Discussion.}
We have investigated the electric properties of topological chiral edge magnons in equilibrium and nonequilibrium.
In an intuitive picture of the vacuum magnetoelectric effect, topological magnons give rise to an electric edge polarization by thermal fluctuations even in centrosymmetric systems. However, the model calculations based on the KNB effect identified further contributions by trivial edge modes, which dominate the overall signal.
Nonetheless, by addressing the topological magnons directly with alternating electric fields of corresponding frequencies, we demonstrated that these modes may be electrically active, as is indicated by peaks in the one-magnon electric susceptibility.
These topological electromagnons have infinite wavelengths and, depending on their specific dispersion relation, might not be present in every nanoribbon.
The electric absorption by topological edge magnons in flakes can be tuned by the edge-to-area ratio, such that additional peaks from magnons with a finite wavelength appear and the signal of magnons with infinite wavelength is increased.
Hence, we believe that THz spectroscopy can be regarded a probe for topological magnons.

Future research in \enquote{topological electromagnonics} could be directed at a \emph{local} THz probe of chiral edge states, as provided by scattering-type scanning near-field optical microscopy~%
\cite{
    cvitkovic_analytical_2007,
    adam_review_2011,
    chen_modern_2019,
    wiecha_terahertz_2021%
}, at a topological electromagnon-polariton formation in THz cavities \cite{Curtis2022}, and at the existence of topological electromagnons beyond the relativistic KNB mechanism \cite{tokura_multiferroics_2014}. We hope that our results provide additional impetus to search for candidate materials with nontrivial magnon band structures and strong magnetoelectric coupling, and to explore the relation of electromagnons to magnon orbitronics~\cite{fishman_orbital_2022, go_intrinsic_2023}.


\begin{acknowledgments}
\paragraph{Acknowledgments.}
This work was funded by the Deutsche Forschungsgemeinschaft (DFG, German Research Foundation) -- Project-ID 328545488 -- TRR 227, project B04; and Project No.~504261060 (Emmy Noether Programme).
\end{acknowledgments} 


\bibliography{short,refs}

\end{document}


\title{
\textit{Supplemental Material}
\vspace{0.25cm}
\hrule
\vspace{0.25cm}
Electrical Activity of Topological Chiral Edge Magnons
}

\author{Robin R. Neumann}
\affiliation{Institut f\"ur Physik, Martin-Luther-Universit\"at Halle-Wittenberg, D-06099 Halle (Saale), Germany}

\author{J\"urgen Henk}
\affiliation{Institut f\"ur Physik, Martin-Luther-Universit\"at Halle-Wittenberg, D-06099 Halle (Saale), Germany}

\author{Ingrid Mertig}
\affiliation{Institut f\"ur Physik, Martin-Luther-Universit\"at Halle-Wittenberg, D-06099 Halle (Saale), Germany}

\author{Alexander Mook}
\affiliation{Institut f\"ur Physik, Johannes Gutenberg-Universit\"at Mainz, D-55128 Mainz, Germany}

\date{\today}

\maketitle

\tableofcontents

\newpage

\section{Theoretical foundations}

\subsection{Electric field of chiral edge magnon by the vacuum magnetoelectric effect}
This section provides the background for the intuitive picture. Moreover, we show how a chiral edge magnon gives rise to an inhomogeneous scalar potential and an electric field.
The generalized Biot-Savart law~%
\cite{
    meier_magnetization_2003,
    schutz_persistent_2003%
}
\begin{align}
    \phi(\vect{r})
    &=
    \frac{\mu_0 I_{\text{m}}}{4 \mathpi}
    \oint
    \qty[
        \dd{\vect{r}'}
        \times
        \unitvect{m}(\vect{r}')
    ]
    \cdot
    \frac{
        \vect{r} - \vect{r}'
    }{
        \abs{\vect{r} - \vect{r}'}^3
    }
    \label{eq:biot}
\end{align}
for a spin current in a thin wire relates the scalar potential $\phi$ with the integral over the path that traces the wire ($\mu_0$ vacuum permeability, $I_{\mathrm{m}}$ magnetization current, $\unitvect{m}$ direction of the magnetic dipole).
$I_{\mathrm{m}}$ satisfies
$
    I_{\mathrm{m}} \dd{\vect{r}}
    =
    \vect{v} \abs{\vect{M}} \dd[3]{r}
$,
where $\vect{v}$ is the velocity of (in this case) magnons, and $\vect{M}$ is the magnetization.
Assuming a constant velocity, the magnetization current carried by magnons can be expressed as
\begin{align}
    I_{\mathrm{m}}
    &=
    \frac{g \mub}{V}
    \sum_{k}
    \abs{\vect{v}}
    \rho(\varepsilon_{k})
    =
    \frac{g \mub \abs{\vect{v}}}{2 \mathpi}
    \int_{-\nicefrac{\mathpi}{a}}^{\nicefrac{\mathpi}{a}}
    \dd{k} \rho(\varepsilon_{k}),
\end{align}
where $V$ is the length of the system, or its circumference for periodic boundary conditions, $g$ is the $g$-factor, $\mub$ is the Bohr magneton, and
$
    \rho(\varepsilon_{k})
    =
    \qty[
        \exp(\beta \varepsilon_{k}) - 1
    ]^{-1}
$
($\beta = \nicefrac{1}{\kb T}$, $\kb$ Boltzmann constant, $T$ temperature)
is the bosonic distribution function with the magnon energy $\varepsilon_{k}$.

In the main text, we assume a circular geometry, such that the path is a loop with radius $R$ parameterized by
\begin{align}
    \gamma(s)
    =
    \begin{pmatrix}
        R \cos s
        \\
        R \sin s
        \\
        0
    \end{pmatrix}
    ,
    \qquad
    s \in [0, 2\mathpi]
    .
\end{align}
Since we focus on ferromagnets in this work, the magnetic dipole of the topological magnon points parallel to the quantization axis, which we define as $\unitvect{z}$ ($\unitvect{m} \parallel \unitvect{z}$).
Then, Eq.~\eqref{eq:biot} reads
\begin{align}
    \phi(\rho, \varphi, z)
    &=
    \frac{\mu_0 I_{\text{m}}}{4 \mathpi}
    \int_0^{2 \mathpi}
    \dd{s}
    \frac{
        \rho R \cos(\varphi - s) - R^2
    }{
        \qty[
            \rho^2 + z^2 + R^2 - 2 \rho R \cos(\varphi - s)
        ]^{\nicefrac{3}{2}}
    }
    ,
\end{align}
with
\begin{align}
    \vect{r}
    =
    \begin{pmatrix}
        \rho \cos \varphi
        \\
        \rho \sin \varphi
        \\
        z
    \end{pmatrix}
\end{align}
in cylindrical coordinates.

Let us consider two special cases.
First, we assume that $\rho = 0$, i.e., $\vect{r} \parallel \unitvect{z}$.
Then, the integral simplifies to
\begin{align}
    \phi(\rho, \varphi, z)
    &=
    -\frac{\mu_0 I_{\mathrm{m}}}{2}
    \frac{R^2}{\qty(R^2 + z^2)^{\nicefrac{3}{2}}},
\end{align}
and the scalar potential scales with $z^{-3}$ for $z \gg R$.

For the second case, we consider $\rho \gg R$ and expand the integrand in terms of
$
    x = \nicefrac{R}{\rho} \ll 1
$:
\begin{align}
    \frac{
        \rho R \cos(\varphi - s) - R^2
    }{
        \qty[
            \rho^2 + z^2 + R^2 - 2 \rho R \cos(\varphi - s)
        ]^{\nicefrac{3}{2}}
    }
    &=
    \frac{
        \rho^2 \cos(\varphi - s)
    }{
        \qty(\rho^2 + z^2)^{\nicefrac{3}{2}}
    }
    x
    +
    \frac{
        \qty[3 \rho^2 \cos^2(\varphi - s) - \rho^2 - z^2] \rho^2
    }{
        \qty(\rho^2 + z^2)^{\nicefrac{5}{2}}
    }
    x^2
    +
    \cdots
\end{align}
While the integral of the linear term in $x$ vanishes, the quadratic term gives
\begin{align}
    \phi(\rho, \varphi, z)
    \approx
    \frac{\mu_0 I_{\text{m}}}{4}
    \frac{R^2 (\rho^2 - 2 z^2)}{\qty(\rho^2 + z^2)^{\nicefrac{5}{2}}}
    .
\end{align}
Therefore, $\phi$ scales with $\rho^{-3}$ for $\rho \gg z$.

\subsection{Linear spin wave theory}
The Holstein-Primakoff transformation~\cite{holstein_field_1940}
\begin{align}
     \frac{\vect{S}_i}{\hbar}
    =
    \qty(S_i - \adj{a}_i a_i) \unitvect{z}_i
    +
    \sqrt{\frac{S_i}{2}}
    \qty(
        \adj{a}_i f_i \unitvect{e}_i^+
        +
        f_i a_i \unitvect{e}_i^-
    )
    ,
    \quad
    f_i
    =
    \sqrt{1 - \frac{\adj{a}_i a_i}{2 S_i}}
    \label{eq:hpt}
\end{align}
($\unitvect{e}_i^\pm = \unitvect{x}_i \pm \iu \unitvect{y}_i$, $\hbar$ reduced Planck constant) maps a spin operator $\vect{S}_i$ with modulus size $S_i$ to bosonic creation and annihilation operators $a_i$ and $\adj{a}_i$ using the local quantization axis $\unitvect{z}_i$.
The latter is defined by the orientation of the spin in its classical magnetic ground state and forms a local tripod together with the perpendicular axes $\unitvect{x}_i$ and $\unitvect{y}_i$.
Using this transformation, any spin Hamiltonian can be transformed into a bosonic Hamiltonian, which can be expressed as a series with increasing number of bosonic operators.

In the harmonic approximation, the Hamiltonian is truncated beyond bilinear order, such that only the classical ground state energy $E_0$, quantum corrections of the classical ground state energy $\varDelta E_0$, and the bilinear Bogoliubov-de Gennes Hamiltonian~%
\cite{
    shindou_topological_2013,
    lein_krein-schrodinger_2019%
}
\begin{align}
    \hamil_2
    =
    \frac{1}{2}
    \sum_{\vect{k}}
    \adjvect{\psi}_{\vect{k}}
    \hmatr_{\vect{k}}
    \vect{\psi}_{\vect{k}}
    \label{eq:bilin_hamil}
\end{align}
with the Nambu spinor
$
    \adjvect{\psi}_{\vect{k}}
    =
    \qty(\mqty{
        \adj{a}_{1\vect{k}} & \cdots & \adj{a}_{\nbands\vect{k}}
        &
        a_{1(-\vect{k})} & \cdots & a_{\nbands(-\vect{k})}
    })
$
remain.
Here, the Hamiltonian and the magnon operators have been written in reciprocal ($\vect{k}$) space and $\nbands$ corresponds to the number of magnon bands or, equivalently, to the number of spins in the magnetic unit cell.

In order to solve the Schrödinger equation, the Bogoliubov diagonalization~\cite{colpa_diagonalization_1978}
\begin{subequations}
\begin{align}
    \adjbogmatr_{\vect{k}} \hmatr_{\vect{k}} \bogmatr_{\vect{k}}
    &=
    \diag
    \mqty(
        \varepsilon_{1\vect{k}} & \cdots & \varepsilon_{\nbands\vect{k}}
        &
        \varepsilon_{1(-\vect{k})} & \cdots & \varepsilon_{\nbands(-\vect{k})}
    )
    ,
    \\
    \adjbogmatr_{\vect{k}} \metricmatr \bogmatr_{\vect{k}}
    &=
    \metricmatr
    ,
    \label{eq:paraunitary}
\end{align}
\end{subequations}
where
$
    \metricmatr
    =
    \diag
    \mqty(
        1 & \cdots & 1
        &
        -1 & \cdots & -1
    )
$
is the bosonic metric, is applied to the Hamiltonian.
Note that $\bogmatr_{\vect{k}}$ does not have to be unitary, but it has to fulfill the para-unitarity condition Eq.~\eqref{eq:paraunitary}, such that it preserves the bosonic commutation relations of the eigenmodes
\begin{align}
    \vect{\varPsi}_{\vect{k}}
    =
    \trp{\mqty(
        \alpha_{1\vect{k}} & \cdots & \alpha_{\nbands\vect{k}}
        &
        \adj{\alpha}_{1(-\vect{k})} & \cdots & \adj{\alpha}_{\nbands(-\vect{k})}
    )}
    =
    \invbogmatr_{\vect{k}}
    \vect{\psi}_{\vect{k}}
    .
    \label{eq:bogtrafo_eigenmodes}
\end{align}
Then, the bilinear part of the Hamiltonian [Eq.~\eqref{eq:bilin_hamil}] reads
\begin{align}
    \hamil_2
    =
    \sum_{\vect{k}}
    \sum_{n = 1}^N
    \varepsilon_{n\vect{k}}
    \qty(
        \adj{\alpha}_{n\vect{k}} \alpha_{n\vect{k}}
        +
        \frac{1}{2}
    )
    .
\end{align}
 
\subsection{Electric dipole operators}
In this section, we discuss the magnon expansion of two mechanisms of spin-driven electric polarization.
We derive electric dipole operators based on the vacuum magnetoelectric (VME) effect and the Katsura-Nagaosa-Balatsky (KNB) effect, which we compare to each other qualitatively and quantitatively.
This lays the groundwork for further derivations.

\subsubsection{Magnon expansion of the vacuum magnetoelectric effect}
In analogy to the classical relativistic electric dipole
$
    \vect{p}
    =
    \vect{m} \times \vect{v} / c^2
$
($c$ speed of light), that is perpendicular to the magnetic dipole $\vect{m}$ and the velocity $\vect{v}$ of the moving (quasi-)particle, we define the electric polarization operator for magnons as
\begin{subequations}
\begin{align}
    \vect{P}^{\ac}
    &=
    \frac{
        \vect{v} \times \vect{m} - \vect{m} \times \vect{v}
    }{
        2 V c^2
    },
    \label{eq:def_p_ac}
    \\
    v_{\gamma}
    &=
    \frac{1}{2}
    \sum_{\vect{k}}
    \adjvect{\psi}_{\vect{k}}
    \matr{v}_{\gamma\vect{k}}
    \vect{\psi}_{\vect{k}}
    ,
    \quad
    \matr{v}_{\gamma\vect{k}}
    =
    \frac{1}{\hbar}
    \pdv{\hmatr_{\vect{k}}}{k_\gamma},
    \\
    m_{\gamma}
    &=
    \frac{1}{2}
    \sum_{\vect{k}}
    \adjvect{\psi}_{\vect{k}}
	\matr{m}_{\gamma\vect{k}}
    \vect{\psi}_{\vect{k}}
    ,
    \quad
    \matr{m}_{\gamma\vect{k}}
    =
    g \mub
    \diag
    \mqty(
        \hat{z}_1^{\gamma} & \cdots & \hat{z}_{\nbands}^{\gamma}
        &
        \hat{z}_1^{\gamma} & \cdots & \hat{z}_{\nbands}^{\gamma}
    )
\end{align}
\end{subequations}
(\ac\ short for vacuum magnetoelectric). Here, the vector product between the group velocity $\vect{v}$ and the operator for the magnetic moment $\vect{m}$ has been symmetrized such that $\vect{P}^{\ac}$ is Hermitian.
$V$ is the total (hyper)volume of the system.

One can show that Eq.~\eqref{eq:def_p_ac} for the polarization evaluates to
\begin{subequations}
\begin{align}
    P_{\gamma}^{\ac}
    &=
    \frac{1}{2}
    \sum_{\vect{k}}
    \adjvect{\psi}_{\vect{k}}
    \matr{P}^{\ac,(2)}_{\gamma\vect{k}}
    \vect{\psi}_{\vect{k}}
    +
    (\text{higher-order terms})
    ,
    \\
    \matr{P}^{\ac,(2)}_{\gamma\vect{k}}
    &=
    \sum_{\alpha, \beta = x, y, z}
    \frac{\epsilon_{\alpha\beta\gamma}}{2 V c^2}
    \qty(
        \matr{v}_{\alpha\vect{k}}
        \metricmatr
        \matr{m}_{\beta\vect{k}}
        +
        \matr{m}_{\beta\vect{k}}
        \metricmatr
        \matr{v}_{\alpha\vect{k}}
    )
\end{align}%
\label{eq:ac_bilin}
\end{subequations}
($\epsilon_{\alpha\beta\gamma}$ Levi-Civita symbol).
The higher-order terms are biquadratic in
$
    \vect{\psi}_{\vect{k}}
$
and
$
    \adjvect{\psi}_{\vect{k}}
$
and neglected in the following.

For later reference, we define the bilinear part of $P^{\ac}_{\gamma}$ as
\begin{align}
    P_{\gamma}^{\ac,(2)}
    &=
    \frac{1}{2}
    \sum_{\vect{k}}
    \adjvect{\psi}_{\vect{k}}
    \matr{P}^{\ac,(2)}_{\gamma\vect{k}}
    \vect{\psi}_{\vect{k}}
    .
\end{align}


\subsubsection{Magnon expansion of the Katsura-Nagaosa-Balatsky mechanism}
As written in the main text, the electric dipole due to the KNB effect is a result of relative spin canting between nearest neighbors,~\cite{katsura_spin_2005}
\begin{align}
    \vect{p}_{ij}
    =
    \frac{q}{\hbar^2}
    \vect{e}_{ij}
    \times
    \qty(\vect{S}_i \times \vect{S}_j)
    \label{eq:knb_pair}
\end{align}
($q$ effective charge, $\vect{e}_{ij}$ bond vector). 
We define the total polarization as
\begin{align}
    \vect{P}^{\knb}
    =
    \frac{1}{V}
    \sum_{\langle ij \rangle}
    \vect{p}_{ij}
    =
    \frac{q}{\hbar^2 V}
    \sum_{\langle ij \rangle}
    \vect{e}_{ij}
    \times
    \qty(\vect{S}_i \times \vect{S}_j)
\end{align}
(\knb\ short for Katsura-Nagaosa-Balatsky). As before, $V$ is the total (hyper)volume of the system.
By employing the Holstein-Primakoff expansion [Eq.~\eqref{eq:hpt}], the three leading orders of the polarization read
\begin{subequations}
\begin{align}
    \vect{P}^{\knb,(0)}
    &=
    \frac{q}{V}
    \sum_{\langle ij \rangle}
    S_i S_j \vect{e}_{ij}
    \times
    \qty(\unitvect{z}_i \times \unitvect{z}_j)
    ,
    \\
    P^{\knb,(1)}_{\gamma}
    &=
    \vect{P}^{\knb,(1)}_{\gamma}
    \cdot
    \vect{\psi}_{\vect{k} = \vect{0}}
    ,
    \\
    P^{\knb,(2)}_{\gamma}
    &=
    \frac{1}{2}
    \sum_{\vect{k}}
    \qty(
	    \adjvect{\psi}_{\vect{k}}
	    \matr{P}^{\knb,(2)}_{\gamma \vect{k}}
	    \vect{\psi}_{\vect{k}}
	    -
	    \Tr \matr{\varLambda}_{\gamma}
    )
    ,
    \label{eq:knb_bilin_oper}
\end{align}
\end{subequations}
($\gamma = x, y, z$ Cartesian components of the polarization)
where
the coefficient tensors are given by
\begin{subequations}
\begin{align}
    \vect{P}^{\knb,(1)}_{\gamma}
    &=
    \begin{pmatrix}
        \vect{R}_{\gamma}
        \\\
        \conjvect{R}_{\gamma}
    \end{pmatrix}
    ,
    \\
    \qty(R_{\gamma})_{m}
    &=
    -\frac{\sqrt{2 S_m \nucs} q}{V}
    \sum_{(n, \vect{\delta}) \in \nbset_m}
    S_n
    \qty(\vect{\delta} + \vect{b}_n - \vect{b}_m)
    \times
    \qty(\unitvect{z}_n \times \unitvect{e}_m^-)
    \label{eq:knb_lin_sub}
\end{align}
\end{subequations}
for the linear order ($\nucs$ total number of unit cells).
The sum in Eq.~\eqref{eq:knb_lin_sub} runs over all tuples $(n, \vect{\delta}) \in \nbset_m$, in which $n$ is the sublattice index and $\vect{\delta}$ is a lattice vector, of neighboring sites with respect to a fixed site of sublattice $m$.
The basis vectors $\vect{b}_i$ represent the positions of the spins on the $i$-th sublattice within their respective unit cells.
For the bilinear order, the operator coefficients
\begin{subequations}
\begin{align}
    \matr{P}^{\knb,(2)}_{\gamma\vect{k}}
    &=
    \begin{pmatrix}
        \matr{X}_{\gamma\vect{k}}
        +
        \matr{\varLambda}_{\gamma}
        &
        \matr{Y}_{\gamma\vect{k}}
        \\
        \conjmatr{Y}_{\gamma(-\vect{k})}
        &
        \conjmatr{X}_{\gamma(-\vect{k})}
        +
        \matr{\varLambda}_{\gamma}
    \end{pmatrix}
    ,
    \\
    {\qty(X_{\gamma\vect{k}})}_{mn}
    &=
    \frac{q \sqrt{S_m S_n}}{V}
    \sum_{\vect{\delta} \in \nblvset_{mn}}
    \qty(\vect{\delta} + \vect{b}_n - \vect{b}_m)
    \times
    \qty(\unitvect{e}^+_m \times \unitvect{e}^-_n)
    \e^{\iu \vect{k} \cdot \qty(\vect{\delta} + \vect{b}_n - \vect{b}_m)}
    ,
    \label{eq:knb_bilin_x}
    \\
    {\qty(Y_{\gamma\vect{k}})}_{mn}
    &=
    \frac{q \sqrt{S_m S_n}}{V}
    \sum_{\vect{\delta} \in \nblvset_{mn}}
    \qty(\vect{\delta} + \vect{b}_n - \vect{b}_m)
    \times
    \qty(\unitvect{e}^+_m \times \unitvect{e}^+_n)
    \e^{\iu \vect{k} \cdot \qty(\vect{\delta} + \vect{b}_n - \vect{b}_m)}
    ,
    \label{eq:knb_bilin_y}
    \\
    {\qty(\varLambda_{\gamma})}_{mn}
    &=
    -\frac{\kron{m}{n} q}{V}
    \sum_{(l, \vect{\delta}) \in \nbset_{m}}
    S_l
    \qty(\vect{\delta} + \vect{b}_l - \vect{b}_m)
    \times
    \qty(\unitvect{z}_m \times \unitvect{z}_l)
\end{align}%
\label{eq:knb_bilin}
\end{subequations}
can be expressed as sums over \emph{unique} lattice vectors $\vect{\delta} \in \nblvset_{mn}$ that connect nearest neighbors on fixed sublattices $m$ and $n$ [cf. Eqs.~\eqref{eq:knb_bilin_x} and \eqref{eq:knb_bilin_y}].

Representing the electric polarization operators in the basis of the magnon eigenmodes by invoking the Bogoliubov transformation [Eq.~\eqref{eq:bogtrafo_eigenmodes}],
\begin{subequations}
\begin{align}
    P^{\knb,(1)}_{\gamma}
    &=
    \vect{\bogp}^{\knb,(1)}_{\gamma}
    \cdot
    \vect{\varPsi}_{\vect{k} = \vect{0}}
    ,
    &
    \vect{\bogp}^{\knb,(1)}_{\gamma}
    &=
    \trpbogmatr_{\vect{k} = \vect{0}} \vect{P}_{\gamma}^{\knb,(1)}
    ,
    \\
    P^{\knb,(2)}_{\gamma}
    &=
    \frac{1}{2}
    \sum_{\vect{k}}
    \qty(
        \adjvect{\varPsi}_{\vect{k}}
        \matr{\bogp}^{\knb,(2)}_{\gamma \vect{k}}
        \vect{\varPsi}_{\vect{k}}
        -
        \Tr \matr{\varLambda}_{\gamma}
    )
    ,
    &
    \matr{\bogp}^{\knb,(2)}_{\gamma \vect{k}}
    &=
    \adjbogmatr_{\vect{k}}
    \matr{P}^{\knb,(2)}_{\gamma\vect{k}}
    \bogmatr_{\vect{k}}
    ,
\end{align}
\end{subequations}
proves useful for calculating the equilibrium electric polarization
\begin{align}
    \expval{\vect{P}^{\knb}}
    =
    \vect{P}^{\knb,(0)}
    +
    \sum_{\gamma = x, y, z}
    \unitvect{\gamma}
    \sum_{\vect{k}}
    \qty[
	    \sum_{n = 1}^{\nbands}
	    \qty(\rho_{n\vect{k}} + \frac{1}{2})
	    \qty(\bogp^{\knb,(2)}_{\gamma\vect{k}})_{nn}
	    -
	    \frac{\Tr \matr{\varLambda}_{\gamma}}{2}
    ]
    ,
\end{align}
to which only $\vect{P}^{\knb,(0)}$ and $\vect{P}^{\knb,(2)}$ contribute. The former represents the classical polarization of the ground state, and the latter comprises both the quantum fluctuations as quantum corrections to the ground state polarization as well as the thermal contributions by magnons.
The quantum corrections can be obtained by setting the Bose function
$
	\rho_{n\vect{k}}
	=
	{\qty[
		\exp(\beta \varepsilon_{n\vect{k}}) - 1	
	]}^{-1}
$
to zero.
Note that $\vect{P}^{\knb,(1)}$ does not change the mean value, but is responsible for equilibrium fluctuations of the polarization, i.e., $\expval{\qty(\varDelta \vect{P}^{\knb})^2}$.

\subsubsection{Estimation of the effective charge}
\label{sec:eff_charge}
We now estimate the value of the parameter $q$, which has the dimension of charge.
Nikolaev and Solovyev~\cite{nikolaev_microscopic_2019} have computed the polarization based on superexchange theory combined with the Berry phase theory of the polarization for the multiferroic material GaV$_4$S$_8$.
They express the antisymmetric part of the inter-site polarization per bond as
\begin{align}
	\vect{P}_{ij}
	=
	\unitvect{e}_{ij}
	\qty[
		\vect{\mathcal{P}}_{ij}
		\cdot
		\qty(\unitvect{S}_i \times \unitvect{S}_j)	
	]
\end{align}
and compute
$
	\abs{\vect{\mathcal{P}}_{ij}}
	=
	\SI{41}{\micro\coulomb\per\meter\squared}
	.
$
Although the mechanism is slightly different, as it predicts a polarization \emph{along} the bond vector direction $\unitvect{e}_{ij}$, we can use this value to estimate $q$ by multiplying $\abs{\vect{\mathcal{P}}_{ij}}$ with the unit cell volume
$
	\vuc = \SI{224}{\angstrom\cubed}
$
and dividing it by the bond length of
$
	a = \SI{4}{\angstrom}
	,
$
which gives
$
	\abs{q} = \SI{2.3e-23}{\coulomb}
	.
$

A similar order of magnitude is obtained by an estimation for YIG by Liu and Vignale~\cite{liu_electric_2011} based on the expression
\begin{align}
	\vect{P}
	=
	-\frac{J e}{E_{\text{SO}} \hbar^2}
	\vect{e}_{ij}
	\times
	\qty(\vect{S}_1 \times \vect{S}_2)
\end{align}
with
$
	E_{\text{SO}} = \SI{3.0}{\electronvolt}
$
and the exchange constant $J$ around \SI{3}{\milli\electronvolt}~\cite{cherepanov_saga_1993} ($e$ elementary charge), as the effective charge is given by
$
	\abs{q}
	=
	\frac{J e}{E_{\text{SO}}}
	=
	10^{-3} e
	=
	\SI{1.6e-22}{\coulomb}
	.
$


According to Fumega and Lado~\cite{fumega_moire-driven_2023}, the effective charge amounts to
$
    q
    \approx
    \num{e-4} e
    \approx
    \SI{1.6e-23}{\coulomb}
$
in CrBr$_3$.
Overall, these numbers suggest that the order of magnitude varies between $\num{e-3} e$ and $\num{e-4} e$ for several materials.


\subsubsection{Comparison of the bilinear electric dipole operators of both mechanisms}
Here, we establish an analogy between $\vect{P}^{\ac,(2)}$ and $\vect{P}^{\knb,(2)}$ for a nearest-neighbor Heisenberg model and estimate their relative magnitude.
The structure of Eq.~\eqref{eq:knb_bilin} reveals that the entries of $\vect{P}^{\knb,(2)}_{\gamma\vect{k}}$ may be rewritten as a derivative with respect to $\vect{k}$ in order to obtain the bond vectors from the derivative of the complex exponentials.
This resembles the definition of the group velocity matrix elements
$
    {\qty(v_{\alpha\vect{k}})}_{mn}
    =
    \partial_{k_\alpha} {(\hmsymb_{\vect{k}})}_{mn} / \hbar
    ,
$
where ${(\hmsymb_{\vect{k}})}_{mn}$ are the matrix elements of the Hamiltonian.

Let us consider a Heisenberg ferromagnet (whose total spin is conserved) with isotropic exchange $J < 0$ between nearest neighbors and a quantization axis along the global $z$ direction.
The Hamiltonian reads
\begin{subequations}
\begin{align}
	\hmatr_{\vect{k}}
    &=
	\begin{pmatrix}
		\matr{A}_{\vect{k}} & \matr{B}_{\vect{k}}
		\\
		\conjmatr{B}_{-\vect{k}} & \conjmatr{A}_{-\vect{k}}
	\end{pmatrix}
    +
    \hmatr_0
    ,
	\\
	\qty(A_{\vect{k}})_{mn}
    &=
	J \sqrt{S_m S_n}
    \sum_{\vect{\delta} \in \nblvset_{mn}}
	\e^{\iu \vect{k} \cdot (\vect{\delta} + \vect{b}_n - \vect{b}_m)}
	,
	\\
	\qty(B_{\vect{k}})_{mn}
    &=
	0
    ,
\end{align}
\end{subequations}
where $\hmatr_0$ is a diagonal matrix that does not depend on the wave vector $\vect{k}$.
We do not make any assumptions on the lattice, the number of sublattices, or the spin lengths $S_m$ on different sublattices.
The bilinear KNB polarization [Eq.~\eqref{eq:knb_bilin}] amounts to
\begin{subequations}
\begin{align}
	{\qty(\vect{X}_{\vect{k}})}_{mn}
    &=
	\frac{2 q \sqrt{S_m S_n}}{V}
    \unitvect{z} \times \grad_{\vect{k}}
	\sum_{\vect{\delta} \in \nblvset_{mn}}
	\e^{\iu \vect{k} \cdot (\vect{\delta} + \vect{b}_n - \vect{b}_m)}
    ,
	\\
	{\qty(\vect{Y}_{\vect{k}})}_{mn}
    &=
    \vect{0}
    ,
    \\
	{\vect{\varLambda}}_{mn}
    &=
    \vect{0}
    ,
\end{align}
\end{subequations}
with the vector of the matrix elements defined as
$
    {\qty(\vect{X}_{\vect{k}})}_{mn}
    =
    \sum_{\gamma = x, y, z}
    \unitvect{\gamma}
    {\qty(X_{\gamma\vect{k}})}_{mn}
$
and analogously for ${(\vect{Y}_{\vect{k}})}_{mn}$ and ${\vect{\varLambda}}_{mn}$.
We thus conclude that
$
    {\qty(\vect{P}^{\knb,(2)}_{\vect{k}})}_{mn}
    =
    \alpha
    \metric_{nn}
    \unitvect{z}
    \times
    \grad_{\vect{k}} {(H_{\vect{k}})}_{mn},
$
and we directly relate the group velocity and the polarization by
$
    {\qty(\vect{P}^{\knb,(2)}_{\vect{k}})}_{mn}
	=
	\hbar \alpha
    \metric_{nn}
    \unitvect{z}
    \times
    {\qty(\vect{v}_{\vect{k}})}_{mn}
    ,
$
with $\alpha = 2 q / (J V)$.

We now compare this result to the matrix element of the VME polarization [Eq.~\eqref{eq:ac_bilin}] defined as
\begin{align}
    {\qty(\vect{P}^{\ac,(2)}_{\vect{k}})}_{mn}
    =
    \sum_{l = 1}^{2 \nbands}
	\frac{
        \metric_{ll}
		{\qty(\vect{v}_{\vect{k}})}_{ml}
        \times
        {\qty(\vect{m}_{\vect{k}})}_{ln}
		-
        \metric_{ll}
		{\qty(\vect{m}_{\vect{k}})}_{ml}
        \times
        {\qty(\vect{v}_{\vect{k}})}_{ln}
    }{2 c^2 V}
	.
\end{align}
It can be simplified to
$
    {\qty(\vect{P}^{\ac,(2)}_{\vect{k}})}_{mn}
    =
	-\frac{g \mub}{c^2 V}
    \metric_{nn}
    \unitvect{z}
	\times
	{\qty(\vect{v}_{\vect{k}})}_{mn}
$
in a Heisenberg ferromagnet, in which the spin is conserved
[$
    {(\vect{m}_{\vect{k}})}_{mn} = g \mub \kron{m}{n} \unitvect{z}
$].

We have used that the matrices are block diagonal, such that $G_{mm}$ and $G_{nn}$ can be interchanged.
Consequently, the two mechanisms are closely related to each other in a Heisenberg ferromagnet:
\begin{align}
	{\qty(\vect{P}^{\knb,(2)}_{\vect{k}})}_{mn}
    =
	\frac{2 q \hbar c^2}{\abs{J} g \mub}
    {\qty(\vect{P}^{\ac,(2)}_{\vect{k}})}_{mn}
    .
\end{align}
The sign of both kinds of electric polarization match for $q > 0$.
For realistic parameters of $q \approx \num{e-4} e$ to $\num{e-3} e$, $g \approx 2$, and $J \approx \SI{1}{\milli\electronvolt}$, the KNB effect is approximately $\num{e5}$ to $\num{e6}$ times larger than the VME effect.

\subsection{Local electric polarization}
A nonzero total electric polarization may only exist in systems with a polar point group.
Here, we define a local polarization, that may be finite even in centrosymmetric systems, based on the KNB effect.
In the following, the KNB superscript in the operator notation is omitted, as we focus on the KNB mechanism throughout.

We define the local electric polarization 
\begin{align}
    \vect{P}_m
    =
    \frac{q}{\hbar^2 L}
    \sum_{\underset{i \in L_m}{\langle ij \rangle}}
    \vect{e}_{ij}
    \times
    \qty(\vect{S}_i \times \vect{S}_j)
    ,
\end{align}
for each sublattice $m$ by adding all bond-specific electric dipoles $\vect{p}_{ij}$ [cf.~Eq.~\eqref{eq:knb_pair}] for adjacent sites $i$ and $j$, where $i \in L_m$ is an element of the set $L_m$ of all sites on the $m$-th sublattice and $j$ may belong to a different sublattice.
Then, one arrives at
\begin{align}
	\vect{P}_m
    =
    \vect{P}_m^{(0)} + \vect{P}_m^{(1)} + \vect{P}_m^{(2)}
    +
    O\qty(S^{\nicefrac{1}{2}})
\end{align}
with the contributions
\begin{subequations}
\begin{align}
	\vect{P}_m^{(0)}
	&=
	\frac{q}{V}
	\sum_{(n, \vect{\delta}) \in \nbset_m}
	S_m S_n
	\qty(\vect{\delta} + \vect{b}_n - \vect{b}_m)
	\times
	\qty(\unitvect{z}_m \times \unitvect{z}_n)
    ,
	\\
	P_{m,\gamma}^{(1)}
	&=
	\vect{P}_{m,\gamma}^{(1)}
	\cdot
	\vect{\psi}_{\vect{k} = \vect{0}}
    ,
	\\
	P_{m,\gamma}^{(2)}
	&=
	\sum_{\vect{k}}
	\adjvect{\psi}_{\vect{k}}
	\matr{P}_{m,\gamma\vect{k}}^{(2)}
	\vect{\psi}_{\vect{k}}
    .
    \label{eq:knb_loc_bilin_oper}
\end{align}
\end{subequations}
The linear-order coefficients read
\begin{subequations}
\begin{align}
	\vect{P}_{m,\gamma}^{(1)}
	&=
	\begin{pmatrix}
		\vect{R}_{m,\gamma}
		\\
		\conjvect{R}_{m,\gamma}
	\end{pmatrix},
	\\
	{\qty(R_{m,\gamma})}_l
	&=
	\frac{q \sqrt{\nucs}}{V}
	\sum_{(n,\vect{\delta}) \in \nbset_m}
	\qty[
		\sqrt{\frac{S_m S_n}{2}}
		\qty(\vect{\delta} + \vect{b}_n - \vect{b}_m)
		\times
		\qty(
			\kron{l}{n} \sqrt{S_m}
			\unitvect{z}_m \times \unitvect{e}_n^-
			-
			\kron{l}{m}
			\sqrt{S_n}
			\unitvect{z}_n \times \unitvect{e}_m^-
		)
	],
\end{align}
\end{subequations}
and for the coefficients of the bilinear operators one obtains
\begin{subequations}
\begin{align}
	\matr{P}^{(2)}_{m,\gamma\vect{k}}
	&=
	\begin{pmatrix}
		\matr{X}_{m,\gamma\vect{k}}
		+
		\matr{\varLambda}_{m,\gamma}
		&
		\matr{Y}_{m,\gamma \vect{k}}
		\\
		\conjmatr{Y}_{m,\gamma(-\vect{k})}
		&
		\conjmatr{X}_{m,\gamma(-\vect{k})}
	\end{pmatrix},
	\\
	{\qty(X_{m,\gamma\vect{k}})}_{ij}
	&=
	\kron{m}{i}
	\sum_{(n,\vect{\delta}) \in \nbset_m}
	\kron{n}{j}
	\frac{q \sqrt{S_m S_n}}{2 V}
	{\qty[
		(\vect{\delta} + \vect{b}_n - \vect{b}_m)
		\times
		\qty(\unitvect{e}_m^+ \times \unitvect{e}_n^-)
	]}_\gamma
	\e^{\iu \vect{k} \cdot (\vect{\delta} + \vect{b}_n - \vect{b}_m)}
    ,
	\\
	{\qty(Y_{m,\gamma\vect{k}})}_{ij}
	&=
	\kron{m}{i}
	\sum_{(n,\vect{\delta}) \in \nbset_m}
	\kron{n}{j}
	\frac{q \sqrt{S_m S_n}}{2 V}
	{\qty[
		(\vect{\delta} + \vect{b}_n - \vect{b}_m)
		\times
		\qty(\unitvect{e}_m^+ \times \unitvect{e}_n^+)
	]}_\gamma
	\e^{\iu \vect{k} \cdot (\vect{\delta} + \vect{b}_n - \vect{b}_m)}
    ,
	\\
	{\qty(\varLambda_{m,\gamma})}_{ij}
	&=
	-\kron{i}{j}
	\sum_{(n,\vect{\delta}) \in \nbset_m}
    \frac{q}{V}
	{\qty[
        (\vect{\delta} + \vect{b}_n - \vect{b}_m)
        \times
        \qty(\unitvect{z}_m \times \unitvect{z}_n)
    ]}_\gamma
	\qty(
		\kron{p}{n} S_m
		+
		\kron{p}{m} S_n	
	)
    .
\end{align}%
\label{eq:knb_loc_bilin}
\end{subequations}
Note that these definitions differ from the total polarization [Eqs.~\eqref{eq:knb_bilin_oper} and~\eqref{eq:knb_bilin}].
In Eq.~\eqref{eq:knb_loc_bilin_oper}, the factor $\nicefrac{1}{2}$ and the trace of the matrix $\matr{\varLambda}_{m,\gamma}$ are missing, and the matrix $\matr{P}^{(2)}_{m,\gamma\vect{k}}$ is neither Hermitian nor particle-hole symmetric.
This is possible because there is no unique form to define the coefficient matrices.
The present form was chosen in favor of compactness of the mathematical expressions.
Note that Hermitian coefficient matrices are not necessary for Hermitian operators and the expectation values do not depend on the chosen representation.

Irrespective of the mathematical form of the coefficient matrices, the local polarization given by the thermal expectation value
\begin{align}
    \expval{\vect{P}_m}
    =
    \vect{P}^{(0)}_m
    +
    \sum_{\gamma = x, y, z}
    \unitvect{\gamma}
    \sum_{\vect{k}}
    \sum_{n = 1}^{\nbands}
    \qty[
        {\qty(\bogp^{(2)}_{m,\gamma\vect{k}})}_{nn}
        \rho_{n\vect{k}}
        +
        {\qty(\bogp^{(2)}_{m,\gamma\vect{k}})}_{n+\nbands,n+\nbands}
        \qty(1 + \rho_{n(-\vect{k})})
    ],
    \label{eq:knb_loc_pol_eq}
\end{align}
of the operator $\vect{P}_m$ is a well-defined observable.
Here, the bilinear part
$
    \matr{\bogp}^{(2)}_{m,\gamma\vect{k}}
    =
    \adjbogmatr_{\vect{k}}
    \matr{P}^{(2)}_{m,\gamma\vect{k}}
    \bogmatr_{\vect{k}}
$
is transformed into the eigenbasis of the Hamiltonian.

\subsection{Electric susceptibility}
In this section, we derive an expression for the electric susceptibility $\chi_{\mu\nu}^{\text{e}}$ that describes the electric response of a system to electric fields:
\begin{align}
    \varDelta \expval{P_{\mu}(\omega)}
    &=
    \sum_{\nu = x, y, z}
    \chi_{\mu\nu}^{\text{e}}(\omega)
    E_{\nu}(\omega)
    .
\end{align}
Here, $\varDelta \expval{P_{\mu}(\omega)}$ is the nonequilibrium contribution of the polarization due to the external stimulus.
It is assumed that (the Fourier components of) the ac electric field $\vect{E}(\omega)$ with angular frequency $\omega$ is sufficiently small, such that the response is linear in $\vect{E}$.
In the framework of this section, we drop the superscript and set
$
    \chi_{\mu\nu}(\omega)
    \coloneqq
    \chi_{\mu\nu}^{\text{e}}(\omega)
$
for ease of notation.
In later sections, in which we compare it to the magnetic susceptibility, superscripts are reintroduced.

The Hamiltonian of the perturbation
\begin{align}
    \hamil'(t)
    =
    - V \vect{P} \cdot \vect{E}(t)
\end{align}
is given as the negative scalar product between the total electric dipole moment $V \vect{P}$ and the time-varying electric field $\vect{E}(t)$.
To derive $\chi_{\mu\nu}$ we employ the Kubo linear response theory formulated in the language of Matsubara correlation functions, following Refs.~%
\onlinecite{
    kubo_statistical-mechanical_1957,
    bruus_many-body_2004%
}.
The electric susceptibility
\begin{align}
    \chi_{\mu\nu}(\omega)
    =
    - V
    C^{\text{R}}_{P_{\mu} P_{\nu}}(\omega)
\end{align}
is related to the retarded electric polarization autocorrelation function $C^{\text{R}}_{P_{\mu} P_{\nu}}(\omega)$, which can be obtained from the Matsubara correlation function as
\begin{align}
    C^{\text{R}}_{P_{\mu} P_{\nu}}(\omega)
    =
    C_{P_{\mu} P_{\nu}}(\iu \omega_l \to \omega + \iu 0^+)
\end{align}
by analytic continuation;
$
    \omega_l = \nicefrac{2 \mathpi l}{\beta \hbar}
$
($l \in \mathbb{Z}$)
are the bosonic Matsubara frequencies.
The Matsubara correlation function is given as
\begin{align}
    C_{P_{\mu} P_{\nu}}(\omega)
    =
    -\int_0^\beta \dd{\lambda}
    \e^{\iu \hbar \omega_l \lambda}
    \expval{\mathcal{T} \hat{P}_\mu(\hbar \lambda) \hat{P}_\nu(0)}_0
    ,
\end{align}
where $\mathcal{T}$ is the time-ordering operator and
$
    \hat{P}_{\mu/\nu}(\tau)
    =
    \exp(\frac{\tau}{\hbar} \hamil_0)
    P_{\mu/\nu}
    \exp(-\frac{\tau}{\hbar} \hamil_0)
$
is the electric polarization in the Dirac or interaction picture for imaginary times ($\hamil_0$ unperturbed Hamiltonian).

Now,
$
    \vect{P} = \vect{P}^{(0)} + \vect{P}^{(1)} + \vect{P}^{(2)} + \cdots
$
can be decomposed into its different orders in magnon creation and annihilation operators, the combinations of which produce various contributions to $\chi_{\mu\nu}(\omega)$.
One can show that the contribution of the ground state polarization $\vect{P}^{(0)}$ vanishes; similarly, one shows that mixing of different orders of $\vect{P}^{(i)}$ and $\vect{P}^{(j)}$ with $i + j$ odd does not lead to nonzero additional contributions as well.
Hence, there remain to consider two important contributions to
$
    \chi_{\mu\nu} = \chi_{\mu\nu}^{(1)} + \chi_{\mu\nu}^{(2)}:
$
the one-magnon $\chi_{\mu\nu}^{(1)}$ and two-magnon processes $\chi_{\mu\nu}^{(2)}$, which we derive in the following.

In the derivations below, we use implicitly the following identities.
First, we define the $2 N \times 2 N$ matrix
\begin{align}
    \matr{\paulix}
    =
    \begin{pmatrix}
        \matr{0} & \idmatr
        \\
        \idmatr & \matr{0}
    \end{pmatrix}
    ,
\end{align}
where $\idmatr$ is the $N \times N$ identity matrix.
Furthermore,
$
    \rho_0 = \e^{-\beta \hamil_0} / Z
$
with 
$
    Z = \Tr \e^{-\beta \hamil_0}
$
denotes the equilibrium density operator of the grand canonical ensemble with zero chemical potential,
$
    \tilde{\varepsilon}_{n\vect{k}}
    =
    \metric_{nn}
    \varepsilon_{n(\metric_{nn}\vect{k})}
$
are the signed magnon eigenenergies, and
$
    \expval{A}_0
    =
    \Tr \rho_0 A
$
represents the thermal equilibrium expectation value of $A$.

By exploiting
$
    \rho(-\varepsilon) = -[1 + \rho(\varepsilon)]
$
for the Bose function, we obtain the following expectation values:
\begin{subequations}
\begin{align}
    \Tr \rho_0
    \adj{\varPsi}_{m\vect{k}}
    {\varPsi}_{n\vect{k}'}
    &=
    \kron{\vect{k}}{\vect{k}'}
    \metric_{mn}
    \rho(\tilde{\varepsilon}_{m\vect{k}})
    ,
    \\
    \Tr \rho_0
    {\varPsi}_{m\vect{k}}
    \adj{\varPsi}_{n\vect{k}'}
    &=
    -\kron{\vect{k}}{\vect{k}'}
    \metric_{mn}
    \rho(-\tilde{\varepsilon}_{m\vect{k}})
    ,
    \\
    \Tr \rho_0
    {\varPsi}_{m\vect{k}}
    {\varPsi}_{n\vect{k}'}
    &=
    -\kron{\vect{k}}{(-\vect{k})}
    \metric_{mm}
    \paulix_{mn}
    \rho(-\tilde{\varepsilon}_{m\vect{k}})
    ,
    \\
    \Tr \rho_0
    \adj{\varPsi}_{m\vect{k}}
    \adj{\varPsi}_{n\vect{k}'}
    &=
    \kron{\vect{k}}{(-\vect{k}')}
    \metric_{mm}
    \paulix_{mn}
    \rho(\tilde{\varepsilon}_{m\vect{k}})
    .
\end{align}%
\label{eq:twopoint_identities}
\end{subequations}
The (imaginary-time) Dirac and Schrödinger pictures of the magnon creation and annihilation operators differ by an exponential factor:
\begin{subequations}
\begin{align}
    \hat{\varPsi}_{n\vect{k}}(\tau)
    &=
    \e^{-\tilde{\omega}_{n\vect{k}} \tau}
    \varPsi_{n\vect{k}}
    ,
    \\
    \adj{\hat{\varPsi}}_{n\vect{k}}(\tau)
    &=
    \e^{\tilde{\omega}_{n\vect{k}} \tau}
    \adj{\varPsi}_{n\vect{k}}
    ,
\end{align}
\end{subequations}
with the signed magnon eigenfrequencies
$
    \tilde{\omega}_{n\vect{k}}
    =
    \tilde{\varepsilon}_{n\vect{k}} / \hbar
    .
$

\subsubsection{One-magnon processes}
Considering one-magnon processes, we confine ourselves to the linear part
\begin{align}
    P^{(1)}_{\gamma}
    =
    \vect{\bogp}_{\gamma}^{(1)}
    \cdot
    \vect{\varPsi}_{\vect{k} = \vect{0}}
\end{align}
 of the polarization operator, where $\vect{\varPsi}_{\vect{k}}$ are the eigenmodes of $\hamil_0$ [cf. Eq.~\eqref{eq:bogtrafo_eigenmodes}].
Note that we do not make any assumption about the mechanism behind $\vect{P}$.

Let us start by evaluating the Matsubara autocorrelation function:
\begin{subequations}
\begin{align}
    C_{P_{\mu}^{(1)} P_{\nu}^{(1)}}(\omega)
    &=
    -\int_0^\beta \dd{\lambda}
    \e^{\iu \hbar \omega_l \lambda}
    \expval{
        \mathcal{T}
        \hat{P}_\mu^{(1)}(\hbar \lambda)
        \hat{P}_\nu^{(1)}(0)}
    ,
    \\
    &=
    -
    \sum_{\vect{k}}
    \sum_{m, n = 1}^{2 \nbands}
    \kron{\vect{k}}{\vect{0}}
    {\qty(\bogp_{\mu}^{(1)})}_{m}
    {\qty(\bogp_{\nu}^{(1)})}_{n}
    \expval{
        \varPsi_{m\vect{k}}
        \varPsi_{n\vect{k}}
    }_0
    \int_0^\beta \dd{\lambda}
    \e^{\lambda (\iu \hbar \omega_l - \tilde{\varepsilon}_{n\vect{k}})}
    ,
    \\
    &=
    \sum_{\vect{k}}
    \sum_{m, n = 1}^{2 \nbands}
    \kron{\vect{k}}{\vect{0}}
    {\qty(\bogp_{\mu}^{(1)})}_{m}
    {\qty(\bogp_{\nu}^{(1)})}_{n}
    \metric_{mm}
    \paulix_{mn}
    \rho(-\tilde{\varepsilon}_{n\vect{k}})
    \frac{
        \e^{-\beta(\tilde{\varepsilon}_{m\vect{k}} - \iu \hbar \omega_l)}
        -
        1
    }{
        \iu \hbar \omega_l - \tilde{\varepsilon}_{m\vect{k}}
    }
    ,
    \\
    &=
    \sum_{\vect{k}}
    \sum_{m = 1}^{2 \nbands}
    \kron{\vect{k}}{\vect{0}}
    \frac{
        \metric_{mm}
        {\qty(\bogp_{\mu}^{(1)})}_{m}
        \conj{\qty(\bogp_{\nu}^{(1)})}_{m}
    }{
        \iu \hbar \omega_l - \tilde{\varepsilon}_{m\vect{k}}
    }
    .
\end{align}
\end{subequations}
We have used
$
    \sum_{n = 1}^{2 \nbands}
    \paulix_{mn}
    {\qty(\bogp_{\nu}^{(1)})}_{n}
    =
    \conj{\qty(\bogp_{\nu}^{(1)})}_{m}
$
due to the hermiticity of $\vect{P}^{(1)}$.
This result allows to compute the one-magnon electric susceptibility:
\begin{subequations}
\begin{align}
    \chi_{\mu\nu}^{(1)}(\omega)
    &=
    -V
    C_{P_{\mu}^{(1)} P_{\nu}^{(1)}}^{\text{R}}(\omega)
    ,
    \\
    &=
    -V
    C_{P_{\mu}^{(1)} P_{\nu}^{(1)}}(\iu \omega_l \to \omega + \iu 0^+)
    ,
    \\
    &=
    -V
    \lim_{\eta \to 0^+}
    \sum_{\vect{k}}
    \sum_{m = 1}^{2 \nbands}
    \kron{\vect{k}}{\vect{0}}
    \frac{
        \metric_{mm}
        {\qty(\bogp_{\mu}^{(1)})}_{m}
        \conj{\qty(\bogp_{\nu}^{(1)})}_{m}
    }{
        \hbar \omega - \tilde{\varepsilon}_{m\vect{k}} + \iu \eta
    }
    ,
    \\
    &=
    -V
    \sum_{m = 1}^{2 \nbands}
    \metric_{mm}
    {\qty(\bogp_{\mu}^{(1)})}_{m}
    \conj{\qty(\bogp_{\nu}^{(1)})}_{m}
    \sokfun{\hbar \omega - \tilde{\varepsilon}_{n \vect{k} = \vect{0}}}
    ,
\end{align}
\end{subequations}
where the Dirac identity
$
    \sokfun{x}
    =
    \pv\qty(\frac{1}{x})
    -
    \iu \mathpi \delta(x)
$
is obtained taking the limit $\eta \to 0^+$ according to the Sokhotski-Plemelj theorem~\cite{weinberg_quantum_1995}.
Here $\pv$ is the principal value and $\delta$ denotes the delta distribution.

For the special case of $\mu = \nu$ and $\omega > 0$, and focusing on the imaginary part, we obtain
\begin{align}
    \Im \chi_{\mu \mu}^{(1)}(\omega)
    &=
    \mathpi V
    \sum_{m = 1}^{\nbands}
    \abs{{\qty(\bogp_{\mu}^{(1)})}_{m}}^2
    \delta\qty(\hbar \omega - \varepsilon_{n \vect{k} = \vect{0}})
    .
    \label{eq:chi_e_omag_imag}
\end{align}
Here, the photon energy $\hbar \omega$ has to match a magnon energy at $\vect{k} = \vect{0}$ in order to conserve both energy and crystal momentum.
Thus, any resonances of $\Im \chi_{\mu \mu}^{(1)}(\omega)$ may only appear at energies with a nonzero magnon density of states.
The electric activity of a magnon mode is governed by ${\qty(\bogp_{\mu}^{(1)})}_{m}$, for which both the lattice geometry and the magnon wave function play a role.
Importantly, the one-magnon processes can be observed at zero temperature, since $\Im \chi_{\mu \mu}^{(1)}(\omega)$ does not depend on temperature.

We now turn to a ferromagnet with centrosymmetric sites or $n$-fold rotational symmetry ($n > 1$) at each site---an example is  the honeycomb lattice with ferromagnetic order---and focus on the KNB effect.
Moreover, we only consider the states at $\vect{k} = \vect{0}$ in the following.
As seen in Eq.~\eqref{eq:knb_lin_sub}, the coefficients of $\vect{P}^{(1)}$ in the Holstein-Primakoff basis vanish at each site, because the nearest-neighbor bond vectors add up to zero due to 3-fold rotational symmetry.
Only at the edges, which break the rotational symmetry locally, the linear electric polarization may be nonzero.
Hence, in order for a photon to create a \emph{single} magnon, this magnon mode has to have a finite probability amplitude at \emph{edge} sites with dangling bonds.

While this is a necessary condition, it is not sufficient, as the phase of the wave functions at the edge sites can cause destructive interference.
Indeed, if a state is an even eigenstate of the parity operator (it has to be an eigenstate if the system is globally centrosymmetric and the state is not degenerate), i.e., if it has the parity eigenvalue +1, its dynamical electric dipole moment associated with $\vect{P}^{(1)}$ interferes destructively with itself between the two edges of a nanoribbon, such that it becomes electrically inactive.
Antisymmetric states  (odd parity, eigenvalue $-1$) may exhibit an electrical response, but could in principle also interfere destructively at a single edge.
This scenario is excluded for nanoribbons whose edges are made up of single sites with dangling bonds per side, e.g., for zigzag- or bearded-terminated honeycomb lattices.

Importantly, pure edge modes are not eigenfunctions of the parity operator and, thus, \emph{can be electrically active}.
They are neither of odd nor even parity.
As a result, centrosymmetry requires the degeneracy of the edge modes, such that a common eigenbasis of both the Hamiltonian and the parity operator can be constructed.
This change of basis, however, does not change the absorption spectra, because one of the constructed states has to have even parity.


For the numerical calculations presented in later sections and in the main text, we replace the delta function by a Lorentzian \begin{align}
    \delta(x)
    \approx
    \frac{1}{\mathpi}
    \frac{\eta}{x^2 + \eta^2}
\end{align}
of finite width $\eta$. This way, the finite linewidth of the magnon modes that may result from many-body interactions is mimicked~\cite{habel_breakdown_2023}.

\subsubsection{Two-magnon processes}
\label{sec:tmag_el_plb}
The two-magnon electric susceptibility is derived similarly to the one-magnon case, although the expressions become more complicated.
As before, the only assumption concerns the form of representation of the electric polarization.
We require that the bilinear electric polarization operator can be represented as
\begin{align}
    \vect{P}_{\gamma}^{(2)}
    =
    \frac{1}{2}
    \sum_{\vect{k}}
    \adjvect{\varPsi}_{\vect{k}}
    \matr{\bogp}_{\gamma\vect{k}}^{(2)}
    \vect{\varPsi}_{\vect{k}}
    ,
\end{align}
but assumptions on the mechanism are not made.
The two-magnon Matsubara autocorrelation function reads
\begin{subequations}
\begin{align}
    C_{P_{\mu}^{(2)} P_{\nu}^{(2)}}(\omega)
    &=
    -\int_0^\beta \dd{\lambda}
    \e^{\iu \hbar \omega_l \lambda}
    \expval{
        \mathcal{T}
        \hat{P}_\mu^{(2)}(\hbar \lambda)
        \hat{P}_\nu^{(2)}(0)
    }
    ,
    \\
    &=
    -\frac{1}{4}
    \sum_{\vect{k} \vect{k}'}
    \sum_{m, n, p, q = 1}^{2 \nbands}
    {\qty(\bogp_{\mu\vect{k}}^{(2)})}_{mn}
    {\qty(\bogp_{\nu\vect{k}'}^{(2)})}_{pq}
    \expval{
        \adj{\varPsi}_{m\vect{k}}
        \varPsi_{n\vect{k}}
        \adj{\varPsi}_{p\vect{k}'}
        \varPsi_{q\vect{k}'}
    }_0
    \int_0^\beta \dd{\lambda}
    \e^{
        \lambda
        (
            \iu \hbar \omega_l
            +
            \tilde{\varepsilon}_{m\vect{k}}
            -
            \tilde{\varepsilon}_{n\vect{k}}
        )
    }
    ,
    \\
    \label{eq:el_tmag_matcorr_fpoint_expval}
    \begin{split}
    &=
    \frac{1}{4}
    \sum_{\vect{k} \vect{k}'}
    \sum_{m, n, p, q = 1}^{2 \nbands}
    {\qty(\bogp_{\mu\vect{k}}^{(2)})}_{mn}
    {\qty(\bogp_{\nu\vect{k}'}^{(2)})}_{pq}
    \frac{
        \e^{
            \beta
            (
                \tilde{\varepsilon}_{m\vect{k}}
                -
                \tilde{\varepsilon}_{n\vect{k}}
            )
        }
        -
        1
    }{
        \iu \hbar \omega_l
        +
        \tilde{\varepsilon}_{m\vect{k}}
        -
        \tilde{\varepsilon}_{n\vect{k}}
    }
    \\
    &\quad
    \times
    \qty[
        \kron{\vect{k}}{\vect{k}'}
        \metric_{mq}
        \metric_{np}
        \rho(\tilde{\varepsilon}_{m\vect{k}})
        \rho(-\tilde{\varepsilon}_{n\vect{k}})
        +
        \kron{\vect{k}}{(-\vect{k}')}
        \metric_{mm}
        \metric_{nn}
        \paulix_{mp}
        \paulix_{nq}
        \rho(\tilde{\varepsilon}_{m\vect{k}})
        \rho(-\tilde{\varepsilon}_{n\vect{k}})
    ]
    ,
    \end{split}
    \\
    \begin{split}
    &=
    \frac{1}{4}
    \sum_{\vect{k}}
    \sum_{m, n, p, q = 1}^{2 \nbands}
    \metric_{mq}
    \metric_{np}
    {\qty(\bogp_{\mu\vect{k}}^{(2)})}_{mn}
    {\qty(\bogp_{\nu\vect{k}}^{(2)})}_{pq}
    \frac{
        \rho(\tilde{\varepsilon}_{m\vect{k}})
        -
        \rho(\tilde{\varepsilon}_{n\vect{k}})
    }{
        \iu \hbar \omega_l
        +
        \tilde{\varepsilon}_{m\vect{k}}
        -
        \tilde{\varepsilon}_{n\vect{k}}
    }
    \\
    &\quad
    +
    \frac{1}{4}
    \sum_{\vect{k}}
    \sum_{m, n, p, q = 1}^{2 \nbands}
    \metric_{mm}
    \metric_{nn}
    \paulix_{mp}
    \paulix_{nq}
    {\qty(\bogp_{\mu\vect{k}}^{(2)})}_{mn}
    {\qty(\bogp_{\nu(-\vect{k})}^{(2)})}_{pq}
    \frac{
        \rho(\tilde{\varepsilon}_{m\vect{k}})
        -
        \rho(\tilde{\varepsilon}_{n\vect{k}})
    }{
        \iu \hbar \omega_l
        +
        \tilde{\varepsilon}_{m\vect{k}}
        -
        \tilde{\varepsilon}_{n\vect{k}}
    }
    ,
    \end{split}
    \\
    &=
    \frac{1}{2}
    \sum_{\vect{k}}
    \sum_{m, n = 1}^{2 \nbands}
    \metric_{mm}
    \metric_{nn}
    {\qty(\bogp_{\mu\vect{k}}^{(2)})}_{mn}
    {\qty(\bogp_{\nu\vect{k}}^{(2)})}_{nm}
    \frac{
        \rho(\tilde{\varepsilon}_{m\vect{k}})
        -
        \rho(\tilde{\varepsilon}_{n\vect{k}})
    }{
        \iu \hbar \omega_l
        +
        \tilde{\varepsilon}_{m\vect{k}}
        -
        \tilde{\varepsilon}_{n\vect{k}}
    }
    .
    \label{eq:el_tmag_matcorr_lstep}
\end{align}
\end{subequations}
The four-point correlator in Eq.~\eqref{eq:el_tmag_matcorr_fpoint_expval} has been reduced to 3 two-point correlators using Wick's theorem~\cite{mahan_many-particle_2000} (one of which is disconnected and vanishes because of the prefactor), the latter being calculated according to the identities in Eq.~\eqref{eq:twopoint_identities}.
For Eq.~\eqref{eq:el_tmag_matcorr_lstep} we have assumed that the bilinear coefficients
$
    \matr{\bogp}_{\mu\vect{k}}^{(2)}
$
have been defined such that they obey particle-hole symmetry,
$
    \matr{\bogp}_{\mu\vect{k}}^{(2)}
    =
    \matr{\paulix}
    \trp{\qty(\matr{\bogp}_{\mu(-\vect{k})}^{(2)})}
    \matr{\paulix}
    .
$


Now, we write the two-magnon electric susceptibility as
\begin{align}
    \chi_{\mu\nu}^{(2)}(\omega)
    &=
    -
    \frac{V}{2}
    \sum_{\vect{k}}
    \sum_{m, n = 1}^{2 \nbands}
    \metric_{mm}
    \metric_{nn}
    {\qty(\bogp_{\mu\vect{k}}^{(2)})}_{mn}
    {\qty(\bogp_{\nu\vect{k}}^{(2)})}_{nm}
    \qty[
        \rho(\tilde{\varepsilon}_{m\vect{k}})
        -
        \rho(\tilde{\varepsilon}_{n\vect{k}})
    ]
    \sokfun{
        \hbar \omega
        +
        \tilde{\varepsilon}_{m\vect{k}}
        -
        \tilde{\varepsilon}_{n\vect{k}}
    }
    .
    \label{eq:chi_e_tmag}
\end{align}
This quite abstract expression is better understood by considering the diagonal imaginary components
\begin{align}
    \begin{split}
    \Im \chi_{\mu\mu}^{(2)}(\omega)
    &=
    \mathpi V
    \sum_{\vect{k}}
    \sum_{m, n = 1}^{\nbands}
    \abs{{\qty(\bogp_{\mu\vect{k}}^{(2)})}_{mn}}^2
    \qty[
        \rho\qty(\varepsilon_{m\vect{k}})
        -
        \rho\qty(\varepsilon_{n\vect{k}})
    ]
    \delta\qty(
        \hbar \omega
        +
        \varepsilon_{m\vect{k}}
        -
        \varepsilon_{n\vect{k}}
    )
    \\
    &\quad
    +
    \frac{\mathpi V}{2}
    \sum_{\vect{k}}
    \sum_{m, n = 1}^{\nbands}
    \abs{{\qty(\bogp_{\mu\vect{k}}^{(2)})}_{m,n+\nbands}}^2
    \qty[
        1
        +
        \rho\qty(\varepsilon_{m\vect{k}})
        +
        \rho\qty(\varepsilon_{n(-\vect{k})})
    ]
    \delta\qty(
        \hbar \omega
        -
        \varepsilon_{m\vect{k}}
        -
        \varepsilon_{n(-\vect{k})}
    )
    \\
    &\quad
    -
    \frac{\mathpi V}{2}
    \sum_{\vect{k}}
    \sum_{m, n = 1}^{\nbands}
    \abs{{\qty(\bogp_{\mu\vect{k}}^{(2)})}_{m,n+\nbands}}^2
    \qty[
        1
        +
        \rho\qty(\varepsilon_{m\vect{k}})
        +
        \rho\qty(\varepsilon_{n(-\vect{k})})
    ]
    \delta\qty(
        \hbar \omega
        +
        \varepsilon_{m\vect{k}}
        +
        \varepsilon_{n(-\vect{k})}
    )
    .
    \end{split}
\end{align}
This result is readily interpreted, as the individual terms correspond to specific kinds of processes. The first line describes an interband transition of magnons due to a photon; an \emph{existing} magnon absorbs a photon and is excited into a state with increased energy without changing its crystal momentum. This process happens only at finite temperatures and, therefore, can be frozen out for gapped magnons if the temperature is small compared to the spin-wave gap.

The second line corresponds to a process that can take place at zero temperature. Here, a photon is annihilated and creates two magnons with lower energies and opposite crystal momenta.
The efficiency of this process increases with temperatures.
Conversely, two magnons can also be destroyed under emission of a photon.
This may only happen if the unperturbed Hamiltonian $\hamil_0$ or $\vect{P}^{(2)}$ do not conserve the magnon number, which is not the case for the Heisenberg ferromagnet with Dzyaloshinskii-Moryia interaction considered in the main text.
However, one-magnon absorption processes are allowed, as any linear-in-magnons operator such as $\vect{P}^{(1)}$ always breaks the magnon-number conservation if being finite.
The third line is analogous to the second, but for negative frequencies.

One- and two-magnon absorptions can be distinguished due to their anisotropic selection rules.
Considering the KNB mechanism in an out-of-plane ($z$) ferromagnet on a two-dimensional lattice, $\Im \chi_{\mu\mu}^{(1)} \propto \kron{\mu}{z}$ is only nonzero if $\vect{E}$ has an out-of-plane component [cf. Eq.~\eqref{eq:knb_lin_sub}], while $\Im \chi_{\mu\mu}^{(2)} \propto (1 - \kron{\mu}{z})$ requires an in-plane component. Hence, one- and two-magnon absorption adhere to the symmetries in different ways.

\subsection{Magnetic susceptibility}
The magnetic susceptibility
\begin{align}
    \varDelta \expval{M_{\mu}(\omega)}
    =
    \sum_{\nu = x, y, z}
    \chi_{\mu\nu}^{\text{m}}(\omega)
    B_{\nu}(\omega)
\end{align}
mediates between the nonequilibrium part of the magnetization
$
    \varDelta \expval{M_{\mu}(\omega)}
$
and the ac external magnetic field $B_{\nu}(\omega)$.
In this section we derive an expression for $\chi_{\mu\nu}^{\text{m}}(\omega)$ analogously to the previous section.
Hereafter we refer to the magnetic susceptibility by the short form
$
    \chi_{\mu\nu}(\omega)
    \coloneqq
    \chi_{\mu\nu}^{\text{m}}(\omega)
    .
$

The total Hamiltonian of the system is perturbed by
\begin{align}
    \hamil'(t)
    =
    - V \vect{M} \cdot \vect{B}(t)
    ,
\end{align}
which is captured in the retarded magnetization autocorrelation function
\begin{align}
    \chi_{\mu\nu}(\omega)
    =
    - V
    C_{M_\mu M_\nu}^{\text{R}}(\omega).
\end{align}
As before, the latter is computed by analytic continuation of the corresponding Matsubara correlation function
\begin{align}
    C_{M_\mu M_\nu}(\omega)
    =
    -\int_0^\beta \dd{\lambda}
    \e^{\iu \hbar \omega_l \lambda}
    \expval{
        \mathcal{T}
        \hat{M}_\mu(\hbar \lambda)
        \hat{M}_\nu(0)
    }_0
\end{align}
$\qty[
    \hat{M}_{\mu/\nu}(\tau)
    =
    \exp\qty(\frac{\tau}{\hbar} \hamil_0)
    M_{\mu/\nu}
    \exp\qty(-\frac{\tau}{\hbar} \hamil_0)
]$
by splitting the magnetization operators into their constant, linear and bilinear parts, respectively.

The magnetization is defined as
\begin{subequations}
\begin{align}
    \vect{M}
    &\coloneqq
    -\frac{g \mub}{\hbar V}
    \sum_{i} \vect{S}_i
    ,
    \label{eq:def_magnetization}
    \\
    &=
    \vect{M}^{(0)}
    +
    \vect{M}^{(1)}
    +
    \vect{M}^{(2)}
    +
    O\qty(S^{\nicefrac{1}{2}})
    .
\end{align}
\end{subequations}
The various orders $\vect{M}^{(i)}$ are derived using the Holstein-Primakoff transformation [Eq.~\eqref{eq:hpt}]:
\begin{subequations}
\begin{align}
    \vect{M}^{(0)}
    &=
    -\frac{g \mub}{\hbar V}
    \sum_{i} S_i \unitvect{z}_i
    ,
    \\
    M_{\gamma}^{(1)}
    &=
    \vect{M}_{\gamma}^{(1)}
    \cdot
    \vect{\psi}_{\vect{k} = \vect{0}}
    ,
    \\
    M_{\gamma}^{(2)}
    &=
    \frac{1}{2}
    \sum_{\vect{k}}
    \qty(
        \adjvect{\psi}_{\vect{k}}
        \matr{M}_{\gamma\vect{k}}^{(2)}
        \vect{\psi}_{\vect{k}}
        -
        \frac{\Tr \matr{M}_{\gamma\vect{k}}^{(2)}}{2}
    )
    ,
\end{align}
\end{subequations}
where the coefficients of the linear part read
\begin{subequations}
\begin{align}
    \vect{M}_{\gamma}^{(1)}
    &=
    \begin{pmatrix}
        \vect{U}_{\gamma}
        \\
        \conjvect{U}_{\gamma}
    \end{pmatrix}
    ,
    \\
    {\qty(U_{\gamma})}_{m}
    &=
    -\sqrt{\frac{\nucs}{2}}
    \frac{g \mub S_m}{V}
    \hat{e}_{m,\gamma}^{-}
    ,
\end{align}%
\label{eq:hpm_lin}
\end{subequations}
and those of the bilinear part read
\begin{align}
    \matr{M}_{\gamma}^{(2)}
    &=
    \frac{g \mub}{V}
    \diag\mqty(
        \hat{z}_{1,\gamma} & \cdots & \hat{z}_{\nbands,\gamma}
        &
        \hat{z}_{1,\gamma} & \cdots & \hat{z}_{\nbands,\gamma}
    )
    .
    \label{eq:hpm_bilin}
\end{align}
For completeness, we also transform the magnetization operators from their Holstein-Primakoff representation to the basis of the magnon eigenmodes:
\begin{subequations}
\begin{align}
    M_{\gamma}^{(1)}
    &=
    \vect{\bogm}_{\gamma}^{(1)}
    \cdot
    \vect{\varPsi}_{\vect{k} = \vect{0}}
    ,
    &
    \vect{\bogm}_{\gamma}^{(1)}
    &=
    \trpbogmatr_{\vect{k} = \vect{0}}
    \vect{M}_{\gamma}^{(1)}
    ,
    \label{eq:bogm_lin}
    \\
    M_{\gamma}^{(2)}
    &=
    \frac{1}{2}
    \sum_{\vect{k}}
    \qty(
        \adjvect{\varPsi}_{\vect{k}}
        \matr{\bogm}_{\gamma\vect{k}}^{(2)}
        \vect{\varPsi}_{\vect{k}}
        -
        \frac{\Tr \matr{M}_{\gamma\vect{k}}^{(2)}}{2}
    )
    ,
    &
    \vect{\bogm}_{\gamma\vect{k}}^{(2)}
    &=
    \adjbogmatr_{\vect{k}}
    \vect{M}_{\gamma\vect{k}}^{(2)}
    \bogmatr_{\vect{k}}
    .
    \label{eq:bogm_bilin}
\end{align}
\end{subequations}
For the following derivations we use the notation and the identities introduced in the previous section. 

\subsubsection{One-magnon processes}
Although having obtained concrete expressions for the magnetization operator, we only use the form in Eq.~\eqref{eq:bogm_lin} for deriving the one-magnon magnetic susceptibility. Since the formulas resemble the one-magnon electric susceptibility, direct transfer of the results yields
\begin{align}
    \chi_{\mu\nu}^{(1)}(\omega)
    &=
    -V
    \sum_{m = 1}^{2 \nbands}
    \metric_{mm}
    {\qty(\bogm_{\mu}^{(1)})}_{m}
    \conj{\qty(\bogm_{\nu}^{(1)})}_{m}
    \sokfun{\hbar \omega - \tilde{\varepsilon}_{n \vect{k} = \vect{0}}}
    .
\end{align}
As for the electric case, only magnons with infinite wavelength contribute to the one-magnon magnetic susceptibility.
Considering a ferromagnet with magnetization along $z$ in an arbitrary lattice and applying the concrete definition Eq.~\eqref{eq:def_magnetization} for $\vect{M}$, $\chi_{\mu\nu}^{(1)}$ may only be nonzero if $\mu \neq z \neq \nu$ [cf. Eq.~\eqref{eq:hpm_lin}].
Pictorially, the magnetic field perpendicular to the magnetization excites a magnon that leads to the precession of the magnetization about its ground state direction, which induces time-varying orthogonal components of the magnetization.
The specific magnetic activity of a magnon mode depends on the phase relation between the spins in a unit cell (the phases between unit cells are synchronized due to $\vect{k} = \vect{0}$), which can lead to interference effects.
In a two-sublattice system, the acoustic magnons embody spin precession with equal phases; they are magnetically active, while the optical magnons imply antiphase spin precession, such that the transversal sublattice magnetizations cancel each other and the optical magnon is magnetically inactive.

The above outlined derivation and discussion is the well-known ferromagnetic resonance of the Kittel mode~\cite{gurevich1996magnetization}.

\subsubsection{Two-magnon processes}
To derive the two-magnon magnetic susceptibility, we use the representation in Eq.~\eqref{eq:bogm_bilin}, which is the same as the bilinear electric polarization, apart from the concrete matrix elements, the details of which do not affect the generality of the derivation.
Hence, we carry over the result from Eq.~\eqref{eq:chi_e_tmag} and substitute the matrix elements:
\begin{align}
    \chi_{\mu\nu}^{(2)}(\omega)
    &=
    -
    \frac{V}{2}
    \sum_{\vect{k}}
    \sum_{m, n = 1}^{2 \nbands}
    \metric_{mm}
    \metric_{nn}
    {\qty(\bogm_{\mu\vect{k}}^{(2)})}_{mn}
    {\qty(\bogm_{\nu\vect{k}}^{(2)})}_{nm}
    \qty[
        \rho(\tilde{\varepsilon}_{m\vect{k}})
        -
        \rho(\tilde{\varepsilon}_{n\vect{k}})
    ]
    \sokfun{
        \hbar \omega
        +
        \tilde{\varepsilon}_{m\vect{k}}
        -
        \tilde{\varepsilon}_{n\vect{k}}
    }
    .
    \label{eq:chi_m_tmag}
\end{align}

To study the shape of this response tensor, we consider a ferromagnet with magnetization along $z$ and $\vect{M}$ as defined in Eq.~\eqref{eq:def_magnetization}. Due to the matrix elements of the bilinear magnetization operator in the Holstein-Primakoff basis [cf. Eq.~\eqref{eq:hpm_bilin}], $\chi_{\mu\nu}^{(2)}(\omega)$ is only nonzero for $\mu = \nu = z$.

As for the two-magnon electric susceptibility, one distinguishes between interband transitions, that may only arise at nonzero temperatures, and pair creation or annihilation of magnons at all temperatures.
The former process should be inefficient for states with opposite spins in a collinear system, because such a spin flip is enabled by spin-orbit coupling (SOC) that breaks spin conservation. However, it can happen in multi-band ferromagnets even without SOC\@.
The latter process is different as it depends on breaking of spin conservation in ferromagnets. It can however appear in collinear antiferromagnets without SOC\@.

\section{Additional results}

\subsection{Additional results for the honeycomb-lattice Heisenberg-DMI model}
In this section, we present supplementary results for the model discussed in the main text. It is based on the spin Hamiltonian
\begin{align}
    \hamil
    &=
    -\sum_{r=1}^3 \frac{J_r}{2 \hbar^2}
    \sum_{\langle ij \rangle_r} \vect{S}_i \cdot \vect{S}_j
    +
    \frac{1}{2 \hbar^2}
    \sum_{\langle ij \rangle_2}
    \vect{D}_{ij}
    \cdot
    \qty(\vect{S}_i \times \vect{S}_j)
    -
    \frac{A}{\hbar^2}
    \sum_i
    {\qty(S_i^z)}^2
    \label{eq:shamil_dmi}
\end{align}
with parameters $J_1 = 1, J_2 = 0.25, J_3 = 0, D_z = -0.1, A = 0.1$, and $S = 1$.
While the exchange interactions parameterized by $J_1$, $J_2$, and $J_3$ govern the total bandwidth as well as the bandwidths of the two bulk bands, the Dzyaloshinskii-Moryia interaction (DMI) opens a topologically nontrivial bulk band gap that is bridged by topological edge states for open boundary conditions (nanoribbon or flake)~%
\cite{
    owerre_first_2016,
    kim_realization_2016%
}.
The easy-axis anisotropy $A$ rigidly shifts the magnon spectrum.

The magnetic ground state is a collinear out-of-plane ferromagnet with a conserved total out-of-plane spin moment and conserved magnon number.
In nanoribbons or flakes the ground state remains collinear~\cite{habel_breakdown_2023}. The collinearity of the ground state ensures that the local electric polarization vanishes in the ground state.
We neglect inhomogeneities in the spin Hamiltonian that might arise from finite size effects~\cite{kvashnin_relativistic_2020}.

\subsubsection{Local electric polarization}
\begin{figure}
    \centering
    \includegraphics[width=\textwidth]{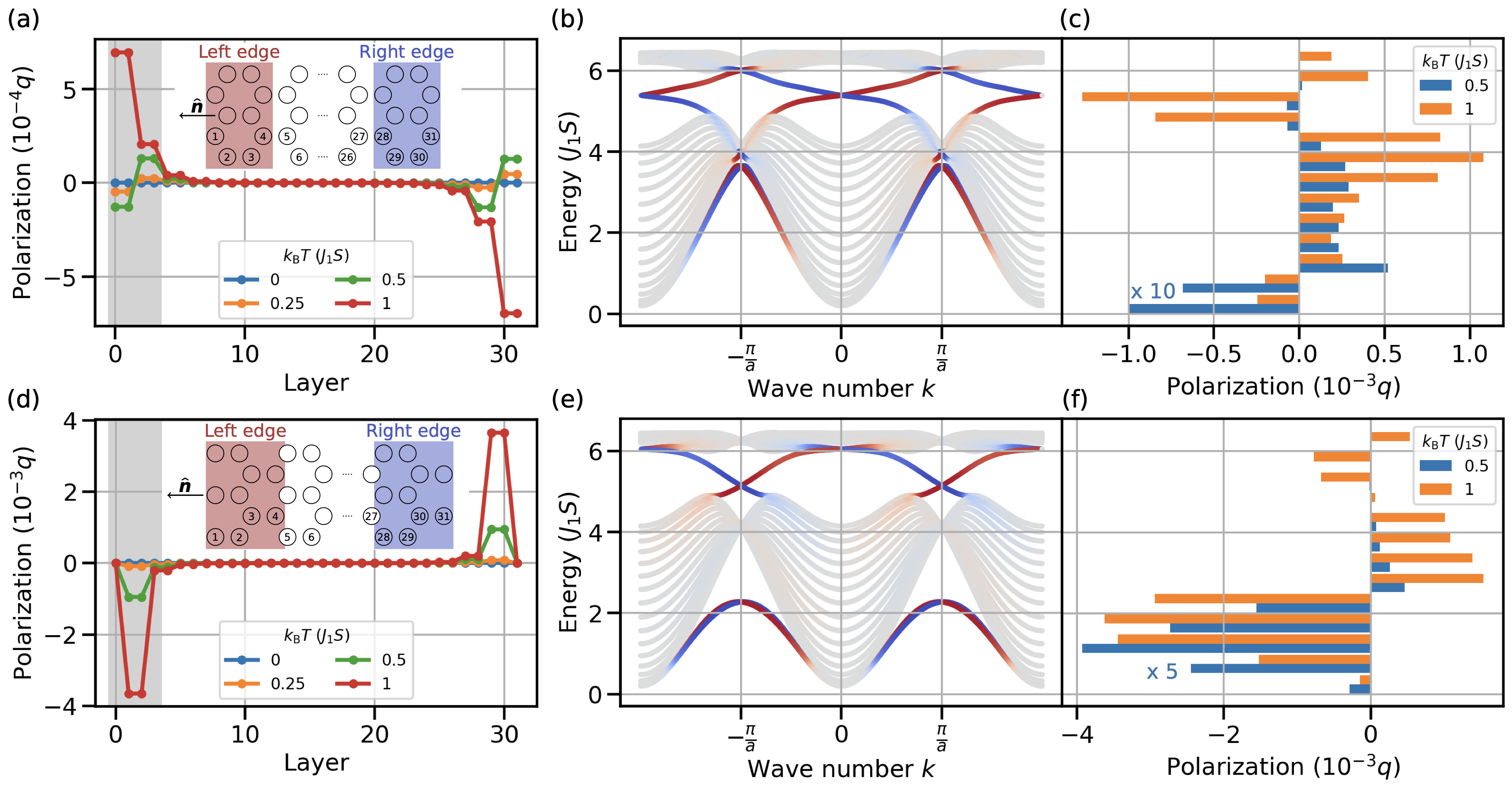}
    \caption{
        Local electric polarization in nanoribbon geometry with 32 layers due to the KNB effect projected onto the outward-facing in-plane normal vector $\unitvect{n}$ of the left edge.
        (a) Layer-resolved electric polarization in zigzag nanoribbon for multiple temperatures (cf.~legend).
        Inset: Section of the nanoribbon including layer labels and the normal vector $\unitvect{n}$.
        The system is finite/infinite along the horizontal/vertical direction, respectively.
        (b) Magnon spectrum of the zigzag nanoribbon with localization on the left (red) and right (blue) edges [defined as the 4 outermost layers per side; cf.~inset in panel~(a)] encoded by color.
        (c) Energy-resolved contributions of the magnon modes to the electric polarization of the zigzag nanoribbon at the left edge [highlighted in panel~(a) by gray background] for two selected temperatures (cf.~legend).
        Each bar comprises the accumulated contributions for an energy interval of $J_1 S / 2$.
        The blue bars have been increased by the indicated factor for better visibility (see inset).
        (d,~e,~f) same as (a,~b,~c) for bearded termination.
        The parameters have been specified as $J_1 = 1, J_2 = 0.25, J_3 = 0, D_z = -0.1, A = 0.1, \text{and } S = 1$.
    }
    \label{fig:rel2_edge_polarization}
\end{figure}

In addition to the local electric polarization for the armchair nanoribbons discussed in the main text, we present similar results for zigzag and bearded nanoribbons in Fig.~\ref{fig:rel2_edge_polarization}, for which Eq.~\eqref{eq:knb_loc_pol_eq} has been numerically evaluated.
The unit of the electric polarization is generally given as the $d$-dimensional electric dipole density, which has the dimension of charge over the ($d - 1$)-th power of length. Here, $d$ equals  $1$, such that the electric polarization is given in units of the effective charge $q$.

Starting with the zigzag termination, the layer-dependent electric polarization vanishes at $T = 0$ due to the collinear order of the classical ground state;  the magnon number is conserved, otherwise quantum corrections would appear [Fig.~\ref{fig:rel2_edge_polarization}(a)].
For $\kb T = 0.25$ and 0.5, there is a sign change between layer 1 and 2, which disappears at higher temperatures ($\kb T = 1$).
In Fig.~\ref{fig:rel2_edge_polarization}(b) and (c), the bands below (above) $J_1 S$ yield a negative (positive) contribution, the magnitudes of which depend on temperature.
For higher temperatures the positive contributions dominate, in agreement with Fig.~\ref{fig:rel2_edge_polarization}(a).
Even though Fig.~\ref{fig:rel2_edge_polarization}(c) shows the energy-resolved cumulated local polarization of the first 4 layers, it still contains information about layers 0 and 1 because the local polarization changes more drastically for layers 0 and 1 than for 2 and 3.

The change of sign of the left-edge electric polarization is absent in the bearded nanoribbon [Fig.~\ref{fig:rel2_edge_polarization}(d)].
The polarization at the left edge is negative, as expected from the VME effect, if the edge polarization was dominated by topological chiral edge magnons.
However, the main contribution does not originate from within the bulk gap, but rather from below [cf.~Fig.~\ref{fig:rel2_edge_polarization}(e,~f)].
Note that the contributions from within the gap between approximately 5 and $6 J_1 S$ are negative for both terminations, as the VME effect predicts.

\subsubsection{Electric susceptibility}
\begin{figure}
    \centering
    \includegraphics[width=\textwidth]{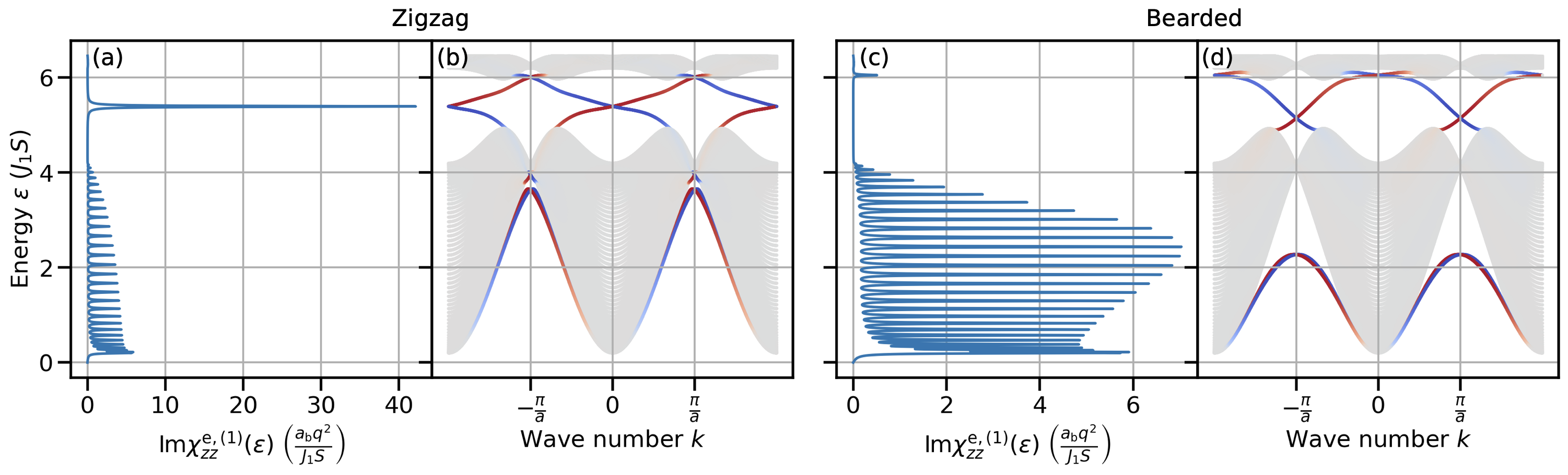}
    \caption{
        (a) Imaginary part of the electric one-magnon susceptibility $\Im \chi_{zz}^{\mathrm{e},(1)}$ of a 120-layers wide nanoribbon with zigzag terminations vs. energy.
        (b) Magnon spectrum of the nanoribbon with wave function amplitude on the left (red) and right (blue) edges encoded by color.
        $a_{\text{b}}$ corresponds to the bulk lattice constant of the underlying honeycomb lattice, while $a$ is the lattice constant of the nanoribbons (here $a = a_{\text{b}}$).
        (c,~d) same as (a,~b) for bearded terminations.
        The parameters have been specified as $J_1 = 1, J_2 = 0.25, J_3 = 0, D_z = -0.1, A = 0.1, \text{and } S = 1$.
        The width of the Lorentzians is $\eta = 0.01 J_1 S$.
    }
    \label{fig:rel2_omag_el_im_plb}
\end{figure}
We have investigated the imaginary one-magnon electric susceptibility for zigzag and bearded edges shown in Fig.~\ref{fig:rel2_omag_el_im_plb}.
Similar to the armchair termination (cf.~main text), the topological modes in the zigzag nanoribbon cross each other in the middle of the gap [Fig.~\ref{fig:rel2_omag_el_im_plb}(b)], which leads to a large in-gap peak in $\Im \chi_{zz}^{\mathrm{e},(1)}$, while the bulk modes below the bulk gap have a small and the bulk modes above have a (almost) nonexistent absorption [Fig.~\ref{fig:rel2_omag_el_im_plb}(a)].

For the bearded edges, the lower bulk bands have a much larger electric response than the topological magnons [Fig.~\ref{fig:rel2_omag_el_im_plb}(c)] because the latter only intersect the $k = 0$ line in the vicinity of the upper bulk band quasi-continuum, where only a direct, but no indirect gap exists [Fig.~\ref{fig:rel2_omag_el_im_plb}(d)].
This finding demonstrates that not only the existence, but also the dispersion of the topological magnons is essential for the electric activity of these chiral modes.

\begin{figure}
    \centering
    \includegraphics[width=\textwidth]{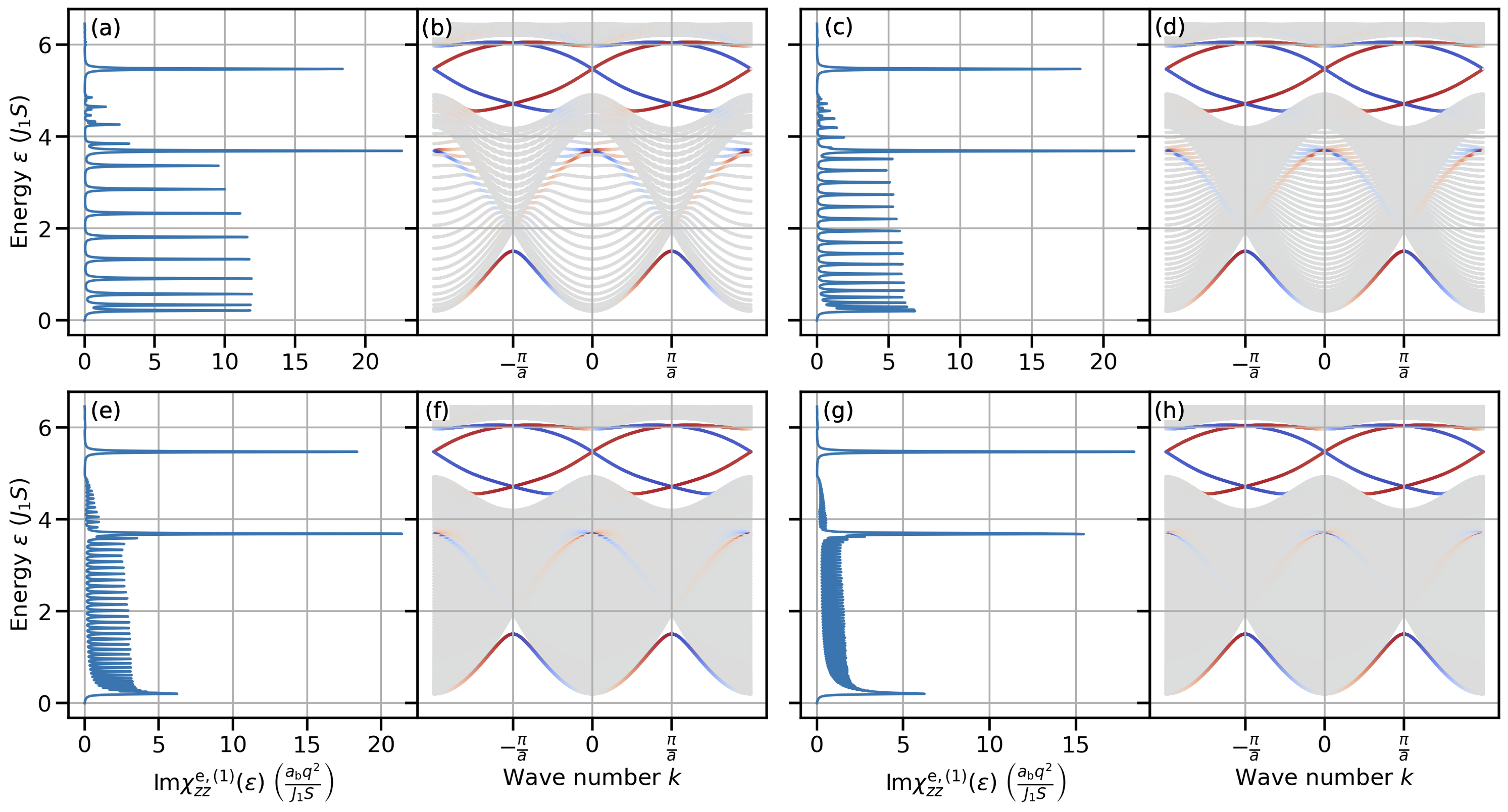}
    \caption{
        Imaginary part of the electric one-magnon susceptibility $\Im \chi_{zz}^{\mathrm{e},(1)}$ vs. energy and magnon spectra of nanoribbons of (a,~b) 40, (c,~d) 80, (e,~f) 160, (g,~h) 320 layers with armchair terminations.
        The magnon spectra encode the wave function amplitudes on the left (red) and right (blue) edges by color.
        $a_{\text{b}}$ is the bulk lattice constant of the underlying honeycomb lattice, while $a = \sqrt{3} a_{\text{b}}$ is the lattice constant of the armchair nanoribbon.
        The parameters have been specified as $J_1 = 1, J_2 = 0.25, J_3 = 0, D_z = -0.1, A = 0.1, \text{and } S = 1$.
        The width of the Lorentzians is $\eta = 0.01 J_1 S$.
    }
    \label{fig:rel2_omag_el_im_plb_width}
\end{figure}
It is important to note two things concerning finite size effects (cf.~Fig.~\ref{fig:rel2_omag_el_im_plb_width}).
First, the width of the nanoribbon determines the number of minibands and thereby the number of peaks.
The larger the nanoribbon, the more bands and peaks exist in the band structure and the absorption spectrum.
Above a certain threshold, the absorption spectrum in the bulk energy region becomes a continuum (depending on the linewidth $\eta$).

Second, the width of the nanoribbon changes the peak sizes depending on the properties of the corresponding modes.
The peak size of an edge mode like a topological one does \emph{not} change (as long as the nanoribbon is sufficiently large compared to its localization length).
Bulk modes, on the other hand, will be dispersed across the nanoribbon weakening their amplitudes at the edges and thereby \emph{reducing} their absorption peaks when the size of the nanoribbon increases.

This statement holds as long as the peaks are sufficiently narrow and do not overlap (i.e., $\eta$ small compared to the splitting).
If the linewidth is larger than the energy splittings between bulk minibands, the reduction of the peak size is compensated by the emergence of additional minibands that contribute to the absorption and the two effects mutually cancel.
Hence, both edge and bulk modes approach a fixed absorption magnitude with a distinguished structure of sharp peaks from the former and broad continua from the latter.

The fact that the in-gap peak from topological magnons do not scale with the size of the sample has important consequences for their experimental observation, since this contribution vanishes in the thermodynamic limit (i.e., the limit of large samples) when one considers three-dimensional susceptibilities, for which the calculated $\chi_{zz}^{\mathrm{e},(1)}$ is divided by the interlayer distance and the width of the nanoribbon.
Hence, the edge contribution would need to be extracted for example by comparing samples of different sizes and extracting the contribution that scales with the inverse sample width when bulk contributions are also present.

\begin{figure}
    \centering
    \includegraphics[width=\textwidth]{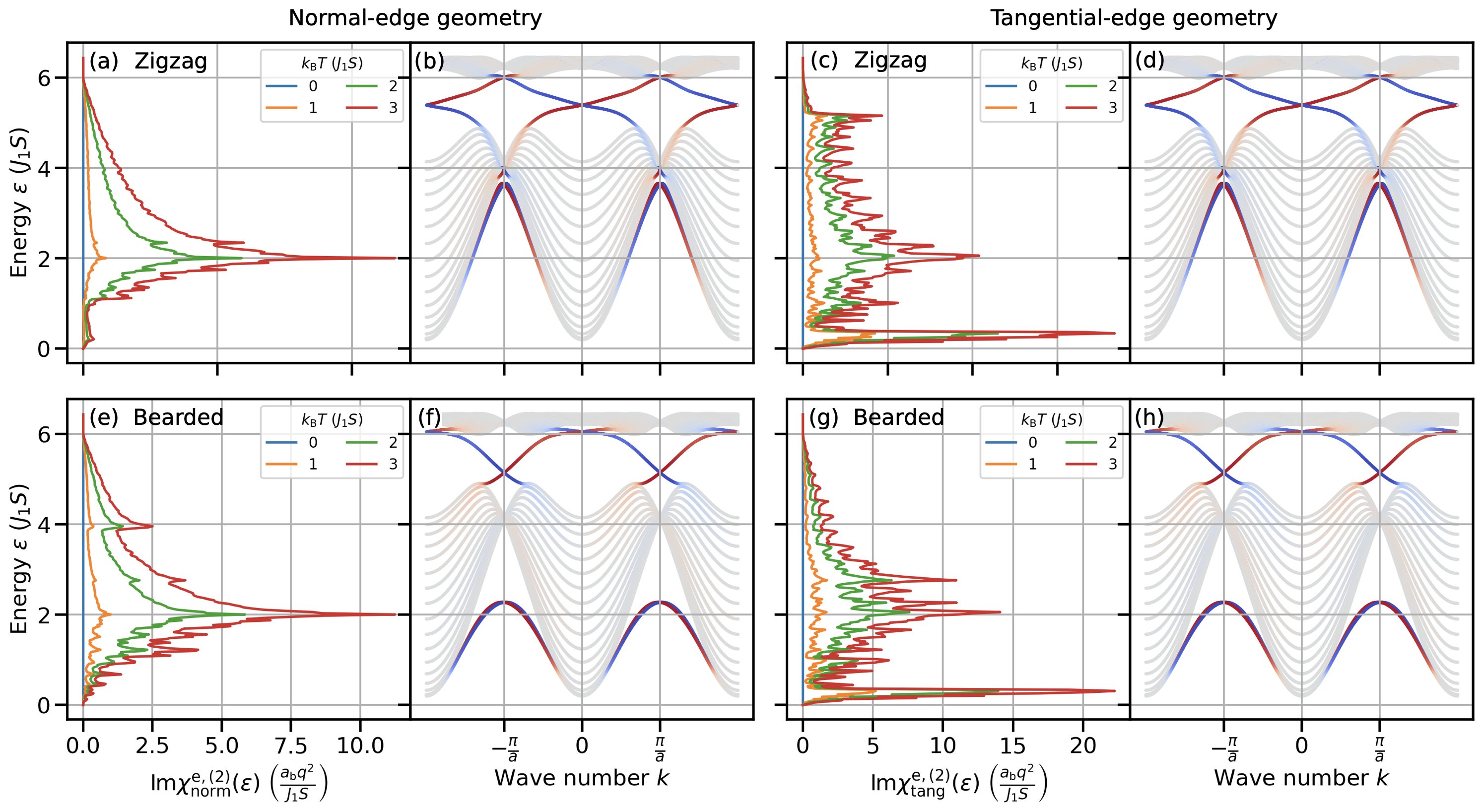}
    \caption{
        (a,~c,~e,~g) Imaginary part of the electric two-magnon susceptibility
        $
            \Im \chi_{zz}^{\mathrm{e},(2)}
        $
        vs. energy for multiple temperatures (see insets) and (b,~d,~f,~h) magnon spectra of 32-layers wide nanoribbons with (a--d) zigzag and (e--h) bearded terminations.
        The red/blue colors of the magnon bands indicates the wave function amplitude at the left/right edges (defined as 4 outermost layers per side).
        $a_{\text{b}}$ corresponds to the bulk lattice constant of the underlying honeycomb lattice, while $a$ is the lattice constant of the nanoribbon (here $a = a_{\text{b}}$).
        The parameters have been specified as $J_1 = 1, J_2 = 0.25, J_3 = 0, D_z = -0.1, A = 0.1, \text{and } S = 1$.
        The width of the Lorentzians is $\eta = 0.01 J_1 S$.
    }
    \label{fig:rel2_tmag_el_im_plb}
\end{figure}
At finite temperatures, the sharp in-gap peaks due to the one-magnon processes are modified due to the temperate-activated two-magnon processes (cf. Sec.~\ref{sec:tmag_el_plb}).
Indeed, the imaginary part of the two-magnon electric susceptibilities vanish for $T = 0$ as displayed in Fig.~\ref{fig:rel2_tmag_el_im_plb} for 32-layers wide nanoribbons with (a--d) zigzag and (e--h) bearded terminations.
If the electric field is oriented normal to the zigzag edges and the component of the resulting electric polarization parallel to the electric field is measured (denoted as \enquote{normal-edge geometry}), the two-magnon electric susceptibility primarily shows a peak at $2 J_1 S$ that increases with temperature [Fig.~\ref{fig:rel2_tmag_el_im_plb}(a)].
Nonzero signals can be observed across the entire energy range of the magnon spectrum [Fig.~\ref{fig:rel2_tmag_el_im_plb}(b)], but above about $6 J_1 S$ the signal drops to zero.
This is to be expected as the two-magnon processes originate from interband transitions of magnons so that the energy absorbed or released by a magnon must be smaller than the maximum magnon energy.
In general, the peaks occur at energies that coincide with energy splittings between the magnon bands.
Furthermore, interband transitions are more likely from low-energy states, which have higher occupation numbers.

Interestingly, by rotating the electric field \emph{tangential} to the edges and measuring the component of the electric polarization also in the tangential direction (denoted as \enquote{tangential-edge geometry}), the electric susceptibility is changed [Fig.~\ref{fig:rel2_tmag_el_im_plb}(c)].
This is contrary to the expectation for the bulk system, where 3-fold rotation symmetry enforces an isotropic in-plane response and must therefore be caused by the symmetry breaking by the edges.
Here, the peak at $2 J_1 S$ is less prominent than for the normal-edge geometry [Fig.~\ref{fig:rel2_tmag_el_im_plb}(a)], but an additional prominent peak emerges at around $0.2 J_1 S$.
Due to the small energies, the interband transition can only be traced back to the minibands in the lower quasicontinuum because excitations across the bulk band gap would require higher energies and interband transitions within or above the bulk band gap are suppressed at low temperatures.

Turning to the bearded-terminated nanoribbons we find mostly similar features in the electric susceptibilities.
In the normal-edge geometry an additional peak is found around $\varepsilon = 4 J_1 S$ [Fig.~\ref{fig:rel2_tmag_el_im_plb}(e)], which must originate from interband transition involving the edge modes since the bulk bands do not sensitively depend on the termination [Fig.~\ref{fig:rel2_tmag_el_im_plb}(f)].
In the tangential-edge geometry the low-energy peak observed in the zigzag nanoribbon [Fig.~\ref{fig:rel2_tmag_el_im_plb}(c)] persists in the bearded nanoribbon [Fig.~\ref{fig:rel2_tmag_el_im_plb}(g)], which suggests that this peak is of bulk origin.

Generally, the two-magnon electric susceptibility has nonzero values at finite temperatures across the bandwidth of the magnon spectrum (including the bulk band gap) and the magnitude can be comparable to the one-magnon electric susceptibility (cf.~Fig.~\ref{fig:rel2_omag_el_im_plb}), although it depends on temperature.
The energy of the peaks cannot be directly linked to the magnon bands, which makes its interpretation difficult.
However, the signals can be frozen out if the temperature is sufficiently low compared to the spin-wave gap.
If the energy splittings in the magnon spectrum are large,
$   
    \Im \chi_{\mu\nu}^{\text{e},(2)}
$
is expected to have fewer, sharper peaks compared to a broader continuum expected for a dense spectrum.
Since the two-magnon electric susceptibilities probes the two-magnon spectral function, in which the in-gap states have a minute contribution, it is unsuitable for their detection.

\subsubsection{Magnetic susceptibility}
\begin{figure}
    \centering
    \includegraphics[width=.5\textwidth]{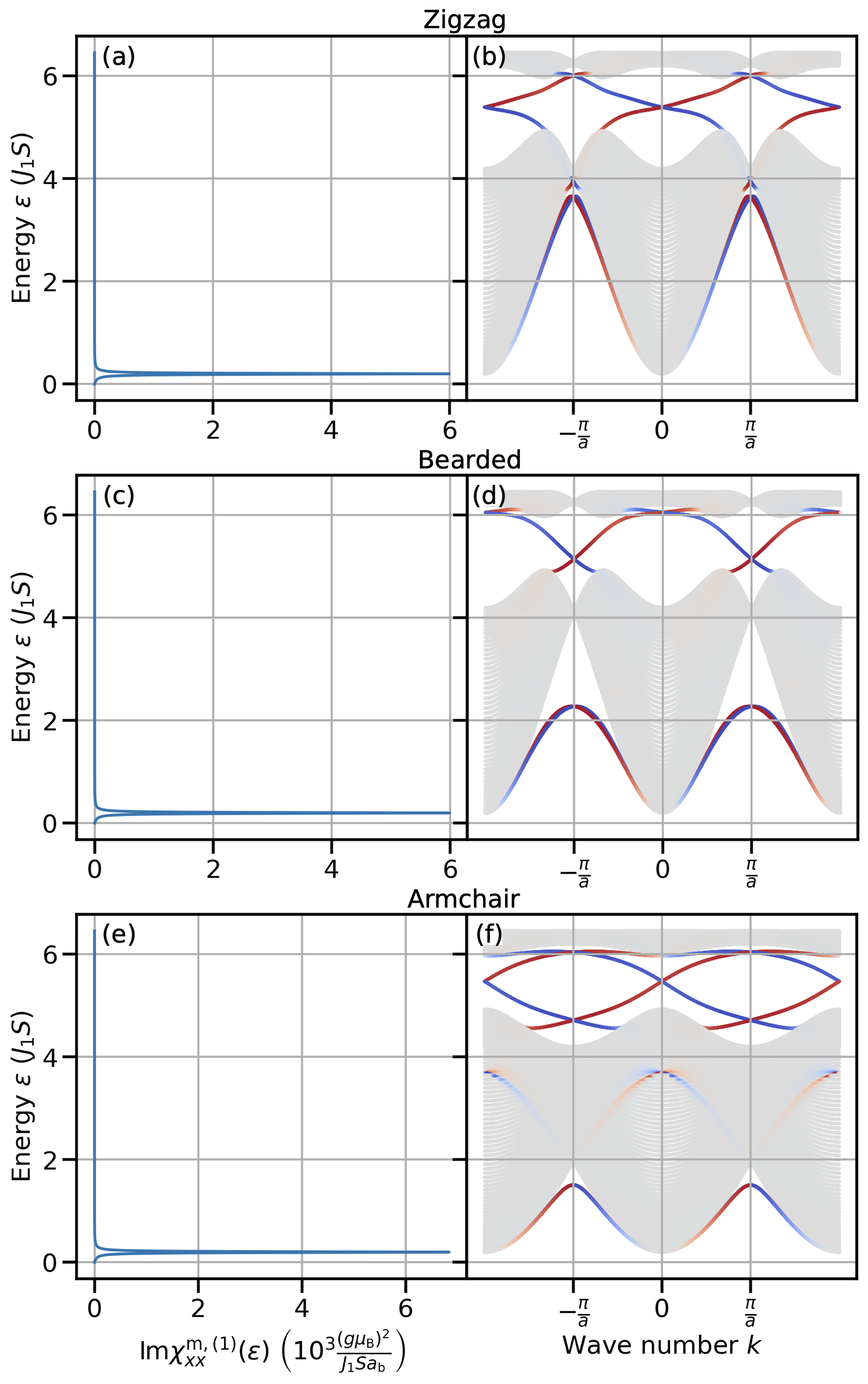}
    \caption{
        (a) Imaginary part of the magnetic one-magnon susceptibility $\Im \chi_{xx}^{\mathrm{m},(1)} = \Im \chi_{yy}^{\mathrm{m},(1)}$ of a 120-layers wide nanoribbon with zigzag terminations vs. energy.
        (b) Magnon spectrum of the nanoribbon with wave function amplitude on the left (red) and right (blue) edges encoded by color.
        $a_{\text{b}}$ corresponds to the bulk lattice constant of the underlying honeycomb lattice, while $a$ is the lattice constant of the nanoribbon.
        (c, d) [(e, f)] same as (a, b) for bearded [armchair] terminations.
        The parameters have been specified as $J_1 = 1, J_2 = 0.25, J_3 = 0, D_z = -0.1, A = 0.1, \text{and } S = 1$.
        The width of the Lorentzians is $\eta = 0.01 J_1 S$.
    }
    \label{fig:rel2_omag_mag_im_plb}
\end{figure}
We have also verified that the magnetic response is unsuitable for the detection of topological magnons as they are magnetically inactive.
This becomes apparent in Fig.~\ref{fig:rel2_omag_mag_im_plb}, where the imaginary magnetic one-magnon susceptibility for (a,~b) zigzag, (c,~d) bearded, and (e,~f) armchair nanoribbons is considered.
In all cases, there is only one low-energy peak at the energy of the spin-wave gap that corresponds to the mode of uniform precession.
The details of the termination are secondary, and the magnetic susceptibility in the nanoribbon behaves similarly to the bulk, where only the acoustic magnon at $k = 0$ is magnetically active.
Even small phase differences between precessing spins (along translationally invariant directions) lead to a cancellation of the transversal moment, such that only the uniform mode can be probed.
The results are consistent with standard ferromagnetic resonance.

Quantitative differences between the terminations can be traced back to the extensive nature of $\Im \chi_{\mu\nu}^{\mathrm{m},(1)}$.
Since $\Im \chi_{\mu\nu}^{\mathrm{m},(1)}$ is a bulk property, it scales with the nanoribbon size and the number of spins per unit cell (not shown) unlike its electrical counterpart.
Since the armchair unit cell comprises twice as many spins for the same nanoribbon width, the expected signal is twice as big.
However, since the armchair lattice constant $a = \sqrt{3} a_{\text{b}}$ is larger than that of the bulk (which is identical to the zigzag and bearded lattice constants), we divide by $\sqrt{3}$.
Hence, the termination only gives rise to a geometric factor, but no physical differences in $\chi_{\mu\nu}^{\mathrm{m},(1)}$.

\subsection{Heisenberg-DMI model with material-specific parameters}
In the following, we use a more realistic parameter set of the same Hamiltonian [Eq.~\ref{eq:shamil_dmi}] obtained from inelastic neutron scattering on the van der Waals honeycomb ferromagnet CrI$_3$:
$
    J_1 = \SI{2.01}{\milli\electronvolt},
    J_2 = \SI{0.16}{\milli\electronvolt},
    J_3 = \SI{-0.08}{\milli\electronvolt},
    D_z = \SI{-0.31}{\milli\electronvolt},
    A = \SI{0.22}{\milli\electronvolt},
    \text{and }
    S = \nicefrac{3}{2}
$
~\cite{chen_topological_2018}.
This material has ferromagnetic intralayer coupling, but its interlayer coupling is antiferromagnetic rendering it a layered antiferromagnet~\cite{huang_layer-dependent_2017}, which would destroy the nontrivial topology as time-reversal invariance would be effectively restored.
Nonetheless, the ferromagnetic order can be effectively retrieved by applying an external magnetic field (around \SI{0.65}{\tesla}~\cite{huang_layer-dependent_2017}).
Here, we focus on the monolayer, which hosts topological magnons provided it can be described with the present Hamiltonian.

\subsubsection{Local electric polarization}
\begin{figure}
    \centering
    \includegraphics[width=\textwidth]{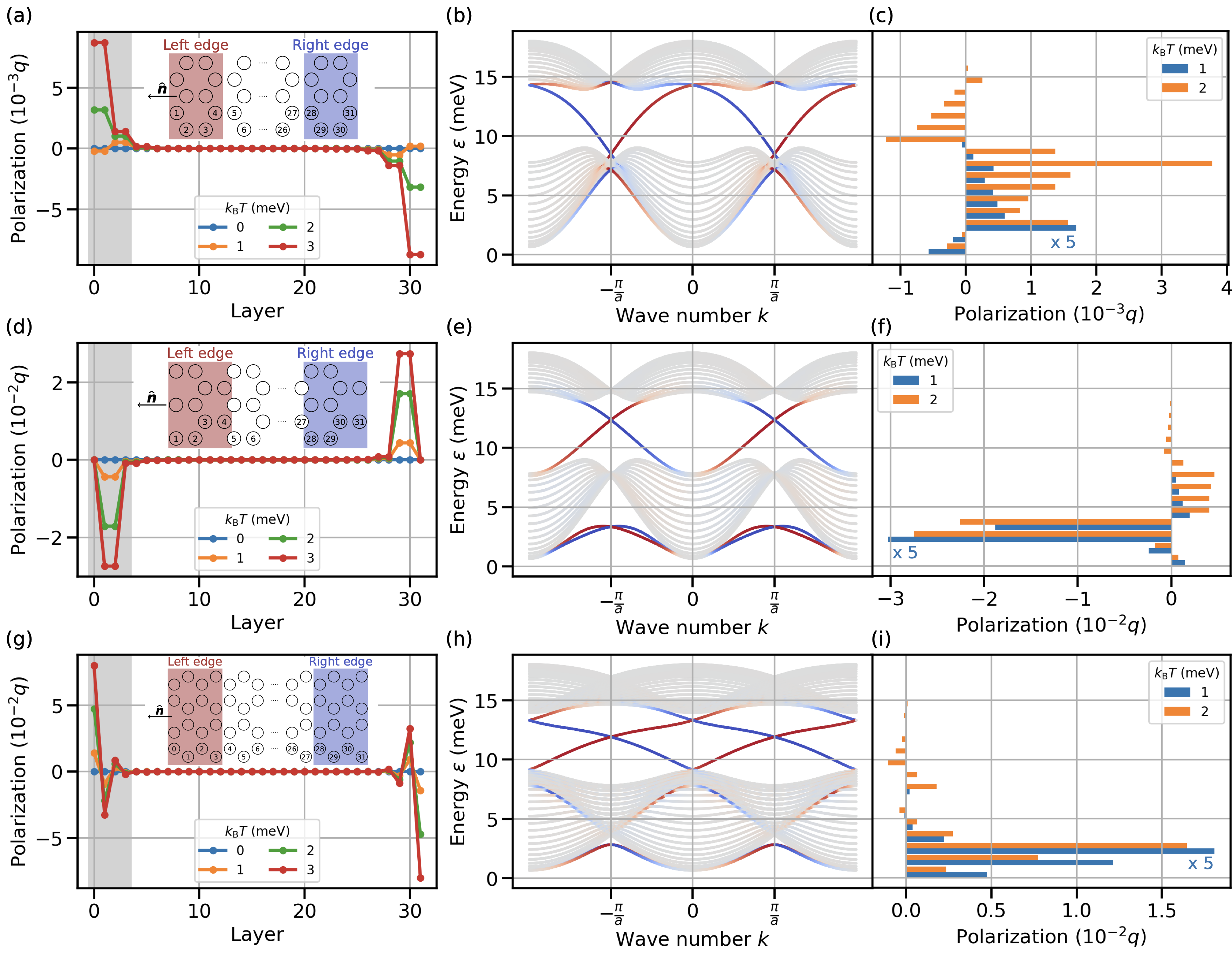}
    \caption{
        Local electric polarization in nanoribbon geometry with 32 layers due to the KNB effect projected onto the outward-facing in-plane normal vector $\unitvect{n}$ of the left edge.
        (a) Layer-resolved electric polarization in zigzag nanoribbon for multiple temperatures (cf.~legend).
        Inset: Section of the nanoribbon including layer labels and the normal vector $\unitvect{n}$.
        The system is finite/infinite along the horizontal/vertical direction, respectively.
        (b) Magnon spectrum of the zigzag nanoribbon with localization on the left (red) and right (blue) edges [defined as the 4 outermost layers per side; cf.~inset in panel~(a)] encoded by color.
        (c) Energy-resolved contributions of the magnon modes to the electric polarization of the zigzag nanoribbon at the left edge [highlighted in panel~(a) by gray background] for two selected temperatures (cf.~legend).
        Each bar comprises the accumulated contributions for an energy interval of \SI{1}{\milli\electronvolt}.
        The blue bars have been increased 5 times for better visibility.
        (d, e, f) [(g,~h,~i)] same as (a, b, c) for bearded [armchair] termination.
        The parameters have been specified as
        $
            J_1 = \SI{2.01}{\milli\electronvolt},
            J_2 = \SI{0.16}{\milli\electronvolt},
            J_3 = \SI{-0.08}{\milli\electronvolt},
            D_z = \SI{-0.31}{\milli\electronvolt},
            A = \SI{0.22}{\milli\electronvolt},
            \text{and }
            S = \nicefrac{3}{2}
            .
        $
    }
    \label{fig:dmi_edge_polarization}
\end{figure}
The local electric polarization for zigzag, bearded, and armchair terminations is displayed in Fig.~\ref{fig:dmi_edge_polarization}.
The results are similar to those for the parameter set used in the main text and above (cf.~Fig.~\ref{fig:rel2_edge_polarization}).
These results confirm that mostly trivial modes govern the sign and the magnitude of the edge polarization.
Only for the bearded edge, the edge polarization has the correct sign to be consistent with the VME effect of topological magnons, but as Fig.~\ref{fig:dmi_edge_polarization}(e) and (f) clearly demonstrate, trivial edge states with weak dispersion are responsible for this observation.

Let us estimate the magnitude of the electric polarization of half of the bearded nanoribbon.
The one-dimensional polarization amounts to $\num{6e-2} q$, which is divided by the interlayer distance of about $\SI{7}{\angstrom}$ for CrI$_3$ and by the width of the nanoribbon, which is approximately \SI{4}{\angstrom} times the number of layers of half of the nanoribbon (here 16)~\cite{mcguire_coupling_2015}.
This results in a three-dimensional electric polarization of about \SI{3}{\micro\coulomb\per\meter\squared}, where we have estimated $q \approx \SI{2e-22}{\coulomb}$ (cf.~Sec.~\ref{sec:eff_charge}).
Since the electrical polarization originates almost exclusively from the edges, this estimate will become smaller for larger samples.

The experimental magnitude of the electric polarization of the edges depends on the weight of the edges in the probed volume.
The larger the number of probed layers, the smaller the weight of the edges and the smaller the averaged polarization.
For example, the electric polarization within the first four layers is \SI{11}{\micro\coulomb\per\meter\squared}, which is larger than for the case above.

Compared to the spin-driven electric polarization of
    \SI{100}{\micro\coulomb\per\meter\squared}
in GaV$_4$S$_8$~\cite{ruff_multiferroicity_2015} and
    \SI{800}{\micro\coulomb\per\meter\squared}
in TbMnO$_3$~\cite{kimura_magnetic_2003, kenzelmann_magnetic_2005} in their cycloidal phases, the magnon-driven electric polarization is at least one order of magnitude smaller.
In these materials, the spin spirals exist at the level of the classical ground state, while we have computed the \emph{thermally activated} electrical polarization of a \emph{collinear} magnet, which explains the quantitative differences.

\subsubsection{Electric susceptibility}
\begin{figure}
    \centering
    \includegraphics[width=0.5\textwidth]{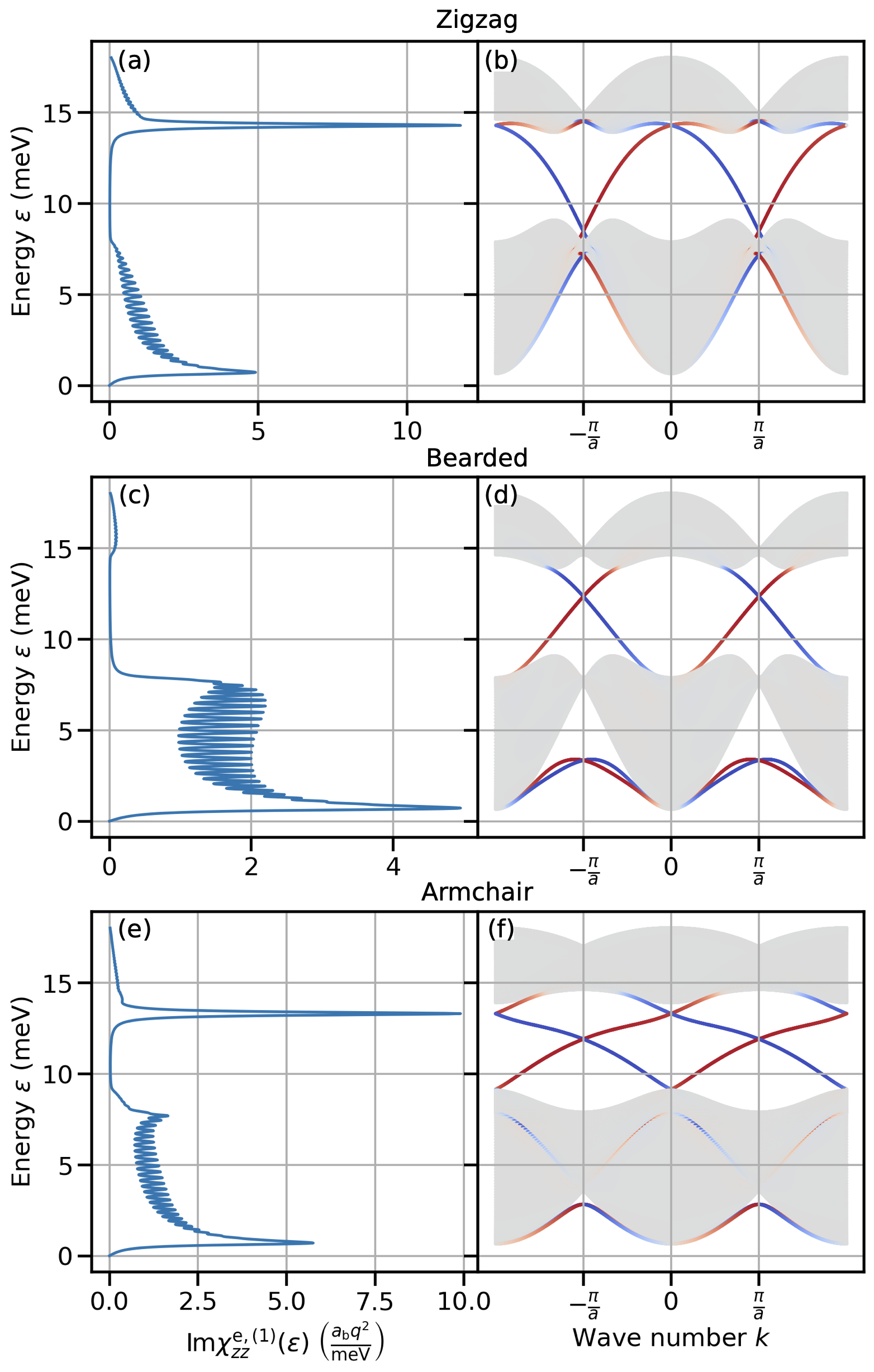}
    \caption{
        (a) Imaginary part of the electric one-magnon susceptibility $\Im \chi_{zz}^{\mathrm{e},(1)}$ of a 120-layers wide nanoribbon with zigzag terminations vs. energy.
        (b) Magnon spectrum of the nanoribbon with wave function amplitude on the left (red) and right (blue) edges encoded by color.
        $a_{\text{b}}$ corresponds to the bulk lattice constant of the underlying honeycomb lattice, while $a$ is the lattice constant of the nanoribbon.
        (c, d) [(e, f)] same as (a, b) for bearded [armchair] terminations.
        The parameters have been specified as
        $
            J_1 = \SI{2.01}{\milli\electronvolt},
            J_2 = \SI{0.16}{\milli\electronvolt},
            J_3 = \SI{-0.08}{\milli\electronvolt},
            D_z = \SI{-0.31}{\milli\electronvolt},
            A = \SI{0.22}{\milli\electronvolt},
            \text{and }
            S = \nicefrac{3}{2}
            .
        $
        The width of the Lorentzians is $\eta = \SI{0.1}{\milli\electronvolt}$.
    }
    \label{fig:dmi_omag_el_im_plb}
\end{figure}
The one-magnon electric susceptibility for electric field and polarization along $z$ is shown in Fig.~\ref{fig:dmi_omag_el_im_plb} for 120-layers wide nanoribbons.
While zigzag [Fig.~\ref{fig:dmi_omag_el_im_plb}(a)] and armchair [Fig.~\ref{fig:dmi_omag_el_im_plb}(c)] edges host topological electromagnons (electrically active topological magnons), the bearded edge produces a smaller signal only from bulk states [Fig.~\ref{fig:dmi_omag_el_im_plb}(b)].
This is because the topological magnons within the bulk band gap only exist at finite wavelengths, i.e., $k \neq 0$, for the latter termination, which cannot be probed by one-magnon processes due to crystal momentum conservation [cf.~Eq.~\eqref{eq:chi_e_omag_imag}].

The estimation of the (three-dimensional) susceptibility entails multiple caveats.
First, the magnitude sensitively depends on the linewidth $\eta$ (here $\eta = \SI{0.1}{\milli\electronvolt}$) of the magnon bands as the peak heights are proportional to $\eta^{-1}$.
Second, the three-dimensional electric susceptibility is obtained by dividing by the width of the nanoribbon and, therefore, depends on the sample size as it is arising from the edge as for the electric edge polarization.
Third, the value depends on the details of the edges.
With these restrictions in mind, we obtain a three-dimensional susceptibility of approximately $\num{6e-3} \varepsilon_0$ ($\varepsilon_0$~vacuum permittivity), where we have estimated the interlayer distance and the bulk lattice constant as \SI{7}{\angstrom}, the width as \SI{4}{\angstrom} per layer, the effective charge as \SI{2e-22}{\coulomb}, and the one-dimensional electric susceptibility as $10 a_{\mathrm{b}} q^2 / \si{\milli\electronvolt}$.


\subsubsection{Magnetic susceptibility}
\begin{figure}
    \centering
    \includegraphics[width=.5\textwidth]{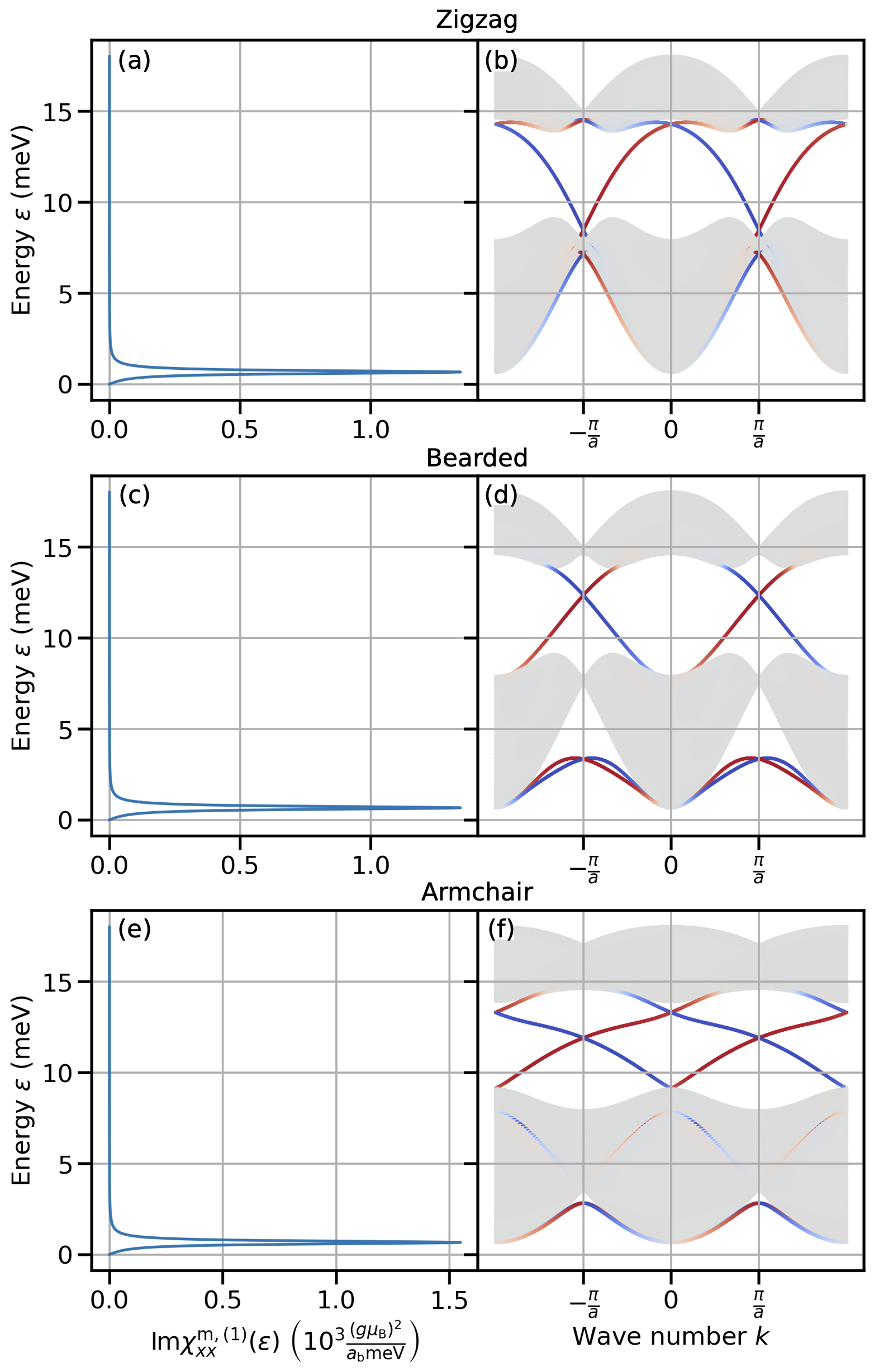}
    \caption{
        (a) Imaginary part of the magnetic one-magnon susceptibility $\Im \chi_{xx}^{\mathrm{m},(1)} = \Im \chi_{yy}^{\mathrm{m},(1)}$ of a 120-layers wide nanoribbon with zigzag terminations vs. energy.
        (b) Magnon spectrum of the zigzag nanoribbon with wave function amplitude on the left (red) and right (blue) edges encoded by color.
        $a_{\text{b}}$ corresponds to the bulk lattice constant of the underlying honeycomb lattice, while $a$ is the lattice constant of the nanoribbon.
        (c, d) [(e, f)] same as (a, b) for bearded [armchair] terminations.
        The parameters have been specified as
        $
            J_1 = \SI{2.01}{\milli\electronvolt},
            J_2 = \SI{0.16}{\milli\electronvolt},
            J_3 = \SI{-0.08}{\milli\electronvolt},
            D_z = \SI{-0.31}{\milli\electronvolt},
            A = \SI{0.22}{\milli\electronvolt},
            \text{and }
            S = \nicefrac{3}{2}
            .
        $
        The width of the Lorentzians is $\eta = \SI{0.1}{\milli\electronvolt}$.
    }
    \label{fig:dmi_omag_mag_im_plb}
\end{figure}
As for the idealized parameters, the magnetic susceptibility for the realistic parameters is only sensitive to the uniform magnon mode, as displayed in Fig.~\ref{fig:dmi_omag_mag_im_plb} for zigzag, bearded, and armchair nanoribbons with 120 layers each.
Analogue to the electric susceptibility we can give an estimate of the three-dimensional magnetic susceptibility based on the parameters of CrI$_3$ and a linewidth of $\eta = \SI{0.1}{\milli\electronvolt}$, for which we obtain \num{0.2} $\mu_0^{-1}$, where $\mu_0$ is the vacuum permeability.

\subsection{Heisenberg-Kitaev model}
Within this section, we consider the Heisenberg-Kitaev Hamiltonian given by~\cite{chaloupka_kitaev-heisenberg_2010,rau_generic_2014,katukuri_kitaev_2014}
\begin{align}
    \hamil
    &=
    -\frac{J_1}{2 \hbar^2}
    \sum_{\langle ij \rangle} \vect{S}_i \cdot \vect{S}_j
    +
    \frac{K}{2 \hbar^2}
    \sum_{\langle ij \rangle_\gamma}
    S_i^\gamma
    S_j^\gamma
    -
    \frac{A}{\hbar^2}
    \sum_i
    {\qty(S_i^z)}^2
    ,
    \label{eq:shamil_kitaev}
\end{align}
which is an alternative description of CrI$_3$ using the parameters
$
    J_1 = \SI{0.53}{\milli\electronvolt},
    K = \SI{4.07}{\milli\electronvolt},
    A = \SI{0.44}{\milli\electronvolt},
    \text{and }
    S = \nicefrac{3}{2}
$
~\cite{aguilera_topological_2020}.
The second term represents the Kitaev interaction~\cite{kitaev_anyons_2006}, where the sum runs over the three nearest neighbor bonds, perpendicular to which one of the three mutually orthogonal spin axes $\gamma = x, y, z$ is oriented.
The spins are projected onto the respective axis that is orthogonal to the bond connecting them.

Note that the notation
$
    \langle ij \rangle_\gamma
$
with $\gamma \in \{x, y, z\}$ is used to denote the set of \emph{nearest} neighbors that also specifies their particular bond direction $\gamma$ and it must not be confused with
$
    \langle ij \rangle_r
$
with $r \in \mathbb{N}$, which is defined as the set of all $r$-th nearest neighbors.

As for DMI, the Kitaev interaction opens a gap in the bulk spectrum that hosts topologically nontrivial chiral edge states in the finite systems.
However, the Kitaev interaction breaks the conservation of the total out-of-plane spin component and induces a spin canting in nanoribbons and flakes~\cite{mcclarty_topological_2018}, which we neglected in the calculations for the following results.

\subsubsection{Local electric polarization}
\begin{figure}
    \centering
    \includegraphics[width=\textwidth]{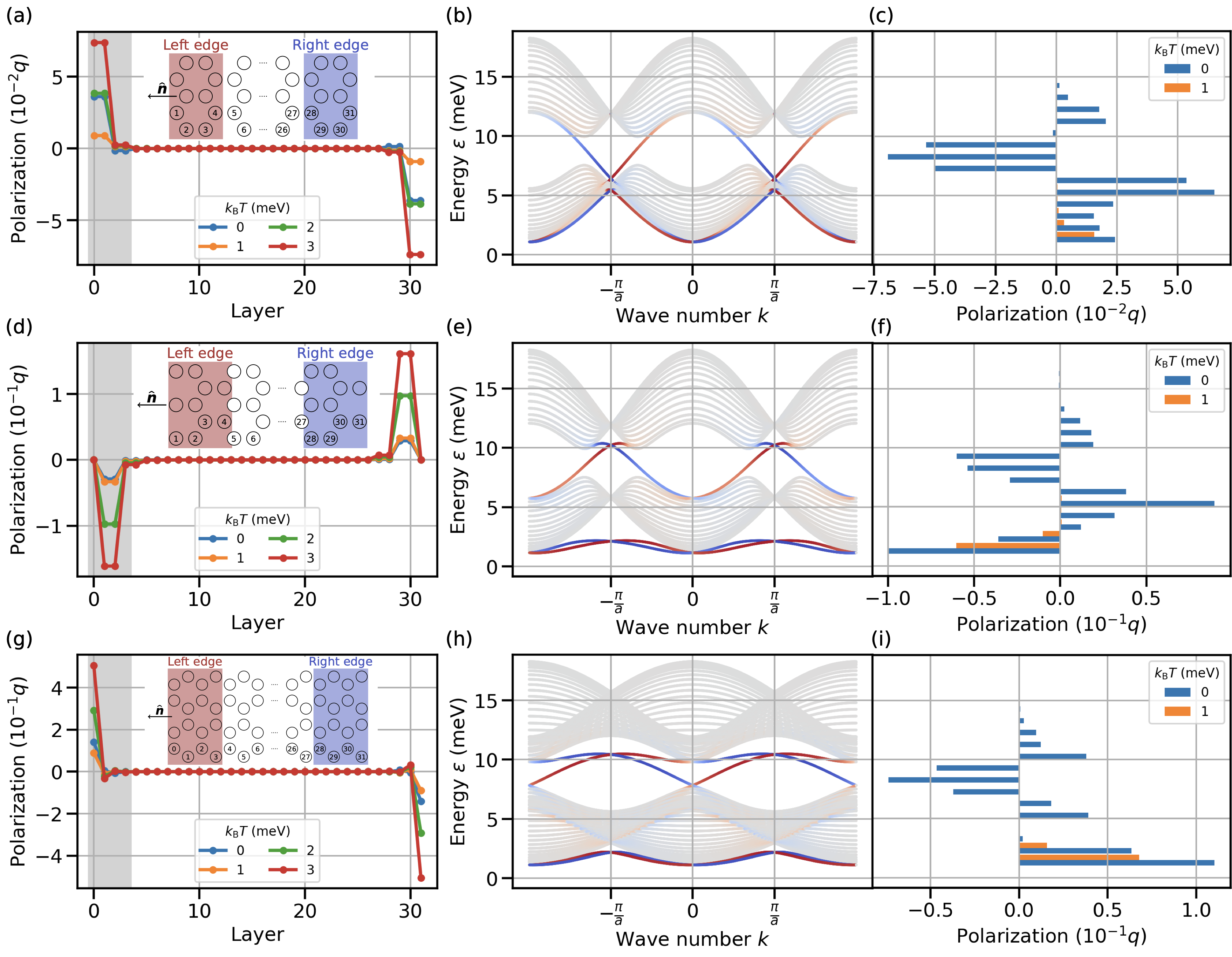}
    \caption{
        Local electric polarization in nanoribbon geometry with 32 layers due to the KNB effect projected onto the outward-facing in-plane normal vector $\unitvect{n}$ of the left edge in the Heisenberg-Kitaev model.
        (a) Layer-resolved electric polarization in zigzag nanoribbon for multiple temperatures (cf.~legend).
        For nonzero temperatures ($\kb T = 1, 2, \SI{3}{\milli\electronvolt}$), the data at $\kb T = 0$ is subtracted.
        Inset: Section of the nanoribbon including layer labels and the normal vector $\unitvect{n}$.
        The system is finite/infinite along the horizontal/vertical direction, respectively.
        (b) Magnon spectrum of the zigzag nanoribbon with localization on the left (red) and right (blue) edges [defined as the 4 outermost layers per side; cf.~inset in panel~(a)] encoded by color.
        (c) Energy-resolved contributions of the magnon modes to the electric polarization of the zigzag nanoribbon at the left edge [highlighted in panel~(a) by gray background] for two selected temperatures (cf.~legend).
        For $\kb T = \SI{1}{\milli\electronvolt}$ (orange bars) only thermal fluctuations are shown; the total electric polarization additionally comprises that at $\kb T = 0$ (blue bars).
        Each bar comprises the accumulated contributions for an energy interval of \SI{1}{\milli\electronvolt}.
        (d,~e,~f) [(g,~h,~i)] same as (a,~b,~c) for bearded [armchair] termination.
        The parameters have been specified as
        $
            J_1 = \SI{0.53}{\milli\electronvolt},
            K = \SI{4.07}{\milli\electronvolt},
            A = \SI{0.44}{\milli\electronvolt},
            \text{and }
            S = \nicefrac{3}{2}
            .
        $
    }
    \label{fig:kitaev_edge_polarization}
\end{figure}
We have calculated the local electric polarization in zigzag, bearded, and armchair nanoribbons using the Hamiltonian Eq.~\eqref{eq:shamil_kitaev} and displayed the result in Fig.~\ref{fig:kitaev_edge_polarization}.
An important difference to the previous model [Eq.~\eqref{eq:shamil_dmi}] is found at zero temperature.
Although the ground state is collinear (within our approximation) and the KNB effect only produces an electric polarization upon spin canting, the electric polarization at the edges is finite even at zero temperature.
At nonzero temperature this is explained by the excitation of spin waves, which, in general, dynamically install coherent spin canting between between different spins in real space.
These \emph{thermal} fluctuations are supplemented by \emph{quantum} fluctuations due to the broken magnon number conservation by the Kitaev interaction.

In contrast to DMI, the anisotropic Kitaev interaction does not conserve the total out-of-plane spin component.
Since the ferromagnetic ground state installs an out-of-plane quantization axis, spin conservation along this axis directly translates to magnon number conservation.
The quantization of the spin is lifted, which results in an unquantized magnon number and quantum corrections to observables in the ground state.
Contrary to thermal fluctuations, where the low-lying excitation usually dominate the low-temperature regime, the quantum contributions of the normal modes do not depend on their energies [cf.~Eq.~\eqref{eq:knb_loc_pol_eq}].

Considering for example the local electric polarization for the zigzag nanoribbon [Fig.~\ref{fig:kitaev_edge_polarization}(a)], we find a ground state electric polarization due to quantum fluctuations (blue line), which is as large as the electric polarization due to thermal fluctuations at $\kb T = \SI{2}{\milli\electronvolt}$ (green line).
Note that the contributions from quantum fluctuations have to be added to the thermal fluctuations in order to obtain the total electric polarization at finite temperatures.
In Fig.~\ref{fig:kitaev_edge_polarization}(b) and (c) we find that the energy-resolved contributions of the quantum fluctuations (blue bars) are much larger than the thermal contributions at $\kb T = 1$.
In particular, there is a sizable contribution from the in-gap states, which are, however, overshadowed by the bulk and trivial edge states, such that the electric polarization at the left edge is still positive.

For the bearded edge, the negative electric polarization at the left edge [Fig.~\ref{fig:kitaev_edge_polarization}(d)] can be traced back to the soft trivial edge modes below the bulk minibands [Fig.~\ref{fig:kitaev_edge_polarization}(e)].
Focusing on the red bands, the modes with positive velocities for $k > 0$ have lower energies than the modes with negative velocities for $k < 0$ and, hence, in total they favor the same sign of the electric polarization as the topological in-gap states.
These weakly-dispersive trivial modes therefore contribute negatively to the left edge electric polarization [Fig.~\ref{fig:kitaev_edge_polarization}(f)].
This is different from the armchair nanoribbon [Fig.~\ref{fig:kitaev_edge_polarization}(g--i)], where $\varepsilon(k < 0) < \varepsilon(k > 0)$ for the red band below the lower bulk quasi-continuum.

\subsubsection{Electric susceptibility}
\begin{figure}
    \centering
    \includegraphics[width=0.5\textwidth]{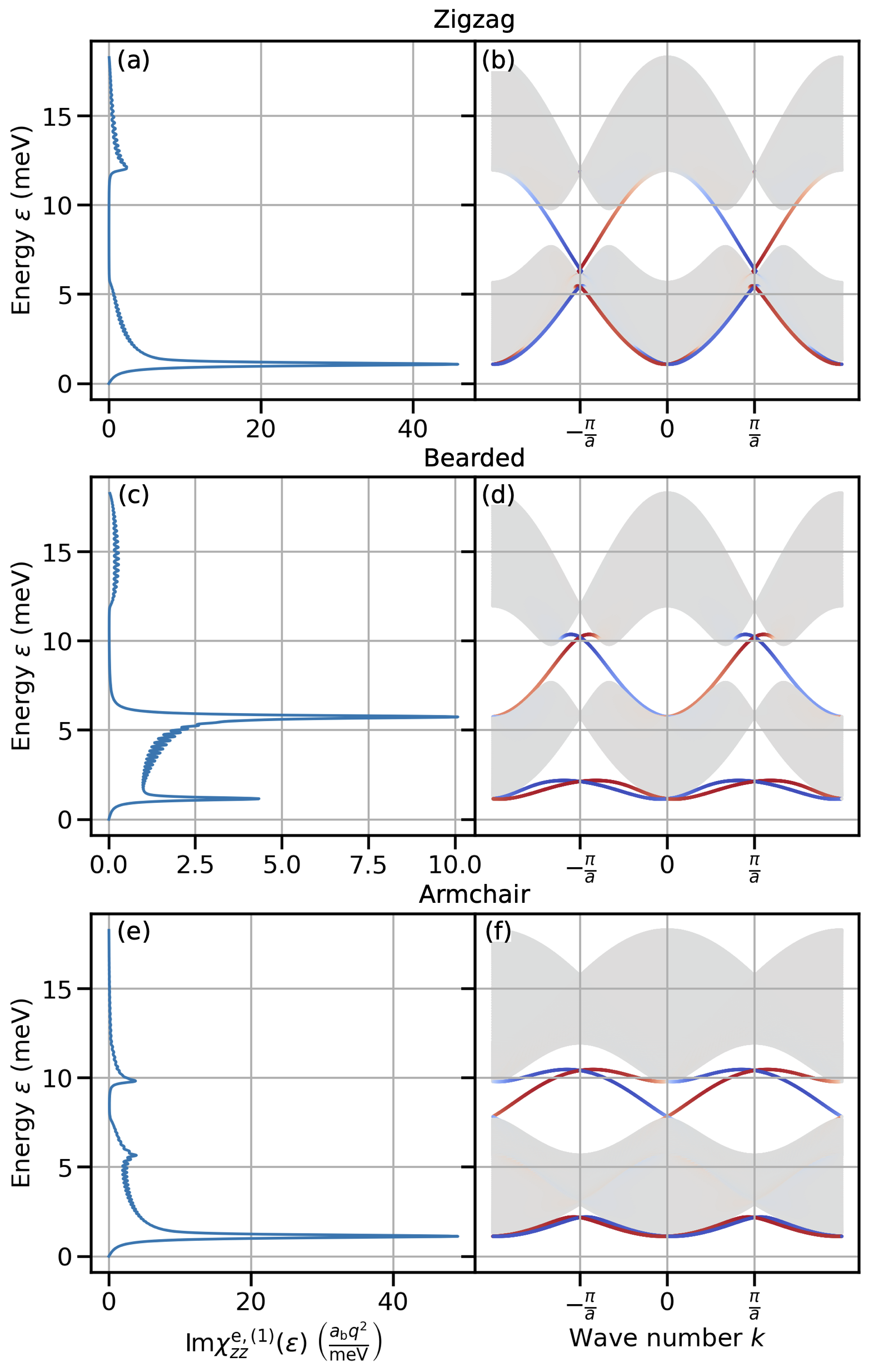}
    \caption{
        (a) Imaginary part of the electric one-magnon susceptibility $\Im \chi_{zz}^{\mathrm{e},(1)}$ of a 120-layers wide nanoribbon with zigzag terminations vs. energy.
        (b) Magnon spectrum of the zigzag nanoribbon with wave function amplitude on the left (red) and right (blue) edges encoded by color.
        $a_{\text{b}}$ corresponds to the bulk lattice constant of the underlying honeycomb lattice, while $a$ is the lattice constant of the nanoribbon.
        (c,~d) [(e,~f)] same as (a,~b) for bearded [armchair] terminations.
        The parameters of the Heisenberg-Kitaev model have been specified as
        $
            J_1 = \SI{0.53}{\milli\electronvolt},
            K = \SI{4.07}{\milli\electronvolt},
            A = \SI{0.44}{\milli\electronvolt},
            \text{and }
            S = \nicefrac{3}{2}
            .
        $
        The width of the Lorentzians is $\eta = \SI{0.1}{\milli\electronvolt}$.
    }
    \label{fig:kitaev_omag_el_im_plb}
\end{figure}
The imaginary part of the one-magnon electric susceptibility shows no or only weak signs of topological magnon modes in the present model (cf.~Fig.~\ref{fig:kitaev_omag_el_im_plb}).
In the zigzag [Fig.~\ref{fig:kitaev_omag_el_im_plb}(a,~b)] and armchair nanoribbons [Fig.~\ref{fig:kitaev_omag_el_im_plb}(e,~f)] there exists only a small signal from the touching point of the topological mode with the upper bulk quasi-continuum, while the bearded nanoribbon shows a similar electrically active touching point at the edge of the lower bulk quasi-continuum [Fig.~\ref{fig:kitaev_omag_el_im_plb}(c,~d)].
In all three cases, however, genuine in-gap peaks are missing.

\subsubsection{Magnetic susceptibility}
\begin{figure}
    \centering
    \includegraphics[width=.5\textwidth]{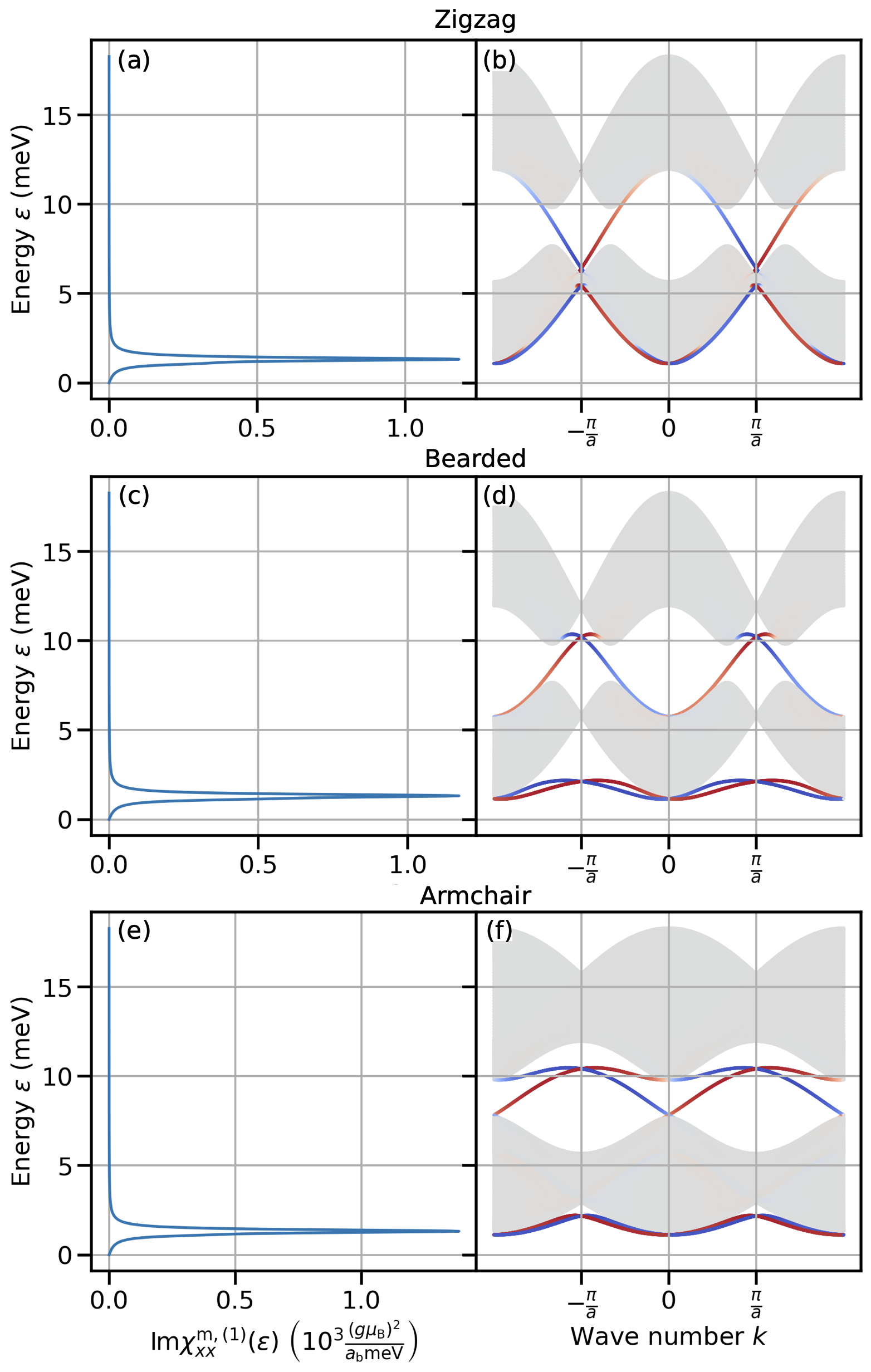}
    \caption{
        (a) Imaginary part of the magnetic one-magnon susceptibility $\Im \chi_{xx}^{\mathrm{m},(1)} = \Im \chi_{yy}^{\mathrm{m},(1)}$ of a 120 layers wide nanoribbon with zigzag terminations vs. energy.
        (b) Magnon spectrum of the zigzag nanoribbon with wave function amplitude on the left (red) and right (blue) edges encoded by color.
        $a_{\text{b}}$ corresponds to the bulk lattice constant of the underlying honeycomb lattice, while $a$ is the lattice constant of the nanoribbon.
        (c,~d) [(e,~f)] same as (a,~b) for bearded [armchair] terminations.
        The parameters of the Heisenberg-Kitaev model have been specified as
        $
            J_1 = \SI{0.53}{\milli\electronvolt},
            K = \SI{4.07}{\milli\electronvolt},
            A = \SI{0.44}{\milli\electronvolt},
            \text{and }
            S = \nicefrac{3}{2}
            .
        $
        The width of the Lorentzians is $\eta = \SI{0.1}{\milli\electronvolt}$.
    }
    \label{fig:kitaev_omag_mag_im_plb}
\end{figure}
The imaginary one-magnon magnetic susceptibility in Fig.~\ref{fig:kitaev_omag_mag_im_plb} shares the same features as for the Heisenberg-DMI model (cf.~Fig.~\ref{fig:dmi_omag_mag_im_plb}).
There is only a small quantitative difference with respect to the height of the peak of the uniform magnon mode.
Pictorially, this is explained by the opening angle of the precessing spins associated with that mode.
For the Heisenberg-DMI mode, the spin is conserved, such that the out-of-plane moment is reduced by $\hbar$ per excited magnon, which is distributed across all sites.
For the Heisenberg-Kitaev model, this out-of-plane moment can be smaller or larger such that the dynamic transverse component is smaller or larger.
Hence, the response of the transverse magnetization depends on the type of spin-orbit coupling in the system.

\subsection{Robustness of local edge polarization}
The previous results have shown that the local polarization is dominated by low-energy trivial edge magnons rather than high-energy topological magnons.
Here, we present results on scenarios in which the contributions of the former may be suppressed.

First, we consider an ordered, but inhomogeneous system with additional easy-axis anisotropies for spins close to the edges.
Second, the effect of disorder of the easy-axis anisotropy on the local polarization is investigated.

\subsubsection{Artificial edge easy-axis anisotropy}
Here, we consider the Heisenberg-Kitaev Hamiltonian of Eq.~\eqref{eq:shamil_kitaev} extended by an artificial easy-axis anisotropy term at the edge sites:
\begin{align}
    \hamil
    &=
    -\frac{J_1}{2 \hbar^2}
    \sum_{\langle ij \rangle} \vect{S}_i \cdot \vect{S}_j
    +
    \frac{K}{2 \hbar^2}
    \sum_{\langle ij \rangle_\gamma}
    S_i^\gamma
    S_j^\gamma
    -
    \frac{A}{\hbar^2}
    \sum_i
    {\qty(S_i^z)}^2
    -
    \frac{A_{\text{e}}}{\hbar^2}
    \sum_{i \in E_{l}}
    {\qty(S_i^z)}^2
    ,
    \label{eq:shamil_kitaev_artpot}
\end{align}
where $E_l$ is the set of all sites that are no further away than $l$ layers from the edges.
This term compensates for the lower effective fields due to the missing neighbors at the edges and, hence, stabilizes the edge spins against fluctuations.
We choose the same parameter values as before 
($
J_1 = \SI{0.53}{\milli\electronvolt},
K = \SI{4.07}{\milli\electronvolt},
A = \SI{0.44}{\milli\electronvolt},
\text{and }
S = \nicefrac{3}{2}
$), while $l = 2$ and $A_{\text{e}}$ is varied.

\begin{figure}
    \centering
    \includegraphics[width=\textwidth]{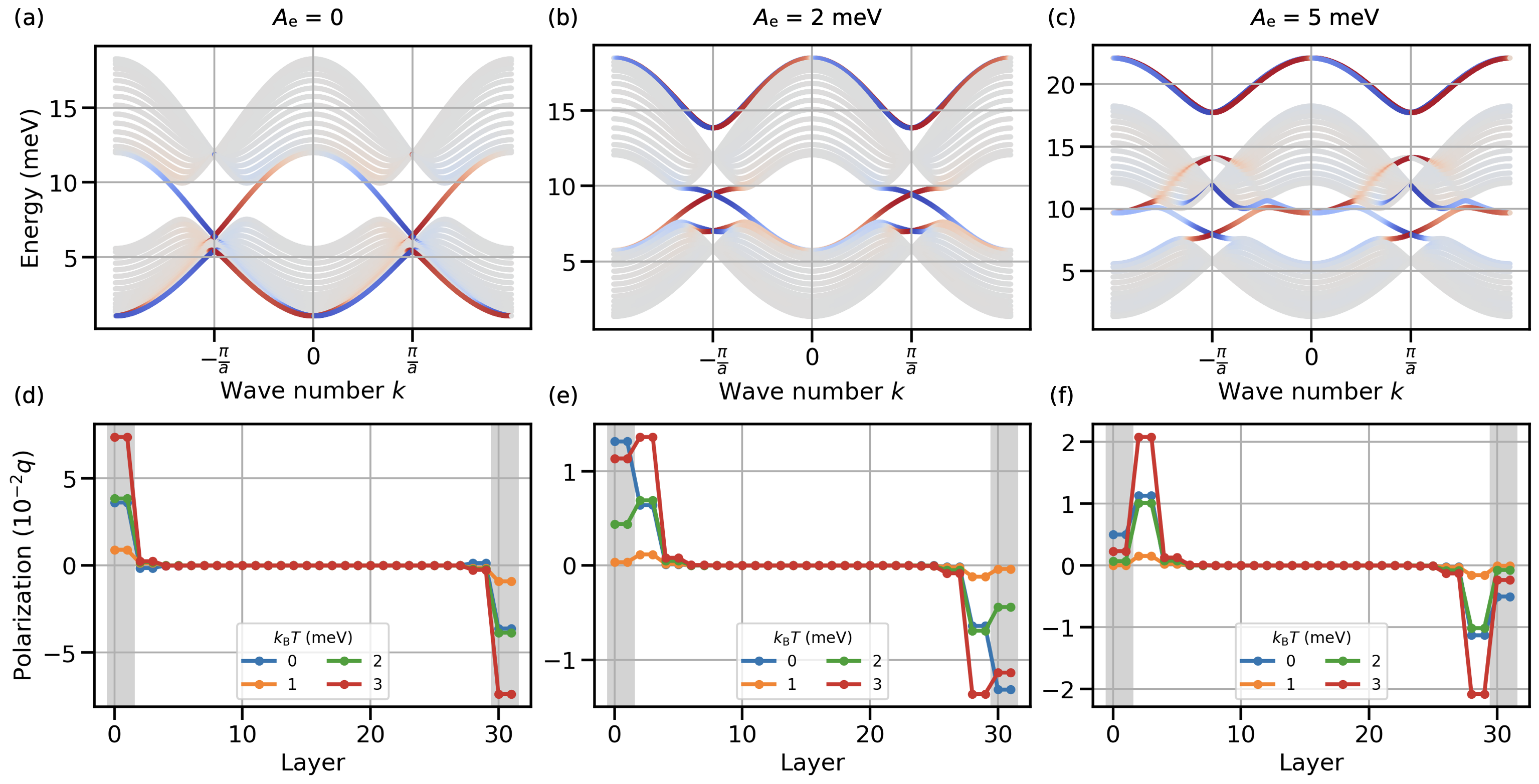}
    \caption{
        Magnon spectra and local electric polarization for zigzag nanoribbons of 32 layers with an artificial easy-axis anisotropies at the two outermost layers.
        (a--c) Magnon spectra with $A_{\text{e}} = 0, \SI{2}{\milli\electronvolt}, \text{and } \SI{5}{\milli\electronvolt}$, respectively (see headings).
        Red/blue color indicates magnon localization at first/last four layers of the nanoribbon.
        (d--f) Layer-resolved electric polarization projected onto the outward-facing in-plane normal vector $\unitvect{n}$ of the left edge with $A_{\text{e}} = 0, \SI{2}{\milli\electronvolt}, \text{and } \SI{5}{\milli\electronvolt}$, respectively, for multiple temperatures (see legends).
        Nonzero temperature curves do not include zero-temperature offset by quantum corrections (blue curves).
        Gray stripes indicate layers with increased easy-axis anisotropies.
        The parameters have been specified as
        $
            J_1 = \SI{0.53}{\milli\electronvolt},
            K = \SI{4.07}{\milli\electronvolt},
            A = \SI{0.44}{\milli\electronvolt},
            S = \nicefrac{3}{2},
            \text{and }
            l = 2
            .
        $
    }
    \label{fig:kitaev_edge_pol_artpot}
\end{figure}
A global easy-axis anisotropy acts like a chemical potential that rigidly shifts the magnon spectrum and opens a spin-wave gap.
In our case, the inhomogeneous easy-axis anisotropy selectively shifts magnon bands that are localized at the affected sites close to the edges [cf.~Fig.~\ref{fig:kitaev_edge_pol_artpot}(a--c)].
The trivial edge modes with energies below the bulk quasi-continua disappear, while a set of bands emerges above the bulk minibands.
In addition, the dispersion of the topologically nontrivial in-gap states is altered by $A_{\text{e}}$.

The edge regions comprising 4 layers at each side, within which the colored modes in Fig.~\ref{fig:kitaev_edge_pol_artpot}(a--c) are largely localized, can be split into two parts.
The first two layers at each side (hereafter called \enquote{outer edge layers}) are affected by the artificial anisotropies and, hence, feature stabilized spins.
The second two layers at each side (hereafter called \enquote{inner edge layers}) are unaffected and may host lower-energy modes.
Therefore, the color does not make any statements about whether a mode is localized at the spins within $E_2$ or within the neighboring two layers.

The artificial anisotropy also has consequences for the edge polarizations displayed in Fig.~\ref{fig:kitaev_edge_pol_artpot}(d--f).
It suppresses the electric edge polarization at the outer edge layers for all temperatures, while the electric polarization at the inner edge layers rises on each side.
Hence, it appears that the local edge polarization is redistributed from the outer to the inner edge layers.
Importantly, the sign of the edge polarization does not change, although the topological magnons alone would cause the opposite sign of the local electric polarization.

\begin{figure}
    \centering
    \includegraphics[width=.6\textwidth]{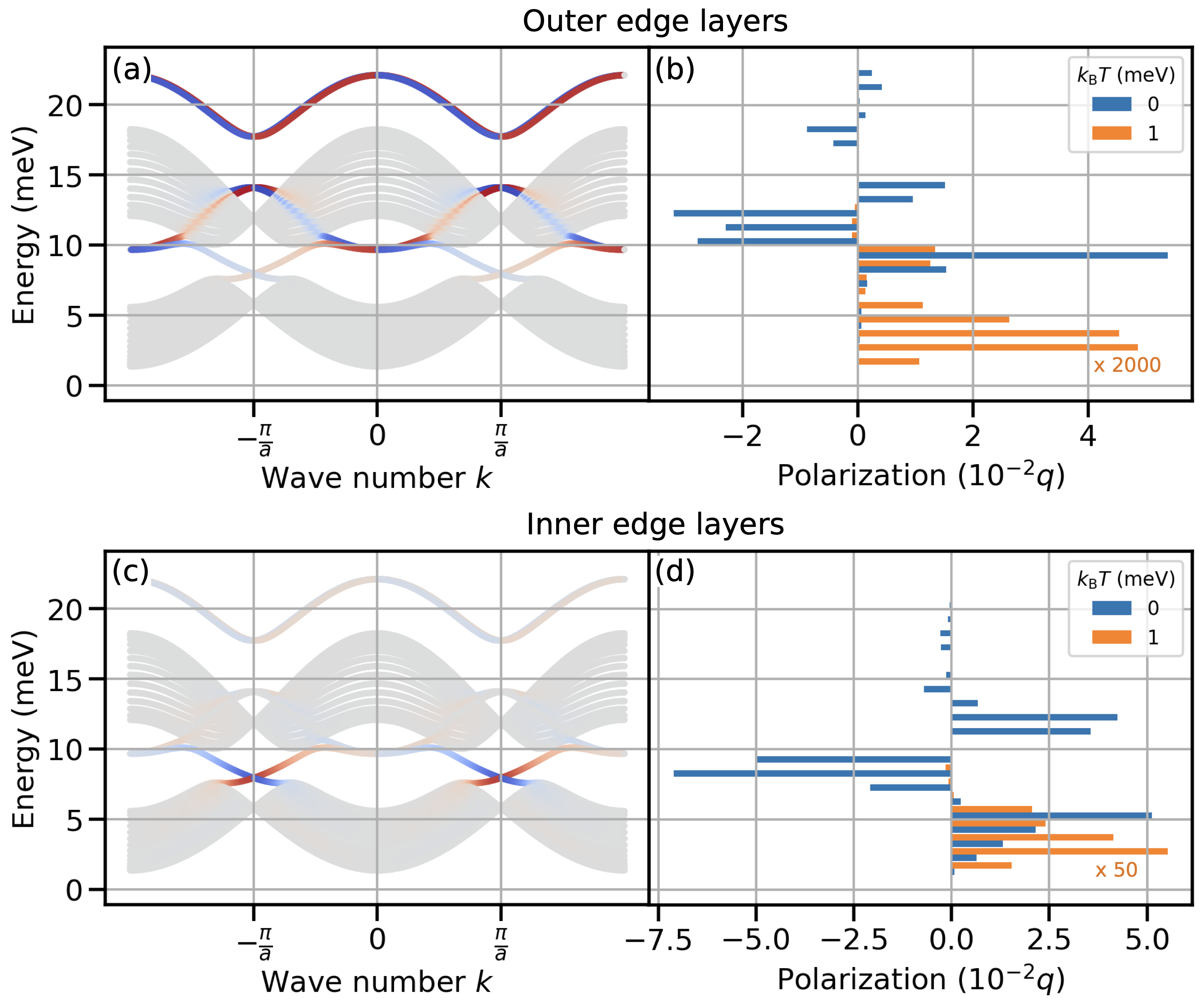}
    \caption{
        Magnon spectra and energy-resolved local edge polarizations for zigzag nanoribbons of 32 layers with artificial easy-axis anisotropies $A_{\text{e}} = \SI{5}{\milli\electronvolt}$ at the two outermost layers.
        (a) Magnon spectrum with magnon wave function amplitudes on the first two left (red) or right (blue) layers (i.e., layers 0, 1 at left edge and 30, 31 at right edge).
        (b) Energy-resolved contributions to the local polarization at the first two left layers for multiple temperatures (see legend).
        Orange bars for $\kb T = \SI{1}{\milli\electronvolt}$ do not include zero-temperature offset by quantum corrections (blue bars).
        (c) Same magnon spectrum as in panel~(a), but color indicates wave function amplitudes on the second two layers (i.e., layers 2, 3 at left edge and 28, 29 at right edge).
        (d) Energy-resolved contributions to the local polarization at the second two left layers.
        Each bar in (b,~d) comprises the accumulated contributions for an energy interval of \SI{1}{\milli\electronvolt}.
        The parameters have been specified as
        $
            J_1 = \SI{0.53}{\milli\electronvolt},
            K = \SI{4.07}{\milli\electronvolt},
            A = \SI{0.44}{\milli\electronvolt},
            S = \nicefrac{3}{2},
            \text{and }
            l = 2
            .
        $
    }
    \label{fig:kitaev_edep_edge_pol_artpot}
\end{figure}
In order to explain the observations, we consider the outer edge layers affected by the artificial anisotropies and the adjacent inner edge layers separately.
In Fig.~\ref{fig:kitaev_edep_edge_pol_artpot}(a) we show the modes at the outer edge layers in color, while Fig.~\ref{fig:kitaev_edep_edge_pol_artpot}(c) highlights modes at the inner edge layers.
Whereas the trivial modes are localized on the stabilized spins and, thus, possess higher energies, the in-gap modes are primarily localized at the inner edge layers.

The layers of stabilized spins feature a small electric edge polarization, as shown for the left side in Fig.~\ref{fig:kitaev_edep_edge_pol_artpot}(b).
At zero temperature the contributions of the various modes partially cancel each other and the trivial bands above the bulk quasi-continua play a negligible role as well as the in-gap states.
At $\kb T = \SI{1}{\milli\electronvolt}$ the contributions are much smaller due to the lack of low-energy trivial edge modes.

At the inner edge layers we have observed the larger polarization.
Indeed, the bulk contributions of the inner layers at $\kb T = \SI{1}{\milli\electronvolt}$ are much larger than the contributions of the outer layers as seen for the left side in Fig.~\ref{fig:kitaev_edep_edge_pol_artpot}(d).
Furthermore, the in-gap states contribute a sizable negative part to the edge polarization.
This suggests that the bulk and in-gap states are expelled from the outer layers due to the anisotropy and only the trivial modes, which are frozen out due to their energy renormalization, remain.
On the other hand, the bulk modes seem to over-compensate the superseded trivial edge modes with respect to the inner edge polarization.
To summarize, although the artificial anisotropies shift the trivial edge modes to higher energies and spatially separate trivial and nontrivial edge modes, the contribution of the nontrivial edge modes to the equilibrium polarization is still overcompensated by bulk modes for the 32-layers wide nanoribbons.

\subsubsection{Easy-axis anisotropy disorder}
Topological magnons are conceived to be robust against disorder.
As long as the energy scale of the disorder is significantly smaller than the size of the bulk gap, the topological modes are protected against local perturbations.
In contrast, trivial modes do not enjoy such a protection as they can be elastically scattered at impurity centers by change of momentum (crystal momentum conservation is broken by disorder), which is kinematically suppressed for in-gap modes.
Here, we study whether disorder can serve as an instrument to suppress trivial contributions to the edge polarization and promote signatures of topological magnons potentially rendering them accessible in experiments.

To describe disorder, one needs to consider configurations, which correspond to specific disorder realizations.
Reverting to the Heisenberg-DMI model, each configuration can be described by the Hamiltonian of Eq.~\eqref{eq:shamil_dmi},
\begin{align}
    \hamil
    &=
    -\sum_{r=1}^3 \frac{J_r}{2 \hbar^2}
    \sum_{\langle ij \rangle_r} \vect{S}_i \cdot \vect{S}_j
    +
    \frac{1}{2 \hbar^2}
    \sum_{\langle ij \rangle_2}
    \vect{D}_{ij}
    \cdot
    \qty(\vect{S}_i \times \vect{S}_j)
    -
    \sum_i
    \frac{A_i}{\hbar^2}
    {\qty(S_i^z)}^2
    ,
    \label{eq:shamil_dmi_disorder}
\end{align}
but with a site-dependent easy-axis anisotropy $A_i$.
The parameters
$
    J_1 = \SI{2.01}{\milli\electronvolt},
    J_2 = \SI{0.16}{\milli\electronvolt},
    J_3 = \SI{-0.08}{\milli\electronvolt},
    D_z = \SI{-0.31}{\milli\electronvolt},
    A = \SI{0.22}{\milli\electronvolt},
    \text{and }
    S = \nicefrac{3}{2}
$
describe the ordered system, while $A_i = A + \varDelta A_i$ with a random variable $\varDelta A_i$ that we draw from a uniform distribution of with zero mean and standard deviation $\sigma$ for a fixed, random fraction $f$ of the sites.
Since disorder breaks translational invariance, we model a flake with open boundary conditions.
We solve the Schrödinger equation of Eq.~\eqref{eq:shamil_dmi_disorder}, compute the local electric polarization according to Eq.~\eqref{eq:knb_loc_pol_eq}, and average the result over many disorder configurations.

\begin{figure}
    \centering
    \includegraphics[width=\textwidth]{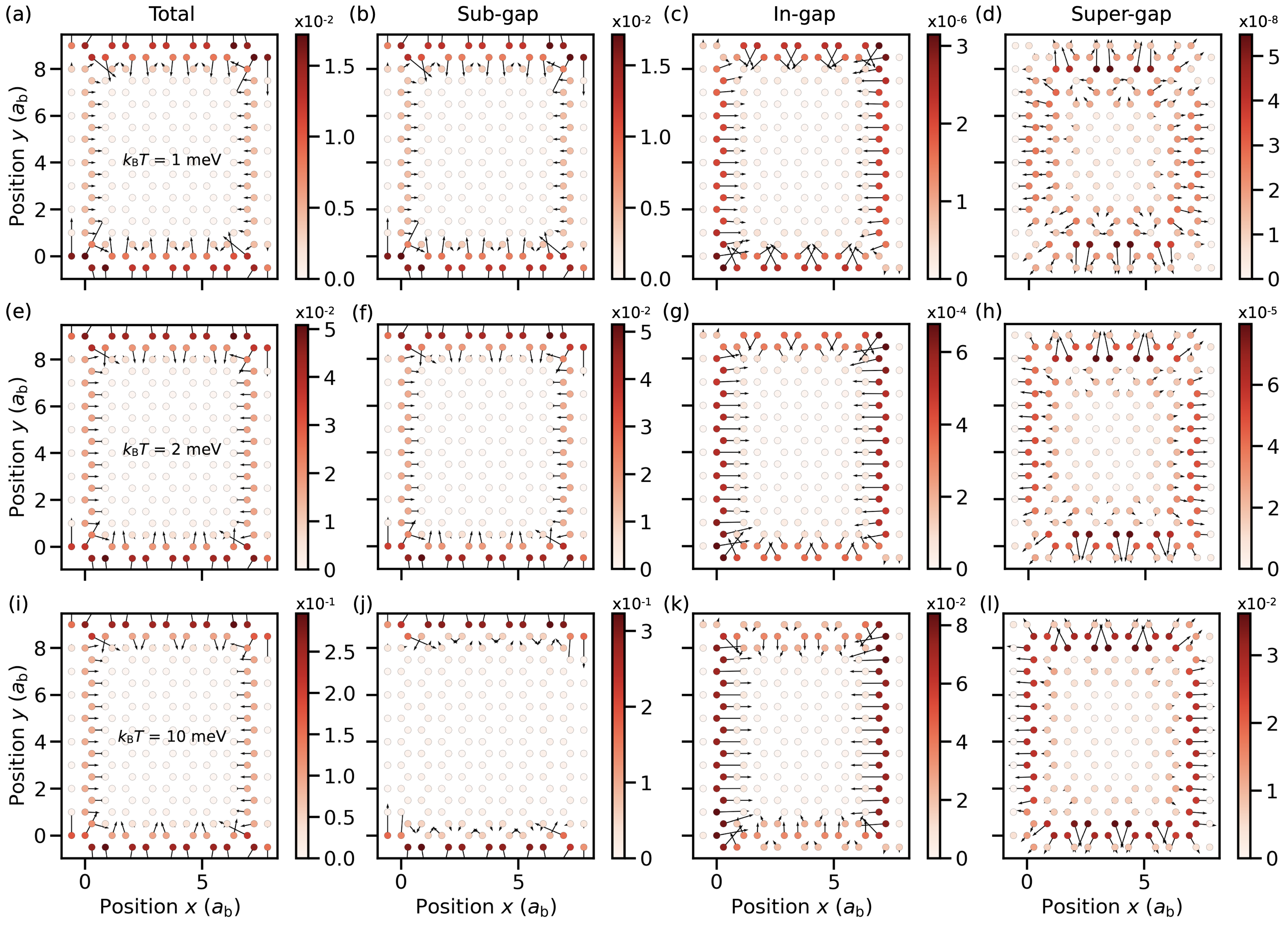}
    \caption{
        Energy-resolved local electric polarization in a clean flake of 200 spins at each site for multiple temperatures (rows) and multiple energy windows (columns).
        In each panel, the orientations of the arrows indicate the direction of the local polarization, while the length of the arrows and the color of the sites represent its magnitude.
        The magnitude in units of $q a_{\text{b}}$ ($q$ effective charge, $a_{\text{b}}$ bulk lattice constant) is given in the color bars.
        Note that the out-of-plane component of the local polarization is zero.
        The temperature has been chosen as (a--d) $\kb T = \SI{1}{\milli\electronvolt}$, (e--h) $\kb T = \SI{2}{\milli\electronvolt}$, (i--l) $\kb T = \SI{10}{\milli\electronvolt}$.
        The local polarization has been filtered to include contributions from (a,~e,~i) all modes and modes (b,~f,~j) below, (c,~g,~k) within, and (d,~h,~l) above the bulk band gap, which is defined as the energy window between \num{9.1} and \SI{13.9}{\milli\electronvolt} (cf.~spectra in Fig.~\ref{fig:dmi_omag_mag_im_plb}).
        The results have been obtained with the parameters
        $
            J_1 = \SI{2.01}{\milli\electronvolt},
            J_2 = \SI{0.16}{\milli\electronvolt},
            J_3 = \SI{-0.08}{\milli\electronvolt},
            D_z = \SI{-0.31}{\milli\electronvolt},
            A = \SI{0.22}{\milli\electronvolt},
            \varDelta A_i = 0,
            \text{and }
            S = \nicefrac{3}{2}
            .
        $
    }
    \label{fig:dmi_flake_edge_polarization_clean}
\end{figure}
We start with a discussion of the local electric polarization for the clean ($\varDelta A_i = 0$ for all $i$) flake with $20 \times 20$ layers presented in Fig.~\ref{fig:dmi_flake_edge_polarization_clean}.
Here, we have calculated the local electric polarization (which corresponds to an electric dipole moment in a zero-dimensional system) at each site for multiple temperatures and filtered the individual contributions from all magnon modes by energy.
The first/second/third rows represent $\kb T = \SI{1}{\milli\electronvolt}/\SI{2}{\milli\electronvolt}/\SI{10}{\milli\electronvolt}$, respectively, while the first/second/third/fourth columns represent the total/sub-gap/in-gap/super-gap contributions to the local electric polarization.
We defined the gap as the energy window between \num{9.1} and \SI{13.9}{\milli\electronvolt} in agreement with the spectra plotted in Fig.~\ref{fig:dmi_omag_mag_im_plb}.

Focusing on $\kb T = 1$, the total local polarization is mostly nonzero at the edges of the bearded and armchair terminations [Fig.~\ref{fig:dmi_flake_edge_polarization_clean}(a)].
On the bearded edges, it is facing inward on the second and the third layers from the edge.
On the armchair edges, the local polarization changes its orientation from pointing outward at the first layer to inward at the second layer back to inward at the third layer.
These findings are in agreement with previous results on the local electric polarization in the nanoribbon for this model (cf.~Fig.~\ref{fig:dmi_edge_polarization}).

It is evident that this polarization is mostly governed by the magnon modes with energies below the bulk band gap [Fig.~\ref{fig:dmi_flake_edge_polarization_clean}(b)].
The contributions from the topological magnons are around 4 orders of magnitude smaller [Fig.~\ref{fig:dmi_flake_edge_polarization_clean}(c)].
Here, we find a local electric polarization always pointing inward at the edges and a vanishing bulk contribution supporting the intuitive picture of the VME effect.
The orientation at the bearded edges is indeed similar to that governed by the sub-gap states, however, the local polarization at the armchair edges does not change its orientation layer by layer.
Above the gap, the contributions are even two orders of magnitude smaller than from within the gap [Fig.~\ref{fig:dmi_flake_edge_polarization_clean}(d)].
Interestingly, the direction of the local electric polarization points almost exclusively radially outward and its magnitude does not decay as fast into the bulk.

At $\kb T = \SI{2}{\milli\electronvolt}$ the magnitude of the local electric polarization increases, but qualitatively we do not find any difference [Fig.~\ref{fig:dmi_flake_edge_polarization_clean}(e--h)].
Going even higher with temperature ($\kb T = \SI{10}{\milli\electronvolt}$), we find that the total polarization has still an edge polarization at the bearded terminations [Fig.~\ref{fig:dmi_flake_edge_polarization_clean}(i)], but the sub-gap contributions are negligible at these sites [Fig.~\ref{fig:dmi_flake_edge_polarization_clean}(j)].
Instead, the in-gap contributions increased in size and seem to dominate the bearded edge polarization [Fig.~\ref{fig:dmi_flake_edge_polarization_clean}(k)], although it seems to compete with the super-gap contributions [Fig.~\ref{fig:dmi_flake_edge_polarization_clean}(l)], which reduces the total magnitude.
The strong preference for the armchair edges in Fig.~\ref{fig:dmi_flake_edge_polarization_clean}(j) hints at trivial edge modes that are only localized at the armchair edges and do not cover all edges like topological magnons~\cite{mook_chiral_2021}.

We have analyzed the local electric polarization in the presence of disorder with $\sigma = \SI{0.1}{\milli\electronvolt}$ and $\sigma = \SI{0.2}{\milli\electronvolt}$ for $f = 0.9$ (not shown).
There are no qualitative but only minute quantitative differences to the local electric polarization in the clean system [Fig.~\ref{fig:dmi_flake_edge_polarization_clean}].
Apparently, the small energy scale of the disorder is insufficient to induce notable changes.
However, the size of $\sigma$ is limited by the requirement that the ground states remains ferromagnetic and, hence, should not close the spin-wave gap induced by $A = \SI{0.22}{\milli\electronvolt}$.
Thus, the tolerance $\sqrt{3} \sigma$ for $\varDelta A_i$ should not be much larger than $A$.
Besides the possibility of a magnetic phase transition, large $\sigma$ may close the bulk band gap that protects the topological magnons.

\begin{figure}
    \centering
    \includegraphics[width=\textwidth]{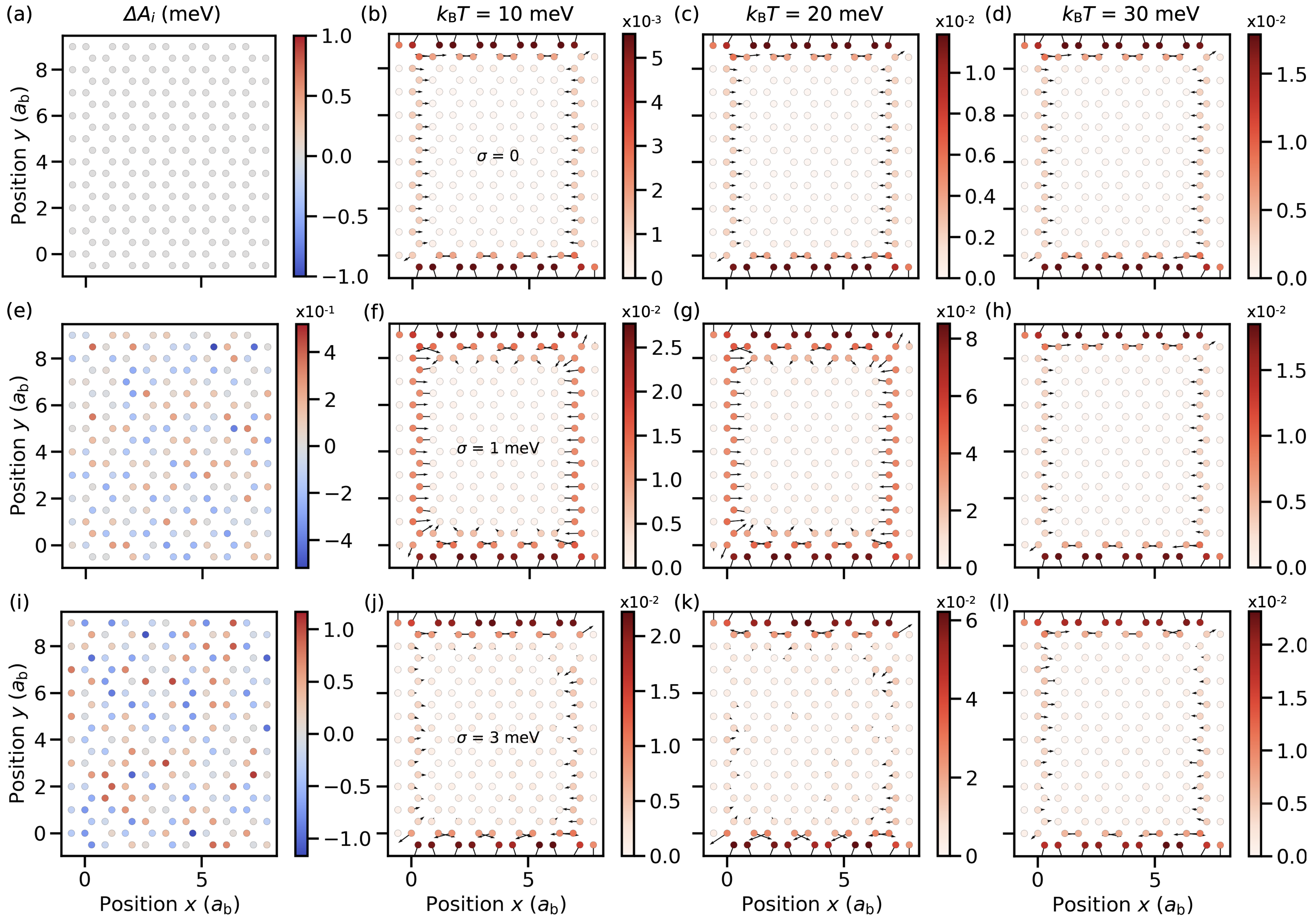}
    \caption{
        Disorder-induced easy-axis anisotropies (first column) and local electric polarization for multiple temperatures (second, third, and fourth columns) in clean (first row) and disordered (second, and third rows) flakes of 200~spins each.
        In panels~(a,~e,~i), each filled circle represents a site within the flake and its red/gray/blue color indicates a positive/zero/negative disorder-induced average easy-axis anisotropy $\expval{\varDelta A_i}_{\text{conf}}$ given in \si{\milli\electronvolt} (see color bar).
        The deviation from zero is an indicator of the degree of convergence.
        In panels~(b--d,~f--h,~j--l), the orientations of the arrows indicate the direction of the local polarization, while the length of the arrows and the color of the sites represent its magnitude.
        The magnitude in units of $q a_{\text{b}}$ ($q$ effective charge, $a_{\text{b}}$ bulk lattice constant) is given in the color bars.
        Note that the out-of-plane component of the local polarization is zero.
        The temperature has been chosen as (b,~f,~j) $\kb T = \SI{10}{\milli\electronvolt}$, (c,~g,~k) $\kb T = \SI{20}{\milli\electronvolt}$, (d,~h,~l) $\kb T = \SI{30}{\milli\electronvolt}$ (cf. column headings).
        The standard deviation of $\varDelta A_i$ is (a--d) $\sigma = 0$, (e--h) $\sigma = \SI{1}{\milli\electronvolt}$, (i--l) $\sigma = \SI{3}{\milli\electronvolt}$ (cf. insets in second column) and its mean is zero.
        All results for $\sigma = \SI{1}{\milli\electronvolt}$ and $\sigma = \SI{3}{\milli\electronvolt}$ have been obtained by averaging over 50 disorder configurations.
        The results have been obtained with the parameters
        $
            J_1 = \SI{2.01}{\milli\electronvolt},
            J_2 = \SI{0.16}{\milli\electronvolt},
            J_3 = \SI{-0.08}{\milli\electronvolt},
            D_z = \SI{-0.31}{\milli\electronvolt},
            A = \SI{5}{\milli\electronvolt},
            S = \nicefrac{3}{2}
            \text{ and }
            f = \num{0.9}
            .
        $
    }
    \label{fig:dmi_flake_edge_polarization_dirty}
\end{figure}
To increase the disorder and maintain the ferromagnetic ground state simultaneously, we have set
$
    A = \SI{5}{\milli\electronvolt}
$ and
$
    \sigma = 1 \text{ and } \SI{3}{\milli\electronvolt}
$,
while the other parameters have been left unchanged.
As a result, the spin-wave gap ($= 2 S A$) has increased from \SI{0.66}{\milli\electronvolt} to \SI{15}{\milli\electronvolt}.
The first [panels~(a--d)], second [panels~(e--h)], third [panels~(i--l)] rows of Fig.~\ref{fig:dmi_flake_edge_polarization_dirty} display the impurity potentials and local electric polarization in the clean flake, its ensemble average for
$
    \sigma = \SI{1}{\milli\electronvolt}
$,
and its ensemble average for
$
    \sigma = \SI{3}{\milli\electronvolt}
$,
respectively.
The first column contains the (ensemble-averaged) $\varDelta A_i$, and the remaining columns comprise the local electric polarization for
$
    \kb T = \num{10},\ \num{20},\ \text{and } \SI{30}{\milli\electronvolt}
$,
respectively (see column labels).
The ensemble averages are taken over 50 configurations, which is insufficient to obtain fully converged results, but sufficient to study the general trend.

Qualitatively, the results for the clean systems are similar to those shown in Fig.~\ref{fig:dmi_flake_edge_polarization_clean} for
$
    A = \SI{0.22}{\milli\electronvolt}
$.
Since the magnon spectrum is shifted towards higher energies, larger temperatures are required to excite the magnons.
Comparing
$
    \sigma = \SI{1}{\milli\electronvolt}
$
with
$
    \sigma = 0
$,
the magnitude of the local polarization increases for
$
    \kb T = \num{10} \text{ and } \SI{20}{\milli\electronvolt}
$,
but remains more or less constant for
$
    \kb T = \SI{30}{\milli\electronvolt}
$.
Overall, the local electric moments change their direction slightly, e.g., in the third layer of the armchair edges, but the reversal of the outer layer does not take place.
Ramping up the disorder strength to
$
    \sigma = \SI{3}{\milli\electronvolt}
$
leads to a partial suppression of the local electric polarization at the bearded edges relative to the armchair edges.
It is apparent that the inversion symmetry is only approximately fulfilled since the local electric moments at opposite sites are no longer opposite everywhere [e.g.,~cf.~Fig.~\ref{fig:dmi_flake_edge_polarization_dirty}(j)].
As mentioned, this indicates that the disorder averaging procedure is not converged. 
Nonetheless, we expect that this trend survives the averaging.
The suppression at the bearded edges may be attributed to the suppression of the trivial modes' contribution to the local polarization at the bearded edges.

\begin{figure}
    \centering
    \includegraphics[width=\textwidth]{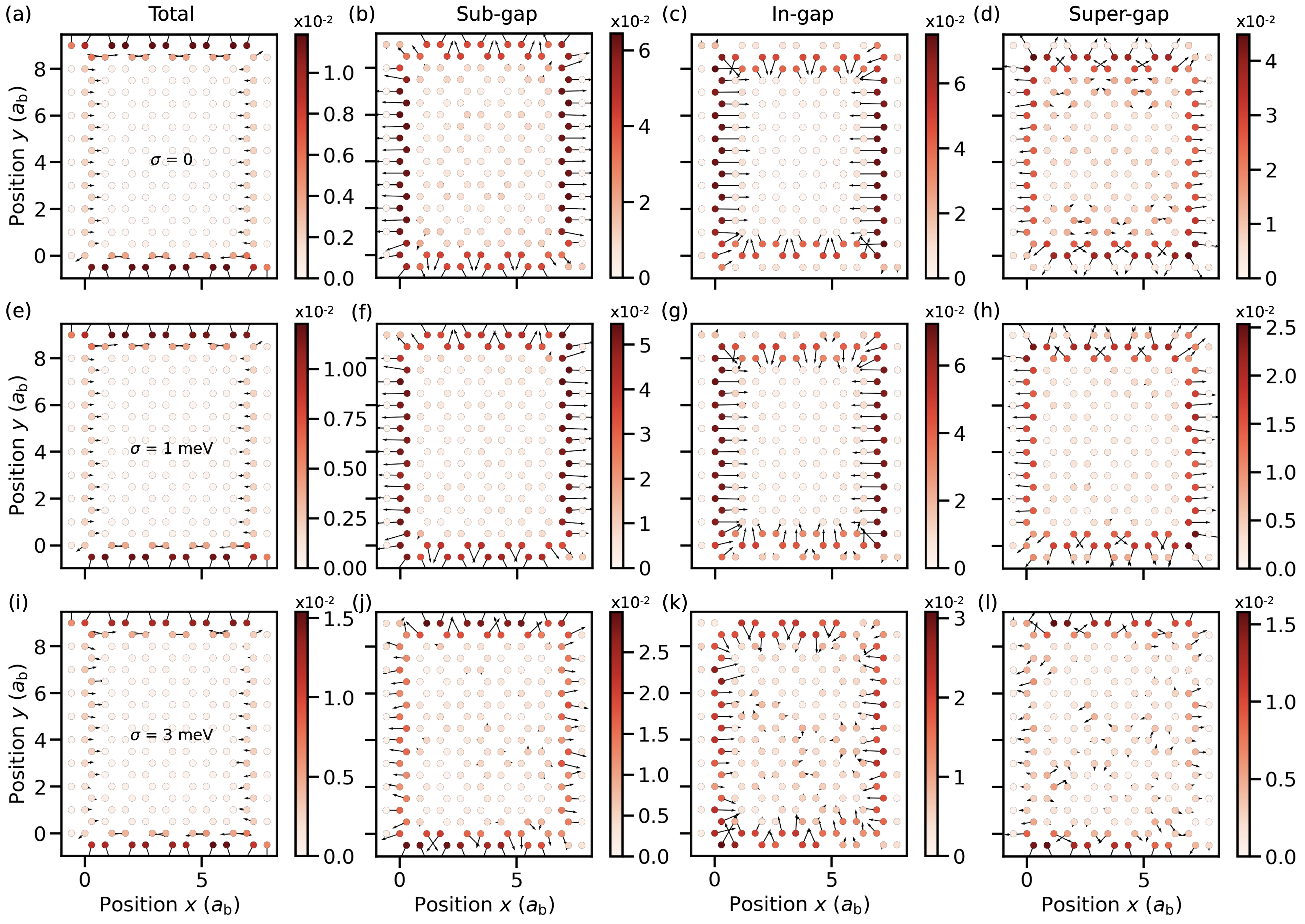}
    \caption{
        Energy-resolved local electric polarization in clean (first row) and disordered (second, and third rows) flakes of 200 spins each for multiple energy windows (columns) at $\kb T = \SI{20}{\milli\electronvolt}$.
        In each panel, the orientations of the arrows indicate the direction of the local polarization, while the length of the arrows and the color of the sites represent its magnitude.
        The magnitude in units of $q a_{\text{b}}$ ($q$ effective charge, $a_{\text{b}}$ bulk lattice constant) is given in the color bars.
        Note that the out-of-plane component of the local polarization is zero.
        The local polarization has been filtered to include contributions from (a,~e,~i) all modes and modes (b,~f,~j) below, (c,~g,~k) within, and (d,~h,~l) above the bulk band gap, which is defined as the energy window between \num{23.4} and \SI{28.3}{\milli\electronvolt}.
        The standard deviation of $\varDelta A_i$ is (a--d) $\sigma = 0$, (e--h) $\sigma = \SI{1}{\milli\electronvolt}$, (i--l) $\sigma = \SI{3}{\milli\electronvolt}$ and its mean is zero.
        All results for $\sigma = \SI{1}{\milli\electronvolt}$ and $\sigma = \SI{3}{\milli\electronvolt}$ have been obtained by averaging over 50 disorder configurations.
        The results have been obtained with the parameters
        $
            J_1 = \SI{2.01}{\milli\electronvolt},
            J_2 = \SI{0.16}{\milli\electronvolt},
            J_3 = \SI{-0.08}{\milli\electronvolt},
            D_z = \SI{-0.31}{\milli\electronvolt},
            A = \SI{5}{\milli\electronvolt},
            S = \nicefrac{3}{2}
            \text{ and }
            f = \num{0.9}
            .
        $
    }
    \label{fig:dmi_flake_edge_polarization_dirty_eres}
\end{figure}
To verify this hypothesis, we have decomposed the total electric polarization at $\kb T = \SI{20}{\milli\electronvolt}$ into its contributions originating from states below
($
    \varepsilon < \SI{23.4}{\milli\electronvolt}
$),
within
($
    \SI{23.4}{\milli\electronvolt}
    \leq
    \varepsilon
    \leq
    \SI{28.3}{\milli\electronvolt}
$),
and above
($  
    \varepsilon > \SI{28.9}{\milli\electronvolt}
$)
the bulk band gap, as before.
While the first row of Fig.~\ref{fig:dmi_flake_edge_polarization_dirty_eres} corresponds to the clean flake, the second (third) row corresponds to the ensemble average with
$
    \sigma = \SI{1}{\milli\electronvolt}
$
($
    \sigma = \SI{3}{\milli\electronvolt}
$),
respectively.
As seen in Fig.~\ref{fig:dmi_flake_edge_polarization_dirty_eres}(a--d), the sub- and super-gap contributions do not govern the orientation of the local polarization at the bearded edges; instead, it is determined by the in-gap states [Fig.~\ref{fig:dmi_flake_edge_polarization_dirty_eres}(c)], since the sub- and super-gap contributions favor the \emph{opposite} orientation [Fig.~\ref{fig:dmi_flake_edge_polarization_dirty_eres}(b,~d)].
Therefore, the difference to the results for
$
    A = \SI{0.22}{\milli\electronvolt}
$
must stem from the interplay between the shifted magnon band structure and the nonlinear, temperature-dependent Bose distribution function.

This finding persists for
$
    \sigma = \SI{1}{\milli\electronvolt}
$,
where we observe a decrease of the sub- and super-gap contributions compared to
$
    \sigma = 0
$
[Fig.~\ref{fig:dmi_flake_edge_polarization_dirty_eres}(f,~h)], but the in-gap electric polarization remains relatively unchanged with small additional electric moments in the outermost armchair layers [Fig.~\ref{fig:dmi_flake_edge_polarization_dirty_eres}(g)].
For
$
    \sigma = \SI{3}{\milli\electronvolt}
$,
the bearded edge electric polarization is also governed by the in-gap states [Fig.~\ref{fig:dmi_flake_edge_polarization_dirty_eres}(k)] and small contributions arise within the bulk in all energy windows [Fig.~\ref{fig:dmi_flake_edge_polarization_dirty_eres}(j--l)] due to the symmetry breaking by the disorder and the not completely converged ensemble averaging procedure.
The magnitude has been decreased by more than a factor of two for all energy-dependent contributions compared to the clean system [Fig.~\ref{fig:dmi_flake_edge_polarization_dirty_eres}(b--d)], but the overall electric polarization has changed only slightly both quantitatively and qualitatively [cf.~Fig.~\ref{fig:dmi_flake_edge_polarization_dirty_eres}(a,~i)].
Hence, our results do not exhibit a promotion of the in-gap contributions relative to sub-gap and super-gap states.
Therefore, we conclude that the local electric polarization seems to be relatively stable against on-site easy-axis disorder.
We speculate that inter-site interactions that are more closely related to the relative phases of the spin precessions between nearest neighbors might be more effective for manipulating the local electric polarization.

\bibliography{short,refs}